\documentclass{jfm}
\usepackage{graphicx}
\usepackage{subcaption}
\usepackage{epstopdf, epsfig}
\usepackage{amsmath}
\graphicspath{{./Figs/}}

\shorttitle{Instability of a low viscosity jet}
\shortauthor{J. Yang and V. Srinivasan}

\title{Instability of a Low Viscosity Jet Emerging into a High Viscosity Medium: Linear Stability Analysis}

\author{Jinwei Yang\aff{1}
 \and Vinod Srinivasan\aff{1}
   \corresp{\email{vinods@umn.edu}}
}
\affiliation{\aff{1}Department of Mechanical Engineering, University of Minnesota,
Minneapolis, MN 55455, USA
}

\begin{document}

\maketitle
\begin{abstract}
Many natural and engineering systems involve the mixing of two fluid streams, in which the effects of density and viscosity gradients play important roles in determining flow stability. We perform linear stability calculations for a jet emerging into an ambient medium of a different viscosity but the same density. These calculations are intended to isolate the effects of viscosity variation alone. We conduct a systematic study of the effect of ambient-to-jet viscosity ratio, jet Reynolds number and the velocity profile specified by the shear layer thickness, the thickness over which the viscosity change occurs, and radial shifts in velocity profiles, on the growth of axisymmetric and helical modes. Additional terms in the disturbance kinetic energy equation that represent the coupling between the velocity fluctuations and the viscosity field are shown to be responsible for the additional destabilization. Radial shifts in velocity profile that represent real effects likely to be encountered in experiments are shown to be strongly destabilizing. In all cases, the temporal growth rates of axisymmetric and helical mode are very close, except at low Reynolds numbers. Spatio-temporal analysis in the complex wavenumber plane suggests that for sufficiently large ambient viscosity, low-viscosity jets become absolutely unstable. Over a wide range of parameters, two modes of absolute instability exist simultaneously, with an axisymmetric mode predicted to dominate a helical mode. Over a certain narrower space,  the helical mode dominates. The transition boundary for absolute/convective instability is compared with recent experiments, and the results are found to in reasonable agreement for the transition of the helical mode, when velocity profiles  are used that correspond to the similarity solution for development of the boundary layer under a spatially variable viscosity. 
\end{abstract}


\section{Introduction}

Mixing of two fluid streams with variable properties and velocity profiles is encountered in many natural systems, such as the flow of estuaries into an ocean, hydrothermal vents, and atmospheric flows. The round flow configuration is of significant importance in engineering applications such chemical reactors and food processing \citep{Cao2003, Pathikonda2021}. The degree and spatial extent of mixing between the two streams is dependent on the strength and nature of the instabilities that develop at the interface between the two fluid streams, which in turn depends on the controlling parameters characterizing the system, such as the Reynolds number, density and viscosity ratios, and profile shapes.

The near-field of the round jet of a fluid issuing into an ambient of the same fluid is subject to the inviscid Kelvin-Helmholtz instability; linear stability analysis with realistic profiles have agreed well with observed trends \citep{Mattingly1974}, with further improvement when the fluid viscosity is considered in a spatial stability analysis \citep{Morris1976}. Several reviews have been written on the subject of such constant property mixing layers, for instance see \cite{Michalke1984, Ho1984a}. The effects of alterations to the basic velocity profile in such mixing layers can have dramatic effects on their stability characteristics. When a primary flow stream of velocity $U_1$ encounters another stream of velocity $U_2$ in the reverse direction, the flow can become absolutely unstable when the parameter $R = \frac{U_2-U_1}{U_2+U_1}$ is greater than 1.315 \citep{Huerre1985}. The presence of absolute instability implies that disturbances have zero group velocity in the laboratory frame of reference and become more amenable to investigation \citep{Strykowski1991a}. Indeed, the onset of absolute instability in countercurrent shear layers has been well-correlated with the experimental observations of discrete frequencies in the power spectrum (`global modes') of velocity fluctuations. Wall confinement of such shear layers may further destabilize the flow \citep{Juniper2006a, Yang2021}.  

Turning to the effects of variable fluid properties, a significant body of literature addresses the instabilities resulting from the presence of density gradients across a shear layer. Low density jets are found to be absolutely unstable when the jet-to-ambient density ratio is below 0.63 \citep{Sreenivasan1989, Monkewitz1990a}. Compared to the case of planar variable-density mixing layers \citep{Monkewitz1988}, the introduction of an additional length scale (jet diameter) that is not too large relative to the interfacial thicknesses characterizing velocity and density gradients can add significant complexity. For example,  \citet{Monkewitz1990a} demonstrated the existence not only of an interfacial 'shear layer' mode produced by the baroclinic torque associated with the density gradient (Mode II in their nomenclature) but also a jet `column' mode (Mode I) in which disturbances do not decay away from the shear layer, but span the entire diameter of the jet. Over a large parameter space determined by the density ratio, Reynolds number and shear layer thickness parameters, the flow is absolutely unstable. Theoretical arguments have been advanced for the close match between the frequency calculated from local base profiles and the experimentally observed modes \citep{Chomaz1991, Pier2001}.  Further, it has also been shown \citep{Srinivasan2010, Raynal1996} that subtle variations in the alignment of the velocity and density profiles can alter the absolute/convective instability transition boundary significantly.

Comparatively, the effects of viscosity contrast in free shear layers, which are primarily relevant to liquid-liquid mixing, are less well-understood. While viscosity is instinctively considered a stabilizing influence, near solid surfaces it produces velocity gradients and therefore creates the conditions for instability. The destabilizing effects of a viscosity jump at an interface were first studied by \cite{Yih1967} who showed that planar two-layer Couette and Poiseuille flows with immiscible fluid layers were unstable to long waves at any Reynolds number. \citet{Hooper1983a} and \cite{Hooper1985} showed the existence of interfacial instabilities in the short and long wavelength limits, while \citet{Hinch1984} suggested a physical mechanism. Subsequently, most work on instabilities in viscosity-stratified flows has focused on internal pressure-driven flows, and is well-summarized in an exhaustive review \citep{Govindarajan2014b}. Full linear stability analyses have been performed for two-layer immiscible flows in planar Poisueille and Couette configurations \citep{Yiantsios1988, Valluri2010, Mohammadi2017} as well as on core-annular flows in the cylindrical geometry \citep{Hickox1971, Hu1989, Joseph1984, Salin2019}. The mechanism driving the instability was explained by \cite{Boomkamp1997a} as being related to the work done at the interface due to the viscosity jump. \cite{Hallberg2006} note a weak influence of viscosity on the frequency of global modes in low-density jets but the driving mechanism remains inviscid. 

The past two decades have seen substantial attention devoted to the effects of weak diffusion at an interface between two miscible fluids of different viscosity. For a planar channel flow with one fluid sandwiched between two layers of another fluid, Govindarajan and co-workers \citep{Ranganathan2001, Govindarajan2004} showed that stabilization(destabilization) occurred when the more(less) viscous fluid was in the inner region, and attributed the instability to an overlap between the mixing zone and the critical layer, where the phase velocity of the disturbance matches the base velocity. For a core-annular flow, \citet{Selvam2007} showed that  the flow becomes unstable beyond a critical viscosity ratio that is dependent on the radial location of the interface. Further, helical modes are more unstable when the annular fluid is more viscous, while axisymmetric modes are dominant for a viscous core. The core-annular flow also displays a transition from convective to absolute instability for miscible fluids, as a function of interface location, viscosity ratio and inertia. \citet{DOlce2008} observed pearl and mushroom-like instabilities at low Re, and associated the transition between these modes with the absolute-convective instability transition \citep{DOlce2008}. 

Compared to wall-bounded flows discussed above, free shear flows with viscosity gradients have received relatively little attention. The classical analyses on the breakdown of capillary jets emerging into a medium of different viscosity or density \citep{Tomotika1935} pertain to low Reynolds numbers. For the related case of buoyant jets with density and viscosity different from the ambient medium, \citet{Chakravarthy2015} and \citet{Chakravarthy2018} performed both local and global linear stability analysis using realistic temperature, velocity and property profiles and concluded that an axisymmetric `puffing' mode is dominant; this mode is convectively unstable at low density ratios and Richardson numbers, and is  globally unstable at high density ratios and Richardson number. Viscosity variation was incorporated but not studied as a separate parameter. For ambient-to-jet density ratios less than 1.05, the flow is globally stable. It must be noted that there is a large body of work on the instabilities at the sharp interface of a mixing layer of gas and liquid streams \citep{Yecko2002, Boeck2005, Matas2011} but the effects of viscosity alone were not isolated due to the context bring the study of liquid atomization. To the authors knowledge, only \citet{Sahu2014} have studied the interaction of an inflection-point velocity profile with a region of viscosity stratification. Such a configuration is characterized by the presence of two inflection-point profiles and two length scales that capture the gradients in velocity and viscosity. The flow was more unstable when the high-speed flow was in a region of low viscosity, and less unstable for the alternate condition. Only two values: $e^1$ and $e^{-1}$ were considered for the viscosity ratios of the two streams. The striking feature of their findings is that unlike the case of wall-bounded flows, species diffusivity in the form of the Schmidt number Sc plays virtually no role in determining the stability behavior. The response is governed primarily by the viscosity profile, the overlap of the variable-viscosity layer with the momentum shear layer, and modifications to the velocity profile induced by the viscosity profile. The viscosity gradients were shown to alter the transition from convective to absolute instability in countercurrent shear layers, relative to the constant property shear layer. 

The corresponding configuration in a cylindrical geometry, namely the interaction of a jet velocity profile with an ambient medium of a different viscosity has not been studied. As with low-density jets, the introduction of an additional length scale (jet diameter) near a shear layer can potentially lead to new unstable modes that are a function of the viscosity ratio, profile shapes and Reynolds number. Another question that arises is whether absolute instability can be triggered in round jets which are weakly miscible with the ambient medium. Knowledge of the transition boundary would have significant implications, including the potential for enhanced mixing in engineering applications. Recent experiments in the authors' laboratory have shown the transition from axisymmetric instabilities to helical instabilities as the ambient-to-jet viscosity ratio is increased beyond a critical Reynolds number-dependent value, accompanied by discrete peaks in the frequency spectrum of velocity fluctuations \citep{Srinivasan2023}. Presumably, the onset of these helical modes also depends on other parameters such as the boundary layer thickness, Schmidt number and degree of diffusion, and Reynolds number. The experiments were performed for a single nozzle geometry, and therefore a fixed relationship between the jet Reynolds number and the boundary layer thickness. However, it is not clear whether these observations can be ascribed to global modes corresponding to absolute instability of profiles in the jet near-field, or whether they are fast-growing convective modes detected against a low level of background noise. The lack of theoretical understanding for this configuration further precludes interpretation. Therefore, a preliminary stability analysis needs to be performed first, before exploring the possibility of absolute instability. This is the focus of the present study.

This paper is organized as follows: Section 2 presents the formulation of an idealized configuration of a viscosity-stratified jet, along with assumed base profiles. Section 3 discusses the linearized stability equations, elements of the numerical solution procedure and code validation results. Section 4 presents the results of a temporal stability analysis for axisymmetric and helical modes that are triggered by a weakly diffusive interface with sharp viscosity gradients, for both high and low viscosity jets. Section 5 discusses additional complexities that may be encountered in practical applications, such as deviations from idealized base profiles, presence of walls, and entraining flow. Finally section 6 presents a summary and conclusions from the results.  

\section{Problem Formulation}
\begin{figure}
\centering
\includegraphics[height=3.0in]{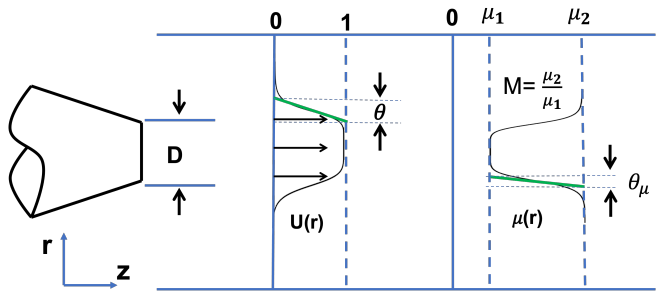}
\caption{Sketch of the flow configuration analyzed: a jet of viscosity $\mu _1$ emerging into an ambient medium of viscosity $\mu _2$, setting up velocity and concentration gradient of thickness $\theta$ and $\theta _\mu$ respectively.  }
\label{fig:jet_sketch}
\end{figure}

We examine the linear stability characteristics of a round jet of one fluid emerging into an ambient medium of the same density but a different viscosity. Both fluids are assumed to be Newtonian and incompressible. The miscibility of the fluids causes a mixing layer to develop around the periphery of the jet, thereby creating three layers: the jet core, the intermediate axisymmetric mixing layer, and an outer region where the velocity profile is the result of entrainment. A schematic diagram of the jet is shown in Fig.~\ref{fig:jet_sketch}. The viscosity of the core fluid $(0<r<1)$ and the annular fluid $(1<r)$ are denoted by $\mu_1$ and $\mu_2$, respectively. 

\subsection{Base Profiles}
For the conventional case of a jet of fluid emerging into an ambient of the same fluid, the near-field axial velocity profiles have often been modeled using tanh-type functions. For the variable viscosity case, several effects need to be accounted for in order to develop realistic profiles. First, we recognize that unlike the planar shear layer case (see \citet{Sahu2014}) no similarity solution exists for the near-field of a jet, and model profiles have to be assumed. A diffusive interface develops downstream of the jet exit, leading to a gradient in concentration of fluid 1 going from a value of unity at the centerline to zero at infinity. The concentration-dependent viscosity profile in the radial direction will also induce changes in the velocity distribution, altering it from the standard tanh-shape. Further, one can expect that when the ambient viscosity is very large, it will retard the fluid in the jet periphery to a certain extent, creating no-slip in the limit of infinite ambient viscosity. We incorporate these ideas into our model profiles, as discussed below. 

The viscosity gradient is assumed to depend on the concentration $c(r)$ of species 2 of the ambient fluid  into the jet fluid (species 1). In specifying a constitutive relation between viscosity and concentration, we follow earlier works by  \cite{tan1986stability},  \cite{goyal2006miscible}, and \cite{selvam2007stability}, and assume the viscosity $\mu$ to be an exponential function of the concentration.  We write 
\begin{equation}
\mu (r)  = \mu _1e^{m c(r)}
\end{equation}
where the concentration varies from a value of 0 along the jet axis (r=0) to a value of 1 at infinity. We further define the viscosity ratio M as 
\begin{equation}
    M=e^m=\frac{\mu _2}{\mu _1}
\end{equation}

The radial profile of concentration  can be approximated as a tanh-profile:          
\begin{equation}
c(r) = \frac{1+tanh[\frac{1}{4\theta _\mu}(r-\frac{1}{r})]}{2}. 
\end{equation}
Here $\theta _\mu$ is a measure of the thickness of the diffusion layer. The velocity profile for the jet core is one of the two profiles used by \cite{Mattingly1974}--- the profile normalized on the centerline velocity $U_c$  can be expressed as
\begin{equation}
\label{eq:velocityprofile}
\frac{U(r)}{U_c} = \frac{1+\tanh\left[\frac{1}{4\theta}(\frac{1}{r}-r)\right]}{2},
\end{equation}
where $U(r)$ is normalized using the jet centerline velocity $U_c$, and $\theta$ is the shear layer momentum thickness evaluated as follows: 
\begin{equation}
\label{theta}
\theta =  \int_{0}^{\infty} \frac{U(r)-U_{\infty}}{U_c-U_{\infty}}[1-\frac{U(r)
-U_{\infty}}{U_c-U_{\infty}}]dr.
\end{equation}

Note that r=1 in the velocity profile corresponds the radial location at which the velocity falls to half its centerline value, and thus the length scale used for non-dimensionalization is the jet radius.  Thus, the radius D/2 of the jet is used as the length scale for non-dimensionalization. The jet Reynolds number is based on this length scale and the centerline values of velocity and viscosity:
\begin{equation}
Re =\frac{\rho U_cD}{2\mu _1}
\end{equation}

For weak diffusion (large Sc), the intermediate layer with a gradient in species concentration is thin, often smaller by an order of magnitude relative to the velocity shear layer. Without recourse to the full solution of the Navier-Stokes equations for specific initial and boundary conditions, one has to choose appropriate modifications to the inviscid profile of \cite{Mattingly1974} that reflect the presence of the viscosity gradient. In this study, we assume that as we move radially away from the jet potential core. the velocity profile varies such that the shear stress is preserved at locations on either side of the concentration shear layer, defined as the region $1-\theta _\mu/2 < r < 1-\theta +\mu/2$.
\begin{equation}
    \mu \frac{dU}{dr}|_ {r=1-\theta _\mu/2} =     \mu \frac{dU}{dr}|_ {r=1 +\theta _\mu/2}
\end{equation}

\begin{figure}
\centering
\begin{subfigure}{0.49\textwidth}
\includegraphics[width= \textwidth]{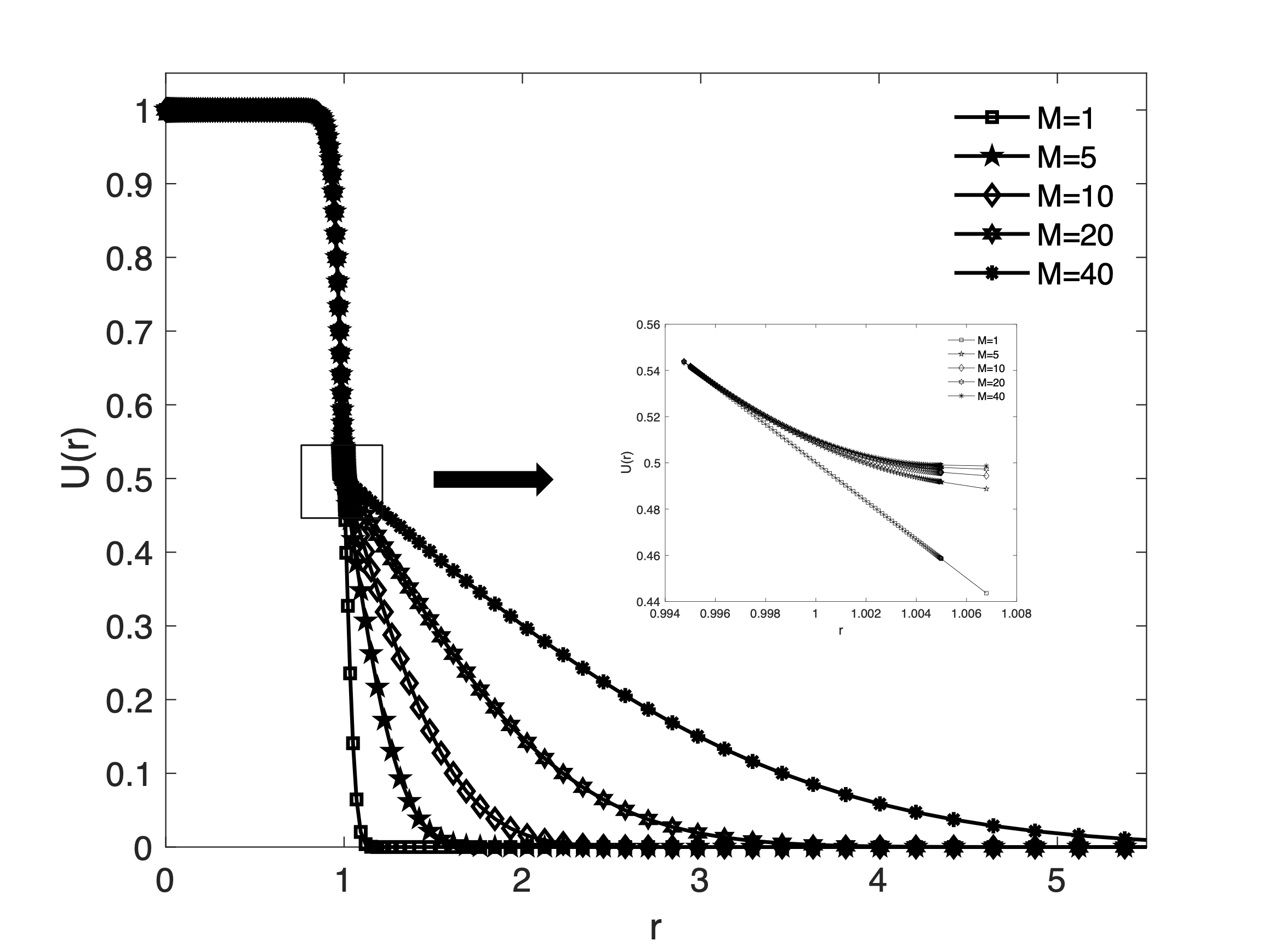}
\caption{}
\end{subfigure}
\begin{subfigure}{0.49\textwidth}
\includegraphics[width= \textwidth]{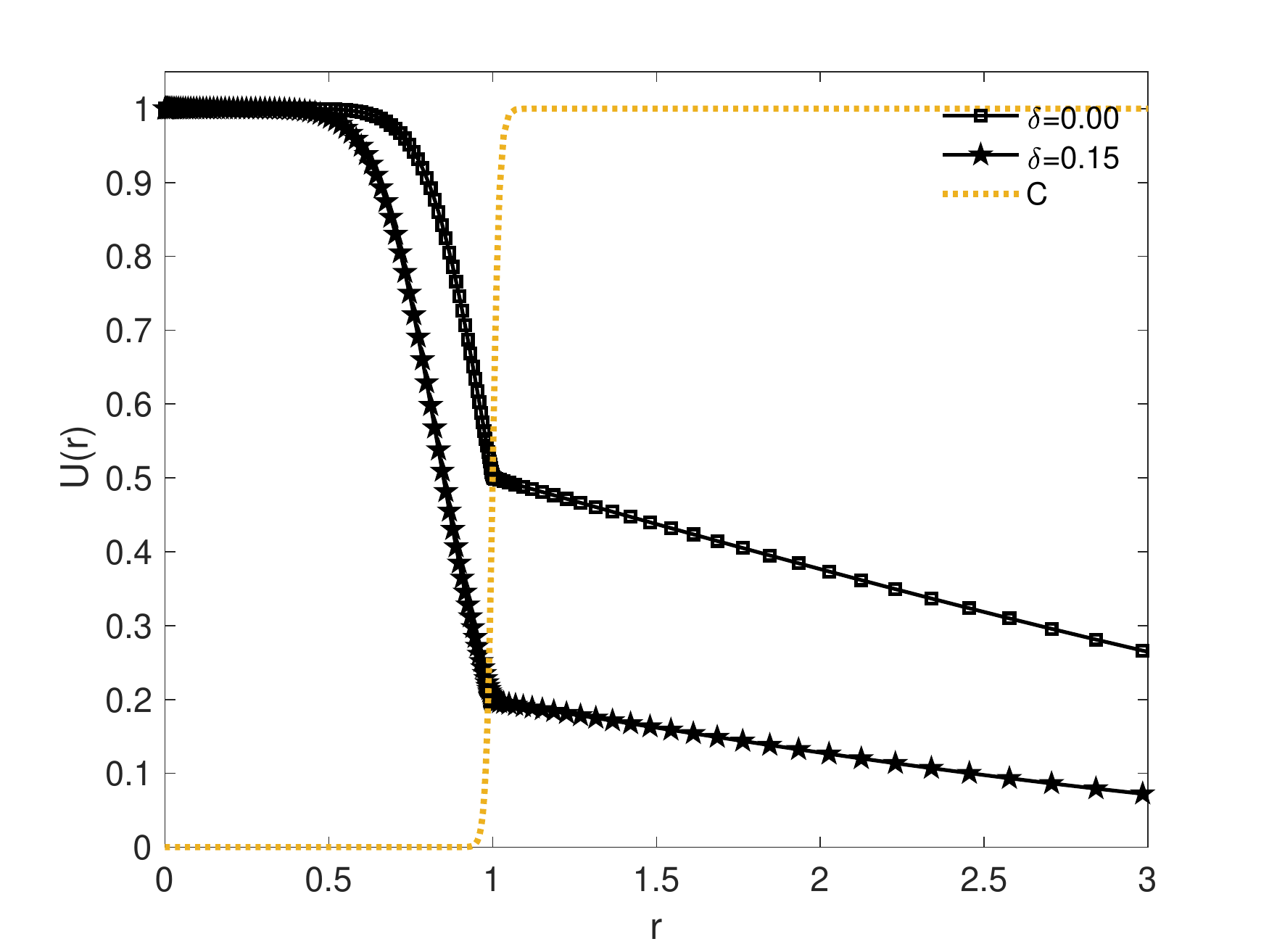}
\caption{}
\end{subfigure}
\caption{ (a) Velocity profiles for different values of viscosity ratio M when momentum thickness $\theta=0.1$ and concentration layer thickness is $\theta_\mu=0.01$; (inset) a magnified view of the velocity profiles in the shear layer (b) Sketch of viscosity profile (dashed line)  as well as unshifted (squares) and shifted (asterisks) velocity  profiles.}
\label{fig:baseprofiles}
\end{figure}

At the inner radial edge of the concentration layer, $r = 1 - \theta _\mu/2$, the velocity and its derivative are known from the core profile enforcing the above equation leads to knowledge of the derivative at the outer radial edge. These three conditions were used to generate a quadratic fit for the velocity profile in the mixing layer. For M=1, this results in the standard tanh- profile with a near linear drop in velocity across the layer. For other values of M, the transmission of shear stress across the layer leads to higher velocities at the outer edge of the concentration layer than for M=1, as shown in the inset in Fig.~\ref{fig:baseprofiles}(a).  We also tested a more stringent technique of enforcing shear stress continuity across the concentration layer to obtain the velocity profile without any curve fitting; this did not lead to appreciable difference for low M but led to substantially less differentiable profiles for the velocity derivative at higher M, and this approach was discarded in favor of the curve fits. At the outer edge of the concentration gradient region, knowledge of the axial velocity and its derivative allows us to calculate an entrained flow velocity in the outer region which satisfies an error-function type decay:  
\begin{equation}
    U(r) = A\left.[1-\text{erf}(B(r-R_1))\right.];\hspace{20pt} r> 1 + \theta _\mu/2
\end{equation}
where A and B are determined by the continuity of the velocity and its derivative at $r =1 + \theta _\mu/2$.
Figure ~\ref{fig:baseprofiles}(a) plots velocity profiles for different values of viscosity ratio M for fixed valued of $\theta$ and $\theta _\mu$. 

Finally, we consider that the dimensionless value of the velocity at r=1 is dependent on the ambient viscosity. It is likely that in experimental realizations of such a configuration, a jet emerging into an ambient with higher viscosity ($M>1$) will have an interfacial velocity that differs from eqn ~\ref{eq:velocityprofile} above, and may have values lower than 0.5 at r=1. This can be modeled by shifting the velocity profile inwards by an amount $\delta$ in eqn. , i.e. by replacing $r$ with $r-\delta$. Such profiles are shown in Fig.~\ref{fig:baseprofiles}(b) for $\theta = 0.1$ and $\theta _\mu = 0.01$. We note that in following the process of generating parametrized base profiles, one may end up with test cases that are physically hard to realize, and may need some discretion in interpretation.

To summarize, the relevant parameters governing the stability of the system are: Re, M, Sc, $\theta$ and $\theta _\mu$ and $\delta$.

\subsection{Linear stability equations}

We employ the incompressible Navier-Stokes equations, along with a species transport equation to describe the jet flow
\begin{equation}\label{NS}
\begin{array}{c}
\nabla \cdot v=0\\
\rho\left(\frac{\partial v}{\partial t}+v \cdot \nabla v\right)=-\nabla p+\nabla \cdot \tau \\
\frac{\partial c}{\partial t}+v \cdot \nabla c=\kappa \nabla^{2}c
\end{array} 
\end{equation}
where $v=(v_r,v_{\theta},v_z)$ denotes the flow velocity, $\tau=\mu(\nabla v + \nabla v^T)$ is the viscous stress tensor, $c$ is the concentration of the jet flow and $\kappa$ denotes the binary diffusion coefficient. 

We chose the radius of the jet $R_0$ as the characteristic length, and the viscosity and velocity in the center-line of the jet as the characteristic viscosity and velocity, respectively. 

Assuming that the flow is nearly parallel, we can examine the stability of the flow to two-dimensional disturbances, assumed to be in the form of traveling waves. Using the standard normal mode analysis with perturbations of infinitesimal amplitude, we admit the possibility of axisymmetric and helical modes with wave number $k$ and $\beta$ as shown below:
\begin{equation}\label{pert}
\left(\begin{array}{l}
v_{r} \\
v_{\theta} \\
v_{z} \\
p \\
c
\end{array}\right)(r, \theta, z, t)=\left(\begin{array}{c}
0 \\
0 \\
\bar{v}_{z}(r) \\
\bar{p}(z) \\
\bar{c}(r)
\end{array}\right)+\left(\begin{array}{c}
i \hat{v}_{r}(r) \\
\hat{v}_{\theta}(r) \\
\hat{v}_{z}(r) \\
\hat{p}(r) \\
\hat{c}(r)
\end{array}\right) \mathrm{e}^{\mathrm{i}(k z+\beta \theta-\omega t)} .
\end{equation}
where, $\omega$ s the wave angular frequency. 

Substituting the equation~(\ref{pert}) into the equations~(\ref{NS}) and linearizing around the base state yields the following perturbation governing equations:
\begin{equation}
\begin{aligned}
\label{linearEQ}
\frac{\mathrm{d} \hat{v}_{r}}{\mathrm{~d} r}+\frac{\hat{v}_{r}}{r}+\frac{\beta \hat{v}_{\theta}}{r}+k \hat{v}_{z}=& 0,\\
\operatorname{Re}\left[-\omega \hat{v}_{r}+k \bar{v}_{z} \hat{v}_{r}\right]=& \frac{\mathrm{d} \hat{p}}{\mathrm{~d} r}-\mathrm{ie}^{M \bar{c}}\left[\frac{\mathrm{d}^{2} \hat{v}_{r}}{\mathrm{~d} r^{2}}+\frac{1}{r} \frac{\mathrm{d} \hat{v}_{r}}{\mathrm{~d} r}-\left(\frac{\beta^{2}+1}{r^{2}}+k^{2}\right) \hat{v}_{r}\right.\\
&\left.-\frac{2 \beta}{r^{2}} \hat{v}_{\theta}+2 M \frac{\mathrm{d} \bar{c}}{\mathrm{~d} r} \frac{\mathrm{d} \hat{v}_{r}}{\mathrm{~d} r}+M k \frac{\mathrm{d} \bar{v}_{z}}{\mathrm{~d} r} \hat{c}\right] \\
\operatorname{Re}\left[-\omega \hat{v}_{\theta}+k \bar{v}_{z} \hat{v}_{\theta}\right]=& \frac{-\beta \hat{p}}{r}-\mathrm{ie}^{M \bar{c}}\left[\frac{\mathrm{d}^{2} \hat{v}_{\theta}}{\mathrm{d} r^{2}}+\frac{1}{r} \frac{\mathrm{d} \hat{v}_{\theta}}{\mathrm{d} r}-\left(\frac{\beta^{2}+1}{r^{2}}+k^{2}\right) \hat{v}_{\theta}\right.\\
&\left.-\frac{2 \beta}{r^{2}} \hat{v}_{r}+M \frac{\mathrm{d} \bar{c}}{\mathrm{~d} r}\left(\frac{\mathrm{d} \hat{v}_{\theta}}{\mathrm{d} r}-\frac{\hat{v}_{\theta}}{r}-\frac{\beta \hat{v}_{r}}{r}\right)\right] \\
\operatorname{Re}\left[-\omega \hat{v}_{z}+k \bar{v}_{z} \hat{v}_{z}+\frac{\mathrm{d} \bar{v}_{z}}{\mathrm{~d} r} \hat{v}_{r}\right]=&-k \hat{p}-\mathrm{ie}{ }^{M \bar{c}}\left[\frac{\mathrm{d}^{2} \hat{v}_{z}}{\mathrm{~d} r^{2}}+\frac{1}{r} \frac{\mathrm{d} \hat{v}_{z}}{\mathrm{~d} r}-\left(\frac{\beta^{2}}{r^{2}}+k^{2}\right) \hat{v}_{z}\right.\\
&+M \frac{\mathrm{d} \bar{c}}{\mathrm{~d} r}\left(\frac{\mathrm{d} \hat{v}_{z}}{\mathrm{~d} r}-k \hat{v}_{r}\right)+M \frac{\mathrm{d} \bar{v}_{z}}{\mathrm{~d} r} \frac{\mathrm{d} \hat{c}}{\mathrm{~d} r}\\
&\left.+M \hat{c}\left(\frac{\mathrm{d}^{2} \bar{v}_{z}}{\mathrm{~d} r^{2}}+\frac{1}{r} \frac{\mathrm{d} \bar{v}_{z}}{\mathrm{~d} r}+M \frac{\mathrm{d} \bar{c}}{\mathrm{~d} r} \frac{\mathrm{d} \bar{v}_{z}}{\mathrm{~d} r}\right)\right]\\
P e\left[-\omega \hat{c}+k \bar{v}_{z} \hat{c}+\frac{\mathrm{d} \bar{c}}{\mathrm{~d} r} \hat{v}_{r}\right]=&-\mathrm{i}\left[\frac{\mathrm{d}^{2} \hat{c}}{\mathrm{~d} r^{2}}+\frac{1\mathrm{~d} \hat{c}}{\mathrm{~d}r}-\left(\frac{\beta^{2}}{r^{2}}+k^{2}\right) \hat{c}\right].
\end{aligned}
\end{equation}
The boundary conditions for the above problem are as follows. For both cases we require all the velocity and the concentration to vanish in the far-field ($r\rightarrow \infty)$. At the center-line of the jet, we consider the single-valuedness of velocity, together with continuity to derive the center-line conditions for different values of azimuthal wavenumber $\beta$  (\cite{khorrami1989application}):

\begin{equation}
\begin{aligned}
\label{boundar_conds}
&\beta=0: \quad \frac{\mathrm{d} \hat{v}_{z}}{\mathrm{~d} r}=0, \quad \hat{v}_{r}=0, \quad \hat{v}_{\theta}=0, \quad \frac{\mathrm{d} \hat{p}}{\mathrm{~d} r}=0, \quad \frac{\mathrm{d} \hat{c}}{\mathrm{~d} r}=0,\\
&\beta=1: \quad \hat{v}_{z}=0, \quad \hat{v}_{r}+\hat{v}_{\theta}=0, \quad 2 \frac{\mathrm{d} \hat{v}_{r}}{\mathrm{~d} r}+\frac{\mathrm{d} \hat{v}_{\theta}}{\mathrm{d} r}=0, \quad \hat{p}=0, \quad \hat{c}=0,\\
&\beta \geqslant 2: \quad \hat{v}_{z}=0, \quad \hat{v}_{r}=0, \quad \hat{v}_{\theta}=0, \quad \hat{p}=0, \quad \hat{c}=0.
\end{aligned}
\end{equation}

Together, the above equations constitute an eigenvalue problem, with the angular frequency representing the eigen value while the velocity and concentration disturbances are the eigen functions. In other words, these equations represent the dispersion relation: 
\begin{equation}
    D(\omega, k, Re, M, Sc, \theta , \theta _\mu , \delta ) =0
\end{equation}

\section{Numerical Solution Procedure and Validation}
\subsection{Chebyshev Spectral Method with Overlapping Domains}

Since gas viscosity does not vary appreciably as a function of species alone, experimental realization of viscosity contrast will rely on liquid flows, which are characterized by high values of the Schmidt number. As a consequence, the concentration layer is likely to be much thinner than the velocity boundary layer thickness. Chebyshev collocation techniques are often preferred for studying shear flows due the exponential convergence rate when approximating polynomial functions, and a high degree of algebraic convergence when weak discontinuities are present. The Chebyshev space [-1,1] is intrinsically advantageous for internal flows, since the Chebyshev modes are clustered towards the boundaries. where the steepest velocity gradients occur. Conventionally, stability analyses of jet flows have been performed using Chebyshev collocation using a suitable single mapping function (see, for example, \citet{Lesshafft2007}) for transforming from the physical domain to the domain [-1, 1]. This implementation is now available in MATLAB as the $chebfun$ set of routines \citep{trefethen2000spectral}. However, we found that such mapping functions did not yield a sufficient number of points in the intermediate mixing region with the largest gradient, especially when the width of this layer was 1\% of the radius. The $chebfun$ routines offer the ability to split domains and have exponentially-convergent Chebyshev interpolation  over each sub-domain, but this causes issues with smoothness of higher order derivatives, which leads to oscillations in the solution \citep{Driscoll2014}. In this study, we adopt a newly-introduced technique of domain-overlapping \citep{Aiton2018}, which offers a compromise between the need for a very high number of grid points in the single domain/single mapping method and the domain splitting technique of $chebfun$ which requires fewer points but suffers from poor convergence. For the present calculations, we typically required 200 polynomials for accurate representation of the eigenfunction solutions. Further details are given in the Appendix. 

\subsection{Code Validation}\textbf{}
We begin the process of verifying our code for the viscosity-stratified case by first comparing predictions the critical Reynolds number and critical wave number and frequency for a uniform-viscosity jet with the spatial instability results of \citet{Morris1976}. Morris studied three velocity profiles: a self-similar profile corresponding to downstream conditions, and two profiles in the near-field. Results are presented here for his `Profile III' which corresponds to the base state in eqn. \ref{eq:velocityprofile}; another profile (`Profile II') will be discussed subsequently in the context of absolute instability. For a shear layer thickness of $\theta =0.16$, the critical Reynolds number $Re_c$,  the corresponding wavenumber $k$  and the frequency $\omega$ are calculated and tabulated in Table 1 for the axisymmetric mode ($\beta =0$) and the helical mode ($\beta =1$).

\begin{figure}
\centering
\includegraphics[height=2.5in]{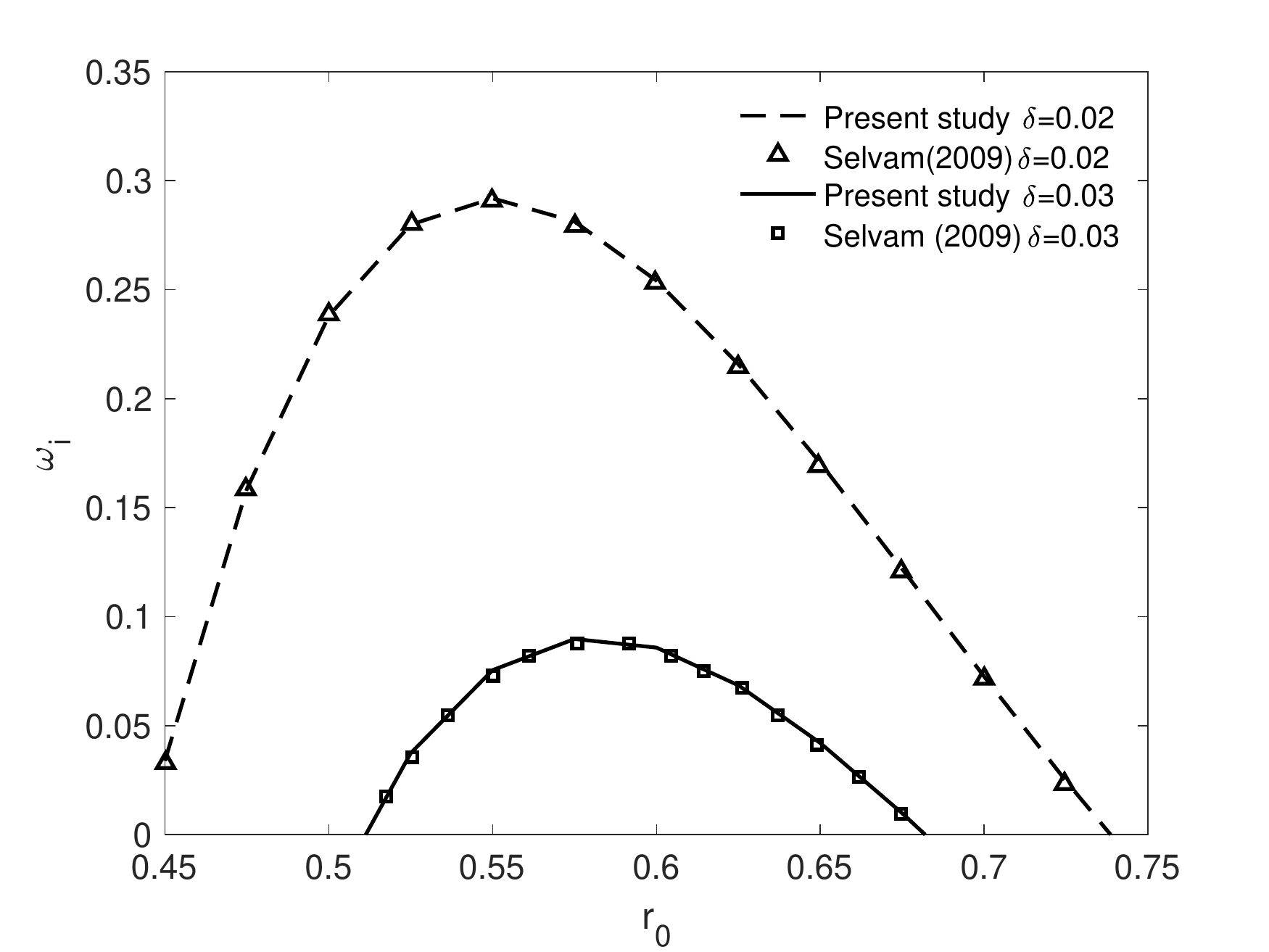}
\caption{Comparison of our calculations with the core-annular pipe flow calculations in Fig.3(a) of \citet{Selvam2009} for the axisymmetric mode. The abscissa $r_0$ corresponds to various locations of the diffusive interface of width $\delta$, for the parameter matrix $(M,Re,Sc,\delta^*)=(25,48,7500,[0.02,0.03])$.  }
\label{M25_Validation}
\end{figure}

For validating the code for situations with viscosity variation due to miscibility of fluids, we replicate the results of \citet{Selvam2009} for miscible core-annular flow in a circular duct. Base profiles corresponding to the laminar quadratic velocities used in that study were employed, along with no-slip conditions at r=1. In Fig. \ref{M25_Validation}, the $R_i$ denotes the location of the diffusive interface normalized by the pipe radius. The parameter $\delta^*= 0.01$ denotes the thickness of the layer with a viscosity gradient. The Schmidt number value is $Sc = 7500$ and the Reynolds number is $Re=48$. The growth rates of the axisymmetric mode agree well with their results.

\begin{table}
\centering
\begin{tabular}{c c c c c c} 
 \hline
 $\beta$ &$Re_c$ & $k $ & $\theta$ &  $\omega $ (Morris1976) & $ \omega $(Present) \\ \hline
0& 55.3125 & 1.0281 & 0.16 & 0.8275 & 0.8277 \\ 
 1&21.7500 & 0.5713 & 0.16 & 0.2181 & 0.2182\\ \hline
\end{tabular}
\caption{Comparison of wavenumber and frequency of the axisymmetric and helical modes at the critical Reynolds numbers for M-1 with the spatial instability analysis results of \cite{Morris1976}.}
\label{table:1}
\end{table}

\section{Results}
\subsection{Temporal instability analysis of low viscosity jets ($M > 1$)}
\begin{figure}
\centering
\begin{subfigure}{0.49\textwidth}
\includegraphics[width= \textwidth]{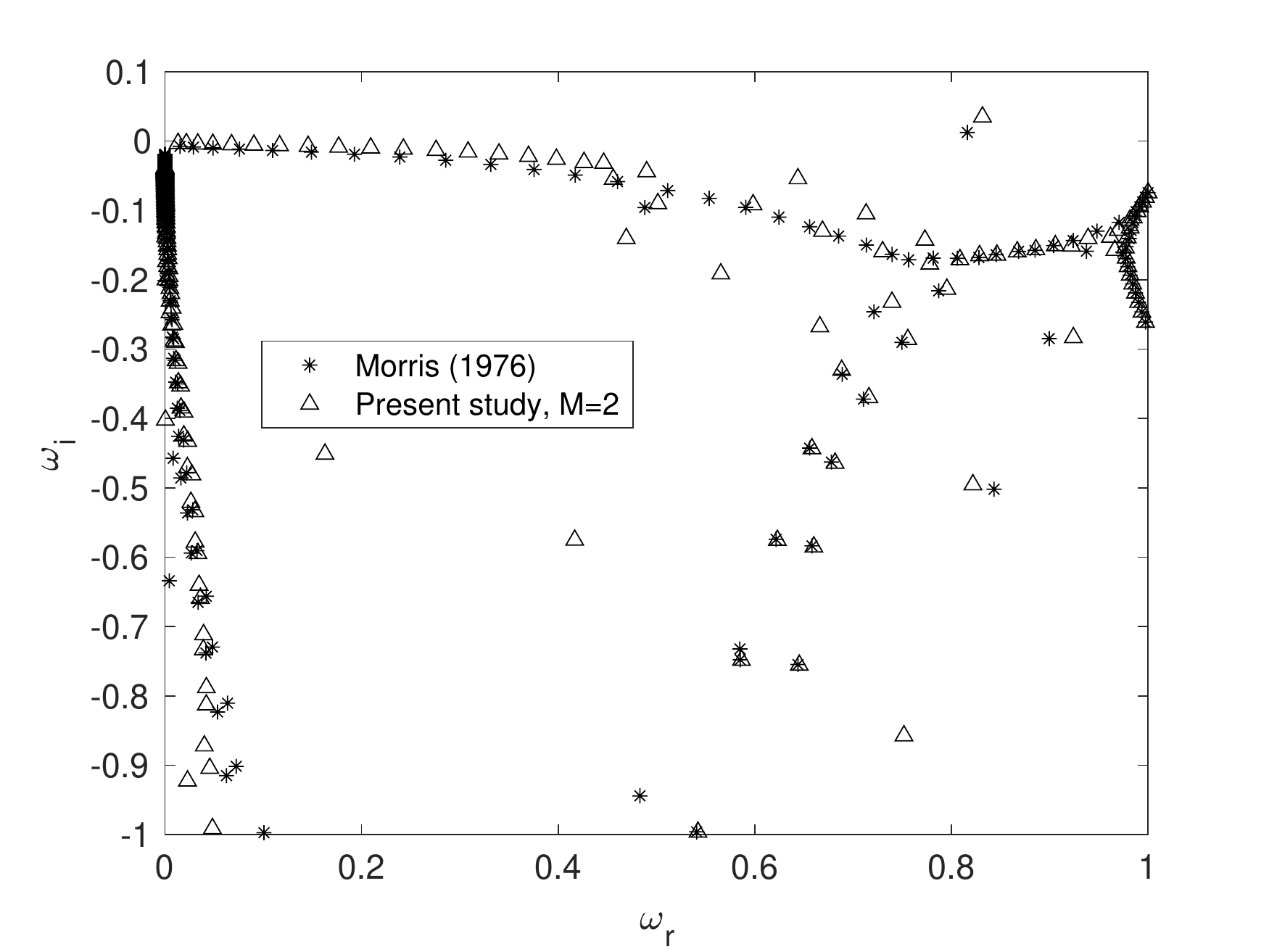}
\caption{}
\end{subfigure}
\begin{subfigure}{0.49\textwidth}
\includegraphics[width= \textwidth]{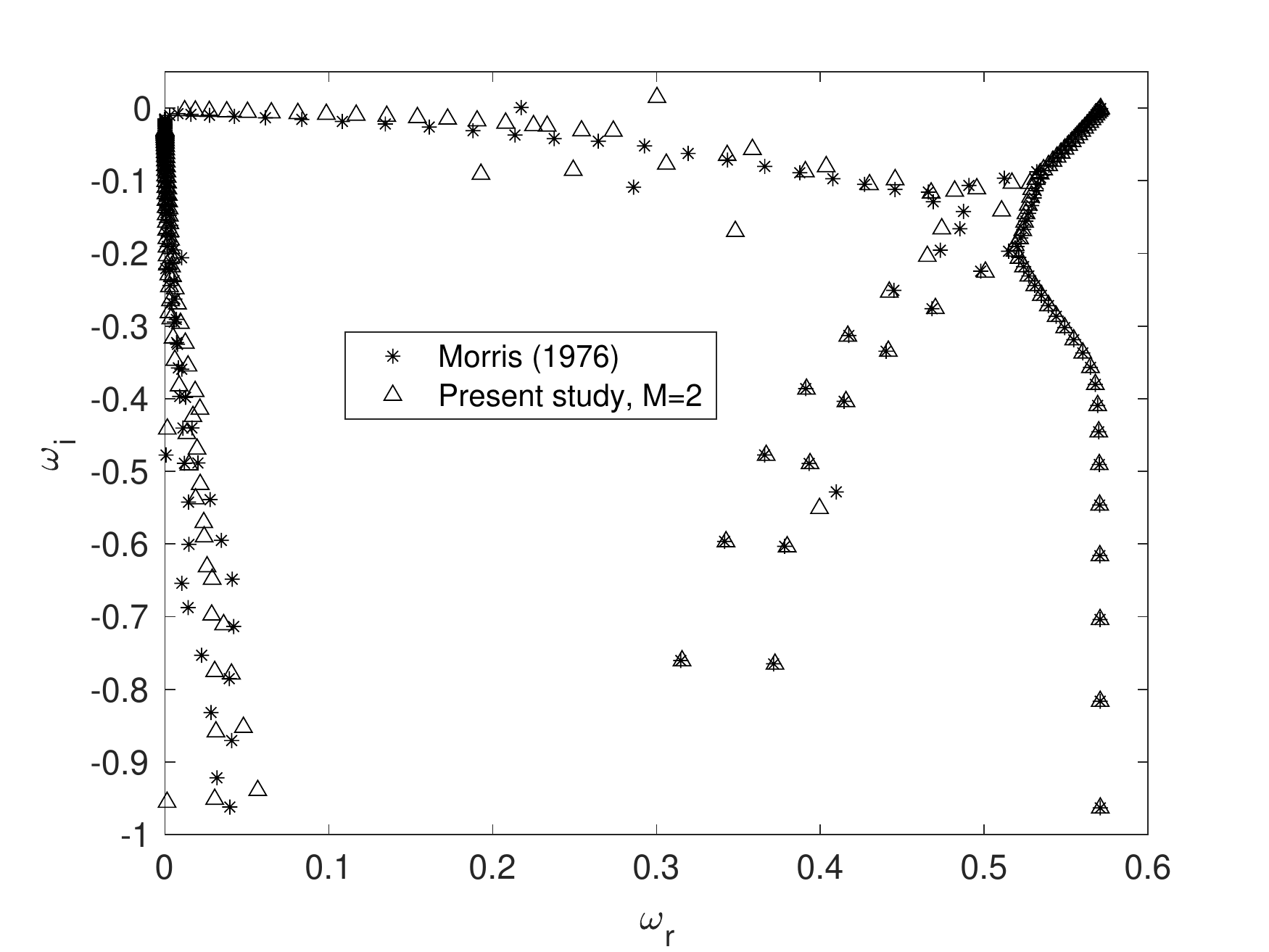}
\caption{}
\end{subfigure}
\caption{Spectra of eigen values for a round jet emerging into an ambient with viscosity ratio M, and with the properties $(Sc, \theta, \theta_\mu)=(100, 0.16, 0.01)$.   (i) Axisymmetric mode: $(Re,\beta,k)=(55.312,0,1.028)$, (ii) Helical mode $(Re,\beta,k)=(21.75,1,0.571)$. The M=1 case corresponds to the critical conditions identified by \citet{Morris1976}.}
\label{fig:spectrum}
\end{figure}

We start by examining the spectrum of eigenvalues for a jet that emerges into an ambient with a viscosity not far from unity, M=2, and compare it with the spectrum for a constant-viscosity jet (M=1) at critical conditions, as calculated by \citet{Morris1976}. Figure ~\ref{fig:spectrum} shows that for $\theta =0.16$, the change in viscosity from M=1 to M=2 does not produce any additional branches in the spectrum. There exists only one unstable mode ($\omega _i >0$)for both M=1 and M=2, suggesting that the M=1 mode can be interpreted as a specific case of a more general viscosity-stratified mode.

\begin{figure}
\centering
\begin{subfigure}{0.49\textwidth}
\includegraphics[width= \textwidth]{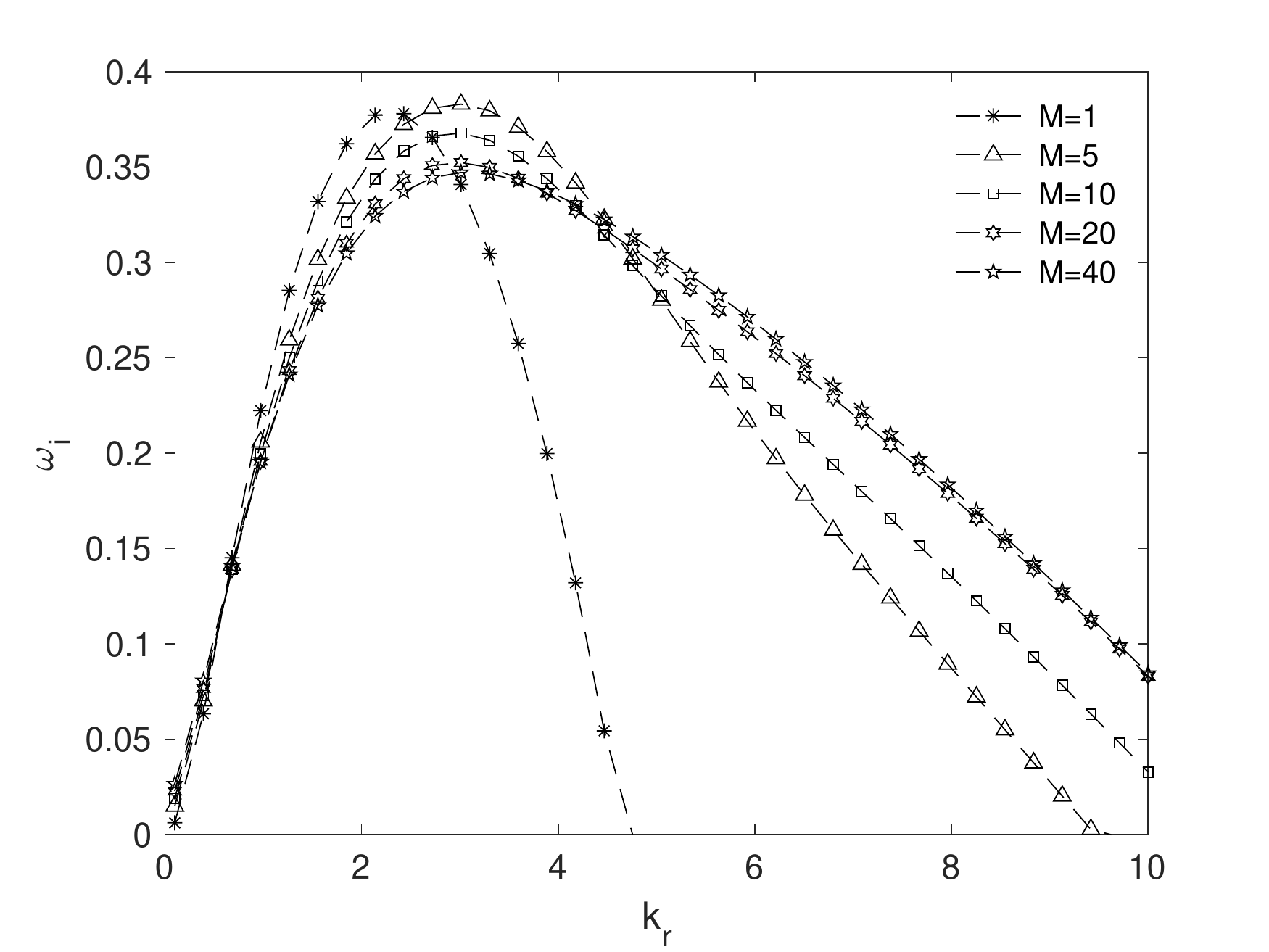}
\caption{}
\end{subfigure}
\begin{subfigure}{0.49\textwidth}
\includegraphics[width= \textwidth]{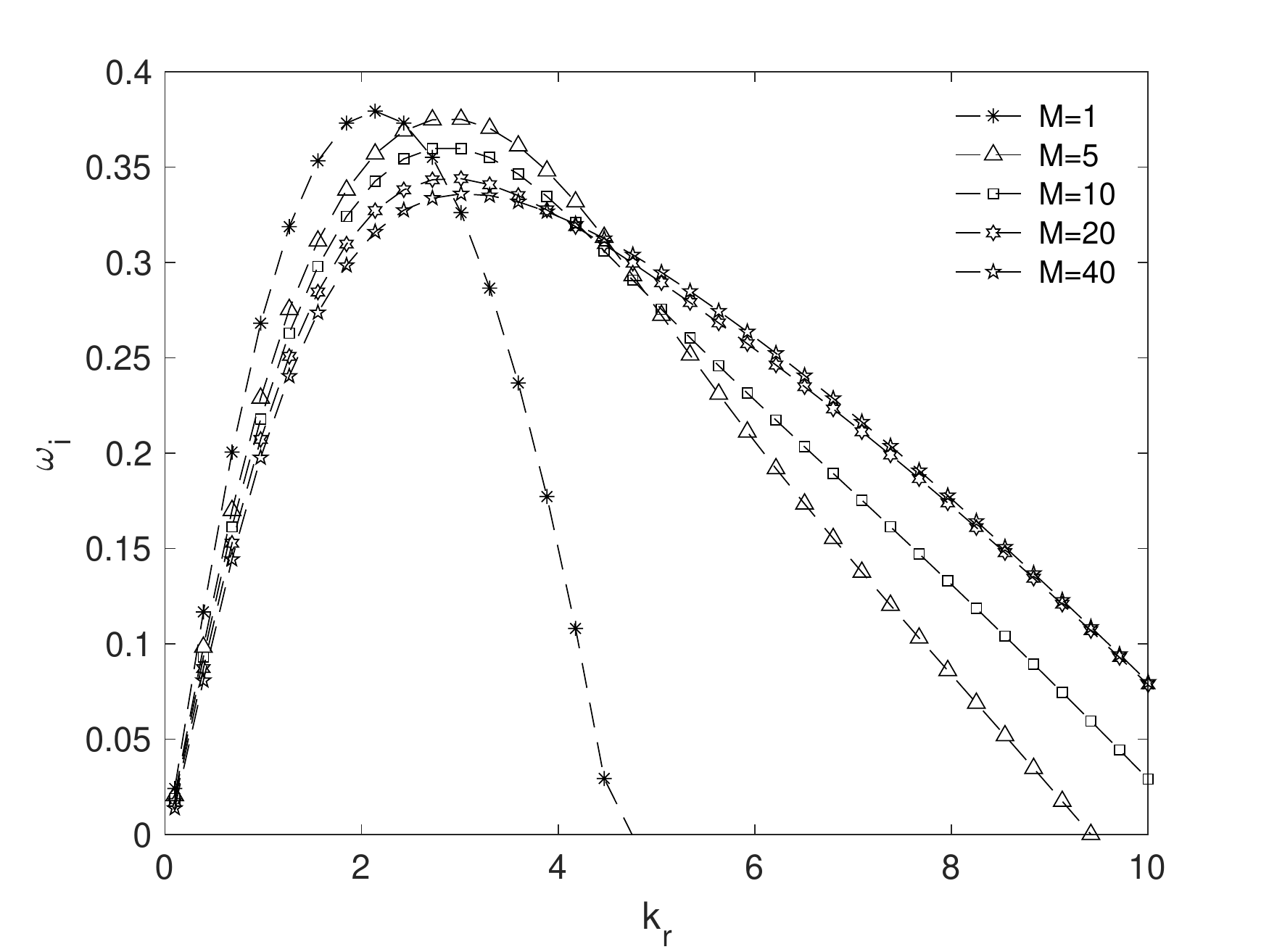}
\caption{}
\end{subfigure}
\caption{ Temporal growth rates for low viscosity jets ($M > 1$) for a round jet with $(Re,Sc,\theta,\theta_\mu)=(1000,100,0.1,0.01)$. (a) The axisymmetric mode, $\beta =0$ and (b) the helical mode, $\beta =1$.}
\label{fig:wi_vs_k_M}
\end{figure}

We now consider in Fig. ~\ref{fig:wi_vs_k_M} the temporal stability of a family of jet profiles specified by constant momentum thickness, $\theta =0.1$ and constant viscosity-gradient thickness $\theta _\mu=0.01$, for a Reynolds number of 1000 and a Schmidt number of Sc=100. Overall, for all viscosity ratios in the range $1 < M <40$, the temporal growth rates of the axisymmetric ($\beta =0$) and helical modes ($\beta =1$) are surprisingly close, with the axisymmetric mode being slightly more unstable. This has been noted previously for constant viscosity jets by other investigators \citep{Morris1976} and offered as a possible explanation for why the breakdown of jets in laboratory experiments (see, for example \citet{Mattingly1974}) appear to start out as axisymmetric disturbances, before progressing to modes with distinct helicity far downstream. Note that there is a distinct difference in the growth rates at lower Reynolds numbers, as exemplified by the difference in critical Reynolds number for the two modes. We shall later see that there is also a strong difference in growth rates of the two modes when conditions admit absolute instability. For the present temporal instability calculations of the two modes, the case of M=1 appears more unstable at small wavenumbers; however, the range of unstable wavenumbers is much lower for M=1 compared to larger M. Both axisymmetric and helical modes remain unstable at larger wavenumbers (short wavelengths) as the viscosity ratio is increased. The fastest growing modes remain those with a wavelength approximately equal to that of the jet diameter ($k \approx 3$).

\begin{figure}
\centering
\begin{subfigure}{0.49\textwidth}
\includegraphics[width= \textwidth]{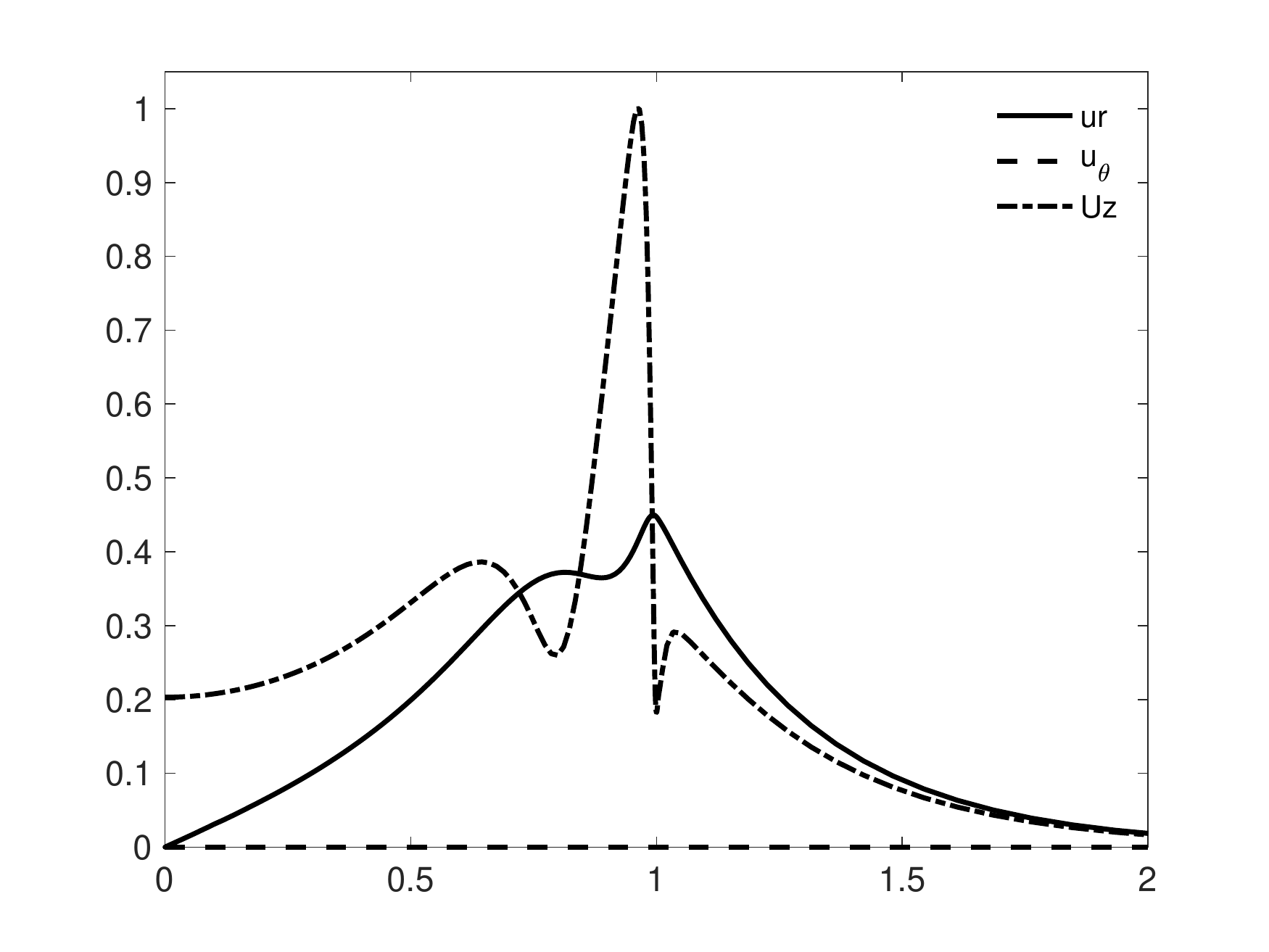}
\caption{}
\end{subfigure}
\begin{subfigure}{0.49\textwidth}
\includegraphics[width= \textwidth]{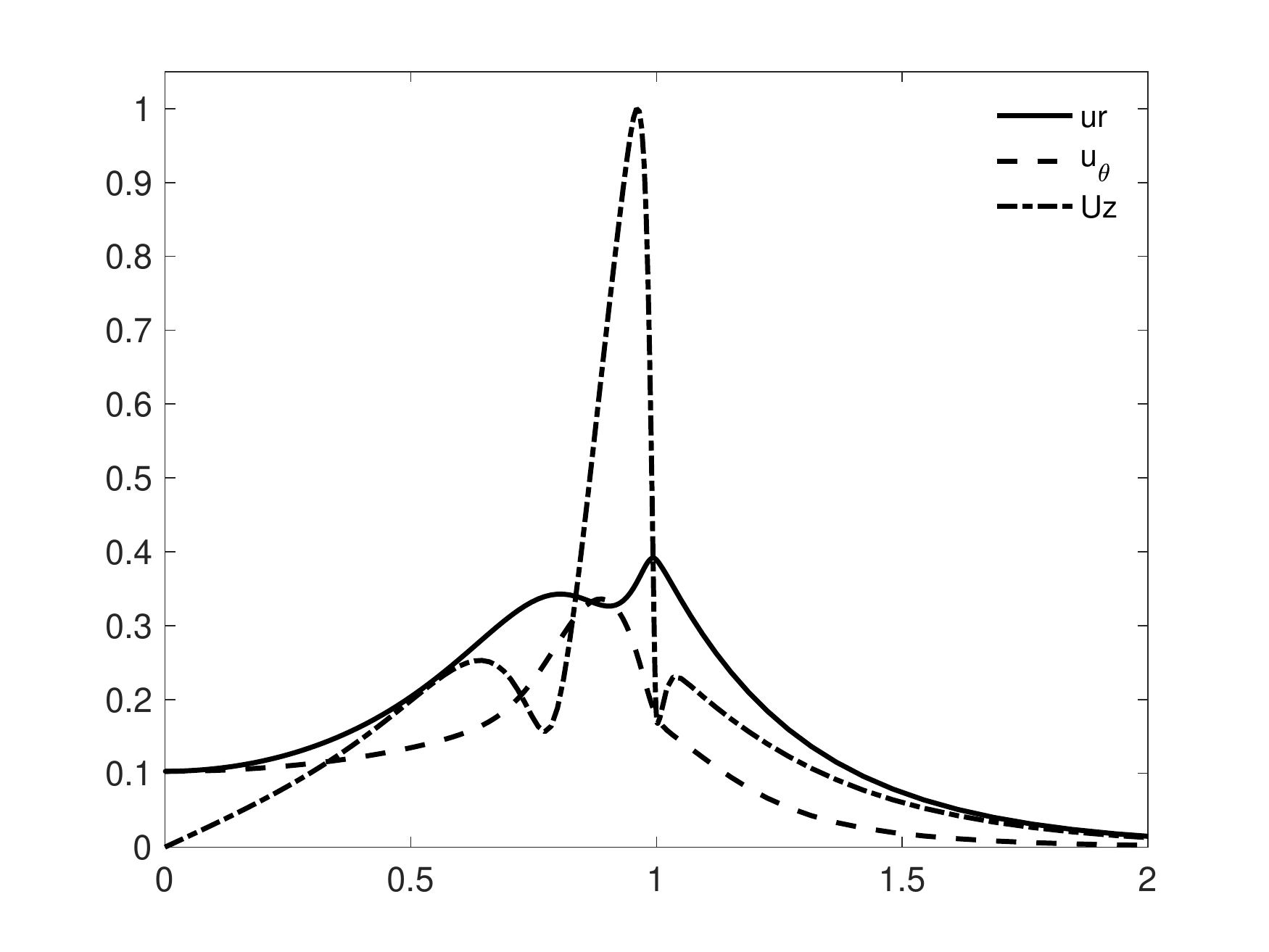}
\caption{}
\end{subfigure}
\caption{ Disturbance velocity functions for low viscosity jets
with $(M,Re,Sc,\theta,\theta_\mu,k)=(5,1000,100,0.1,0.01,3)$. (a) The axisymmetric mode, $\beta =0$ and (b) the helical mode, $\beta =1$.}
\label{fig:eigenfunctions_high_M}. 
\end{figure}

The velocity disturbance functions for a wavenumber ($k=3$) close to the maximum growth rate case is shown in Fig. \ref{fig:eigenfunctions_high_M}. The striking feature of the disturbance profiles is that even though the instability arises in the shear layer, and the disturbance peaks in the shear layer, the axial disturbance velocity does not decay to zero at the jet centerline, and therefore the disturbance spans the diameter of the jet. 

\begin{figure}
\centering
\begin{subfigure}{0.49\textwidth}
\includegraphics[width= \textwidth]{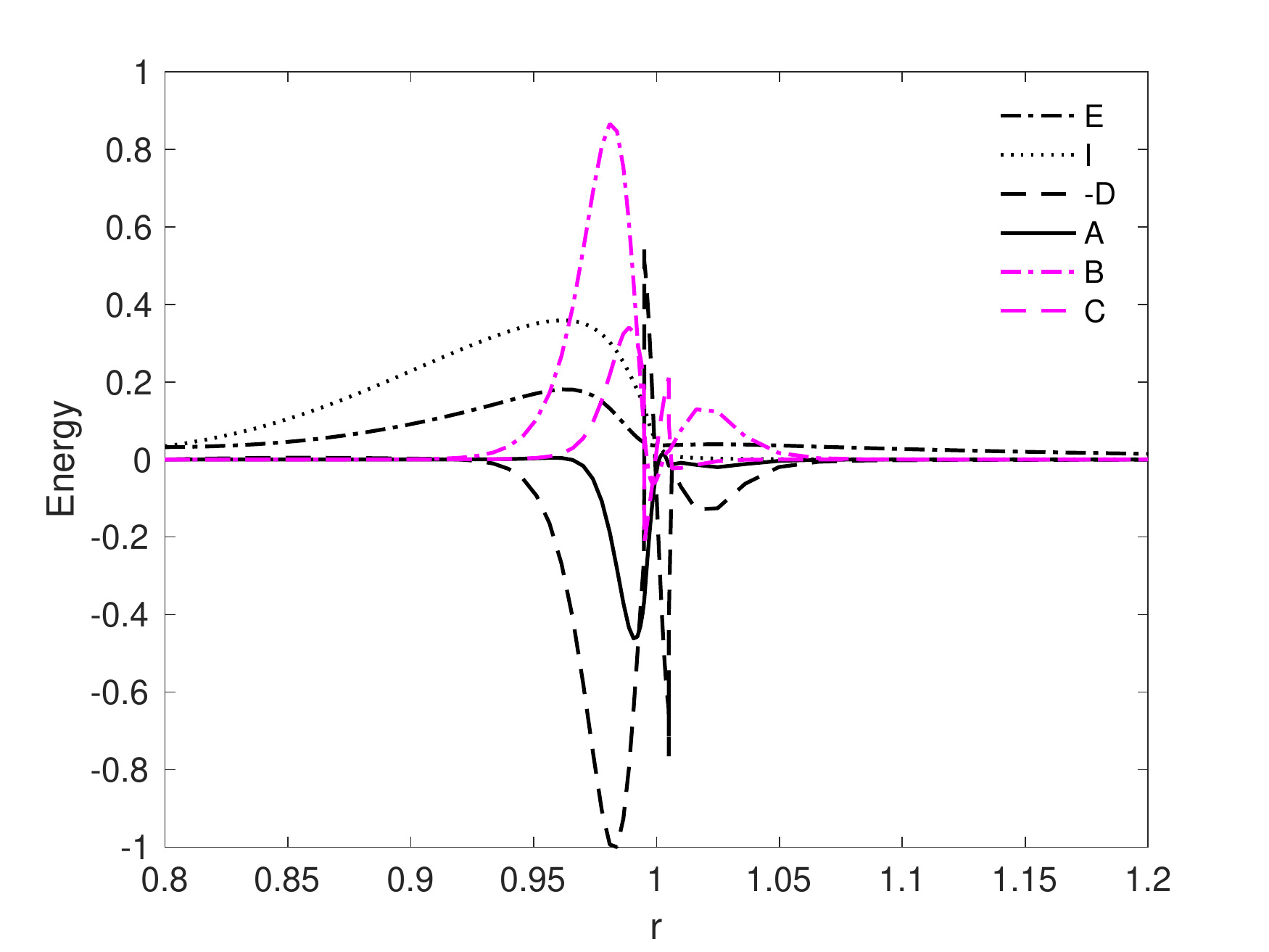}
\caption{}
\end{subfigure}
\begin{subfigure}{0.49\textwidth}
\includegraphics[width= \textwidth]{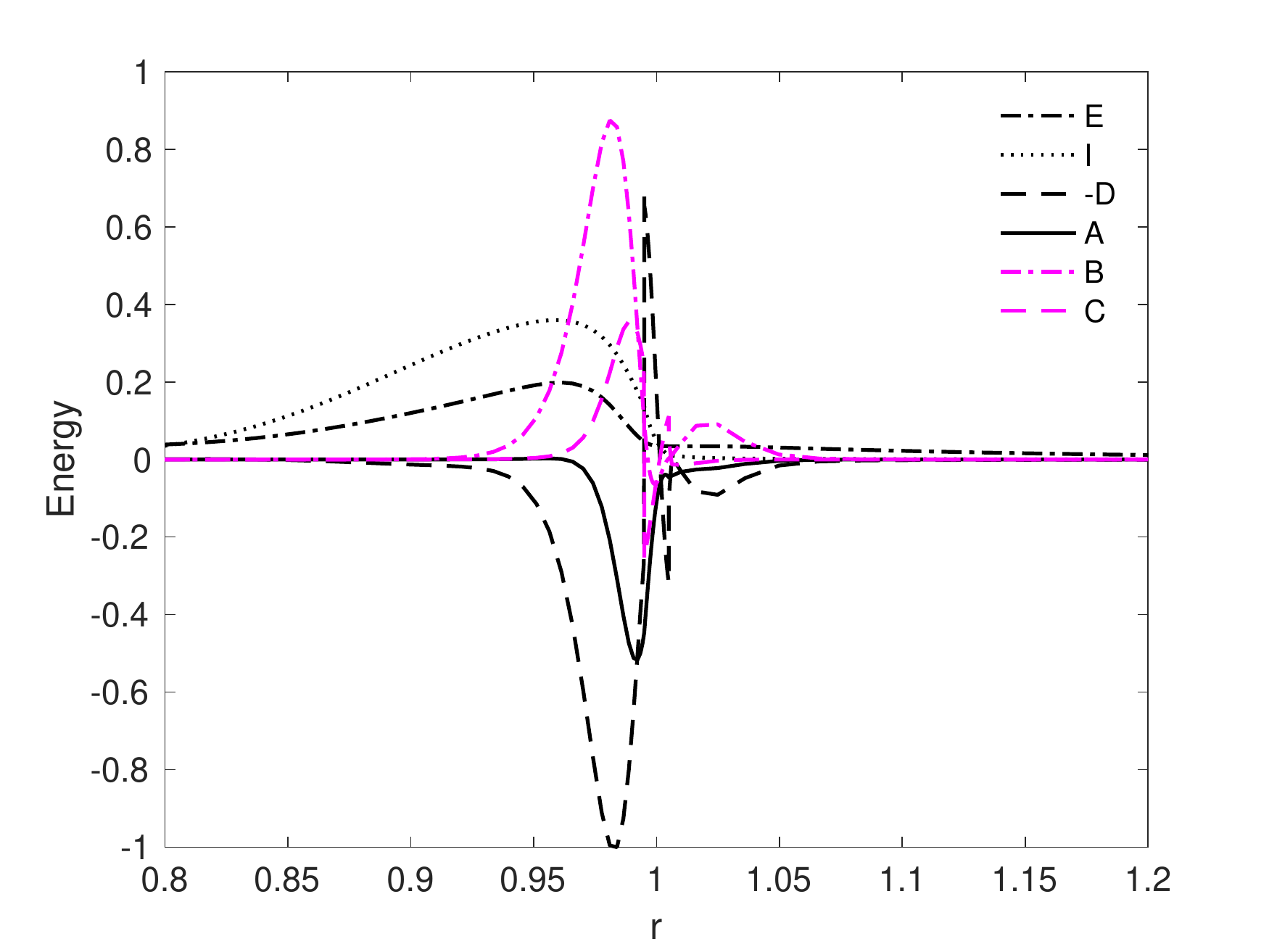}
\caption{}
\end{subfigure}
\caption{ The disturbance quantities contributing to the kinetic energy budget, for the same conditions as in Fig. 6. }
\label{figLTKE_spatial_high_M}
\end{figure}

\begin{figure}
\centering
\begin{subfigure}{0.49\textwidth}
\includegraphics[width= \textwidth]{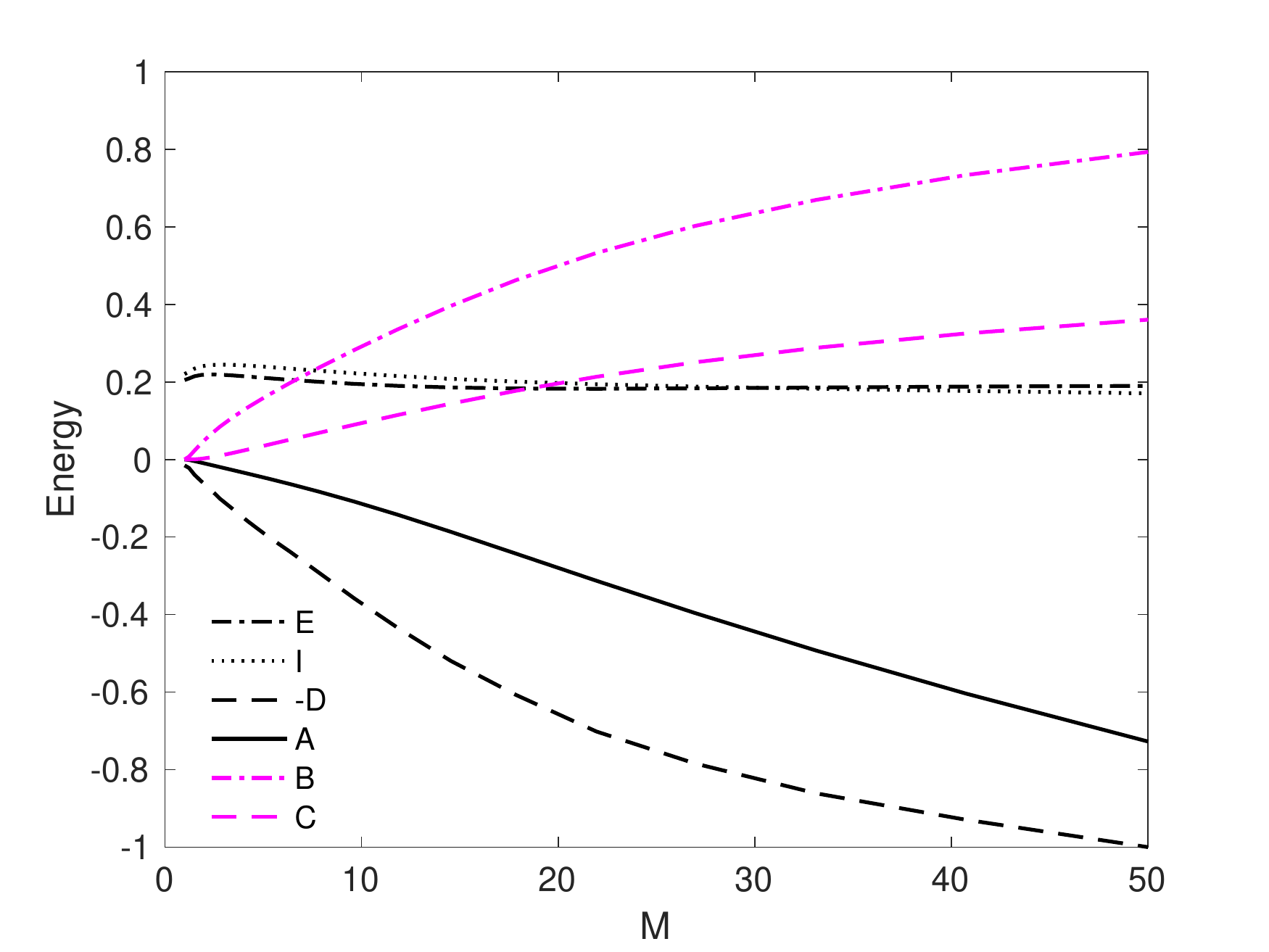}
\caption{}
\end{subfigure}
\begin{subfigure}{0.49\textwidth}
\includegraphics[width= \textwidth]{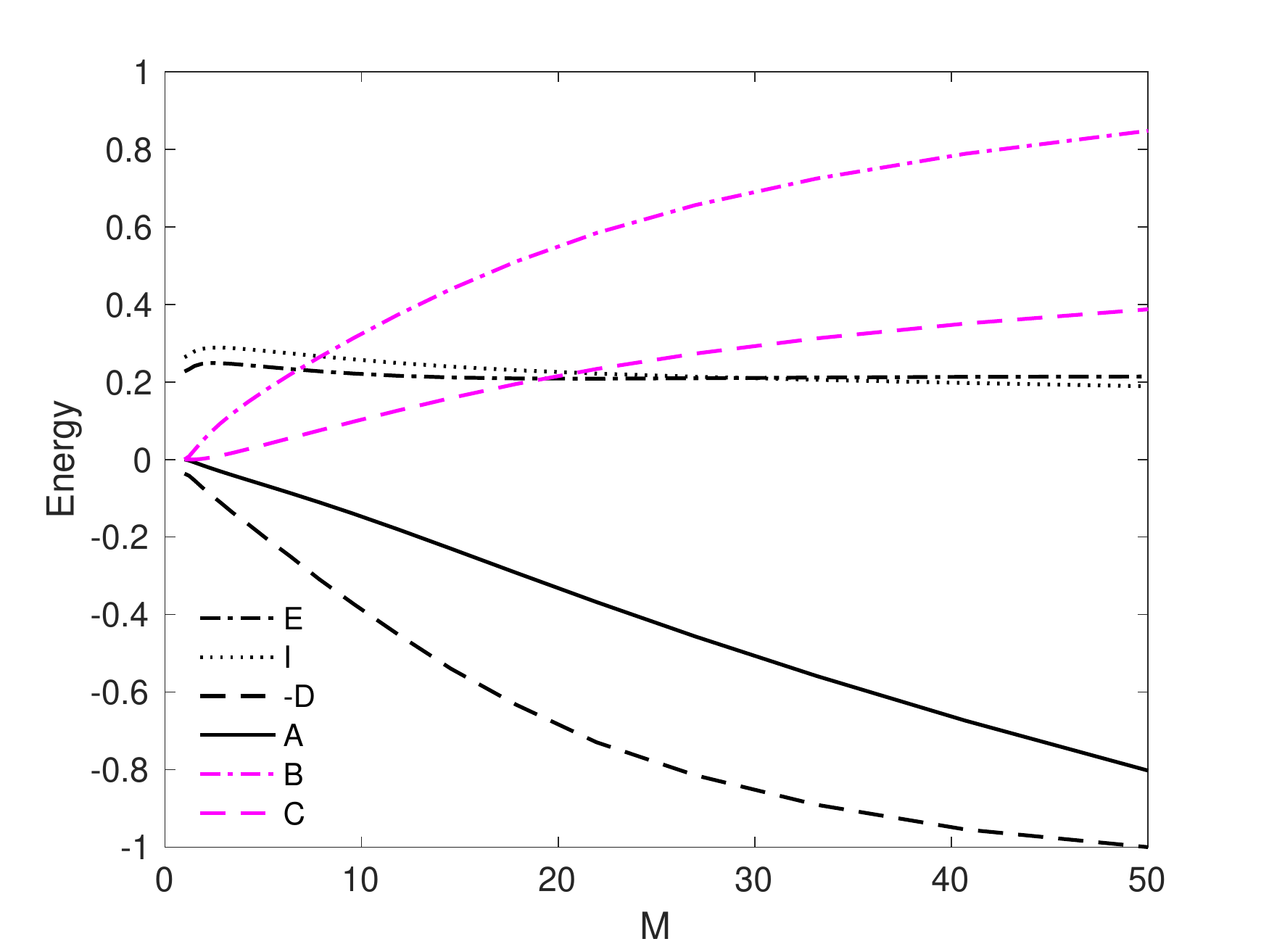}
\caption{}
\end{subfigure}
 \caption{ The disturbance kinetic energy budget (integral) for a jet with $(Re, Sc,\theta,\theta_\mu,k)=(1000,100,0.1,0.01,3)$, as a function of M. (a) Symmetric mode (b) Helical mode. }
\label{fig:TKE_budget_high_M}
 \end{figure}

To further understand the mechanisms responsible for driving the instability, we examine the disturbance kinetic energy equation, obtained by multiplying the governing each equation for each velocity perturbation with its complex conjugate and summing all three equations. Following \citet{Selvam2007}, we use the disturbance velocities $(v_r,v_{\theta},v_z)$, we evaluate the terms of the disturbance kinetic energy equation:

\begin{equation}
\dot{E}=I+P-D+A+B+C
\end{equation}
where, 
\begin{equation}
\dot{E} =\omega_{i} \int_{0}^{\infty}\left(\left|\hat{v}_{r}\right|^{2}+\left|\hat{v}_{\theta}\right|^{2}+\left|\hat{v}_{z}\right|^{2}\right) r \mathrm{~d} r
\end{equation}

\begin{equation}
I =\int_{0}^{\infty} \frac{\mathrm{d} \bar{v}_{z}}{\mathrm{~d} r} \operatorname{Im}\left\{\hat{v}_{r} \hat{v}_{z}^{*}\right\} r \mathrm{~d} r
\end{equation}

\begin{equation}
P = -\frac{1}{Re}\int_{0}^{\infty} \operatorname{Im} \left( \frac{d\hat{p}}{dr}\hat{v}_r^* -\frac{\beta}{r}\hat{p}\hat{v}_{\theta}^* -k\hat{p}\hat{v}_{z}^* \right) rdr
\end{equation}

\begin{equation}
\begin{aligned}
D &=\frac{1}{Re}\int_{0}^{\infty}\mathrm{e}^{M \bar{c}}\left[ \left(|\frac{d\hat{u_r}}{dr}|^2+|\frac{d\hat{u_\theta}}{dr}|^2+|\frac{d\hat{u_z}}{dr}|^2\right)-\frac{1}{r}\operatorname{Real}\left(\frac{d}{dr}(r\hat{u}_r^*\frac{d\hat{u}_r}{dr} 
+r\hat{u}_\theta^*\frac{d\hat{u}_\theta}{dr} 
+r\hat{u}_z^*\frac{d\hat{u}_z}{dr})\right)\right]rdr\\
+&\frac{1}{Re}\int_{0}^{\infty}\mathrm{e}^{M \bar{c}}\left[ (\frac{\beta^2}{r^2}+k^2)(|\hat{v}_{r}|^2 +|\hat{v}_{\theta}|^2 +|\hat{v}_{z}|^2) +\frac{|\hat{v}_{r}|^2 +|\hat{v}_{\theta}|^2 + 4\beta\operatorname{Real}(\hat{v}_{\theta}\hat{v}_r^*)}{r^2} \right]rdr
\end{aligned}
\end{equation}

\begin{equation}
\begin{split}
A = & \frac{M}{Re}\int_{0}^{\infty}\mathrm{e}^{M \bar{c}}\left[\frac{d\bar{c}}{dr}\left(\operatorname{Real}\left(\frac{d\hat{v}_r}{dr}\hat{v}_r^* +\frac{\mathrm{d}\hat{v}_{\theta}}{\mathrm{d} r}\hat{v}_{\theta}^*+ \frac{d\hat{v}_z}{dr}\hat{v}_z^*\right)+ \frac{1}{r}\left(\frac{d(r|\hat{u}_r|^2)}{dr}-|\hat{u}_{\theta}|^2\right)\right)\right]rdr
\end{split}
\end{equation}

\begin{equation}
B=B_{r}+B_{z}=\frac{M}{\operatorname{Re}}\left[\int_{0}^{\infty} \mathrm{e}^{M \bar{c}} \frac{\mathrm{d} \bar{v}_{z}}{\mathrm{~d} r} \operatorname{Real}\left\{\frac{\mathrm{d} \hat{c}}{\mathrm{~d} r} \hat{v}_{z}^{*}\right\} r \mathrm{~d} r+\int_{0}^{1} \mathrm{e}^{M \bar{c}} \frac{\mathrm{d} \bar{v}_{z}}{\mathrm{~d} r} \operatorname{Real}\left\{k \hat{c} \hat{v_{r}^{*}}\right\} r \mathrm{~d} r\right]
\end{equation}

\begin{equation}
C=\frac{M}{\operatorname{Re}} \int_{0}^{\infty} \left(\frac{\mathrm{d}^{2} \bar{v}_{z}}{\mathrm{~d} r^{2}}+\frac{1}{r} \frac{\mathrm{d} \bar{v}_{z}}{\mathrm{~d} r}+M \frac{\mathrm{d} \bar{c}}{\mathrm{~d} r} \frac{\mathrm{d} \bar{v}_{z}}{\mathrm{~d} r}\right) \operatorname{Real}\left\{\hat{c} \hat{v}_{z}^{*}\right\} r\mathrm{d}r
\end{equation}
The left hand side of the equation is the rate of change of disturbance kinetic energy; the first time on the right hand side is the usual kinetic energy generation term for constant property flows, while the second term is the viscous dissipation. Additionally, the variable viscosity field gives rise to terms that couple the mean viscosity gradient with the velocity perturbations (A), the mean velocity gradient with the variable viscosity field (B) and the fluctuating velocity and viscosity fields (C). Plotting the spatial distributions of the terms (Fig.\ref{figLTKE_spatial_high_M} clearly shows that the term associated with the viscosity gradient is responsible for the instability, peaking in the shear layer and causing an increase in the disturbance energy production term. Integrating these terms from $r=0$ to $r=\infty$, we construct the budget for disturbance kinetic energy and plot as a function of M and Re in Fig. \ref{fig:TKE_budget_high_M}(a) and (b). It is evident that for both axisymmetric and helical modes as M increases, the major source of the disturbance energy is the term B, the coupling of the mean velocity gradient with the mean viscosity field. 

\subsubsection{Parametric Study}

\begin{figure}
\centering
\begin{subfigure}{0.49\textwidth}
\includegraphics[width= \textwidth]{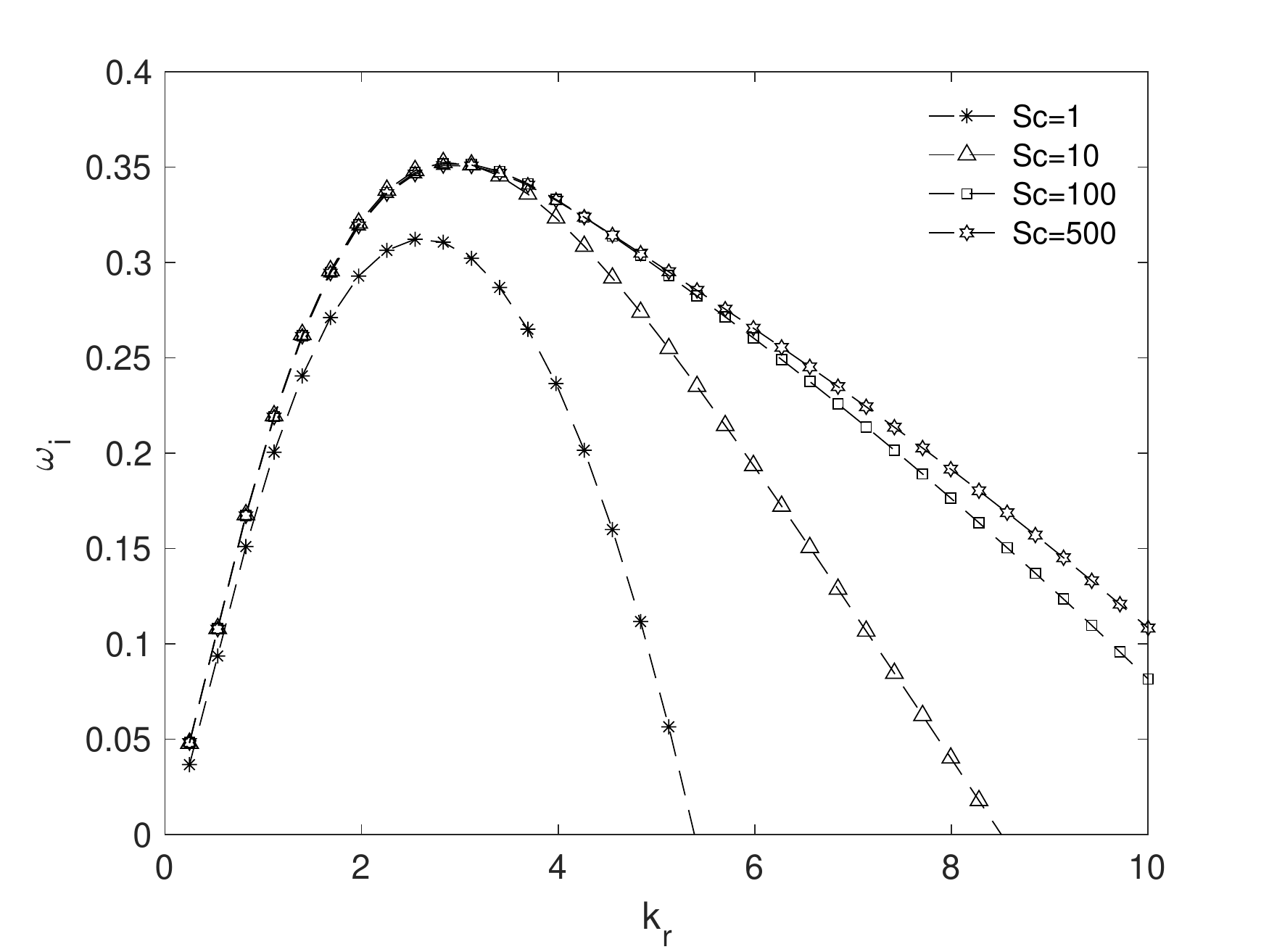}
\caption{}
\end{subfigure}
\begin{subfigure}{0.49\textwidth}
\includegraphics[width= \textwidth]{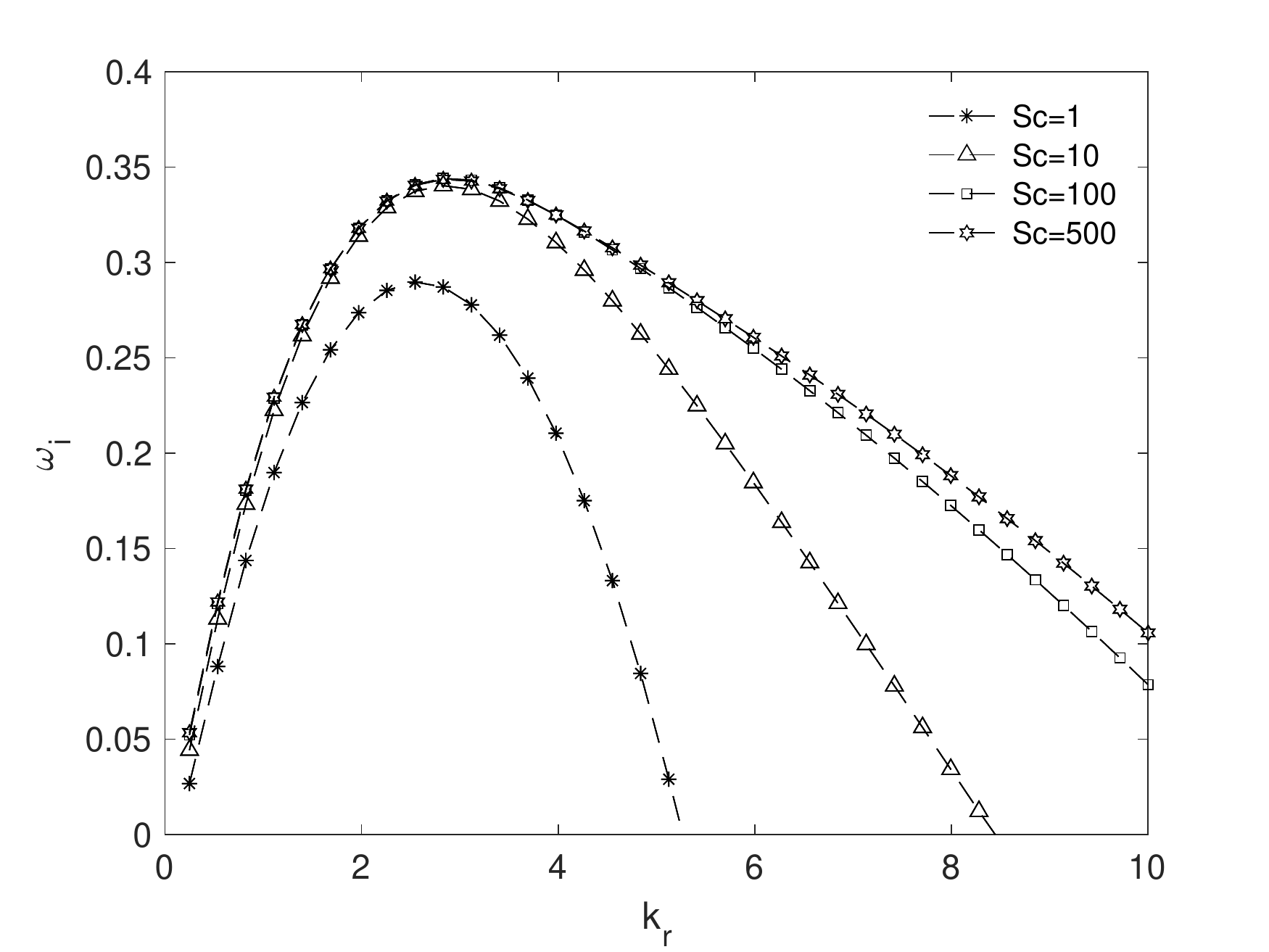}
\caption{}
\end{subfigure}
\caption{ The effects of Schmidt number $Sc$ on the temporal stability for $(M,Re,\theta,\theta_\mu)=(20,1000,0.1,0.01)$. (a) The axisymmetric mode, $\beta =0$ and (b) the helical mode, $\beta =1$.}
\label{fig:wi_vs_k_Sc}
\end{figure}

We now turn to the effects of other parameters. The Schmidt number Sc controls the diffusion of viscosity fluctuations induced by the instability; strong diffusion would be expected to weaken the instability. A look at the temporal growth rates for fixed velocity and viscosity profiles (Fig. ~\ref{fig:wi_vs_k_Sc}) by and large supports this expectation; however the effects of Sc are not felt beyond Sc=10. Further, while the effects of weak diffusion play a substantial role in determining growth rates at high wavenumbers (short waves), the fastest growing mode is nearly unaffected by Sc except for Sc$\approx 1$, for both axisymmetric and helical modes. This is consistent with the findings from the disturbance kinetic energy equation. The main source of instability is the presence of a velocity fluctuation in a variable viscosity field; viscosity fluctuations do not play a significant role. \citet{Sahu2014}, studying the planar counterpart of the present study, terms this a quasi-inviscid instability, in the sense that viscosity plays a role in generating the mean velocity profile but otherwise does not play a direct role. We also note that large disparities in $\theta$ and $\theta _\mu$ are unlikely to be experimentally realizable for Sc=1, and this case is used only as a point of reference.    

\begin{figure}
\centering
\begin{subfigure}{0.49\textwidth}
\includegraphics[width= \textwidth]{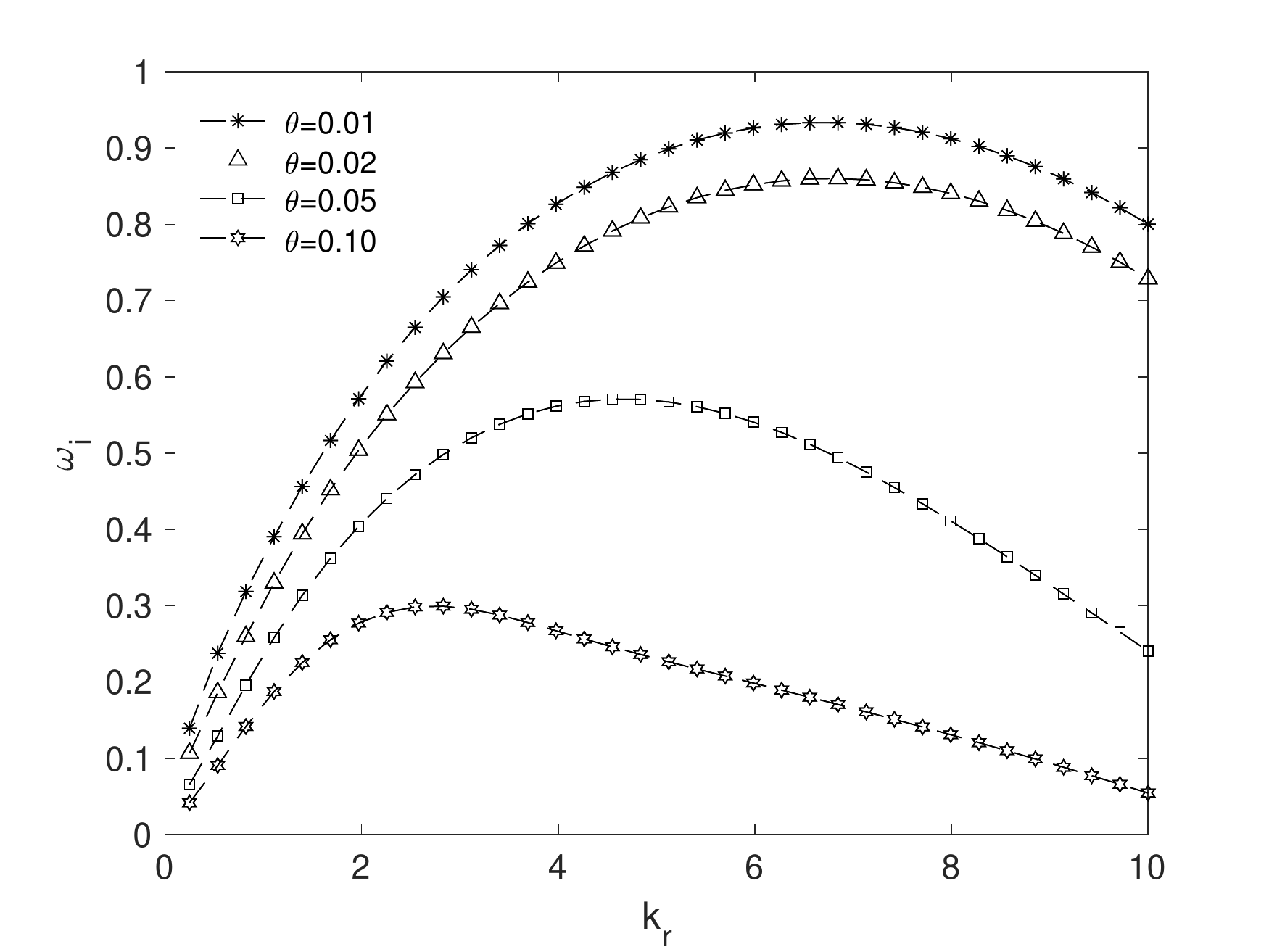}
\caption{}
\end{subfigure}
\begin{subfigure}{0.49\textwidth}
\includegraphics[width= \textwidth]{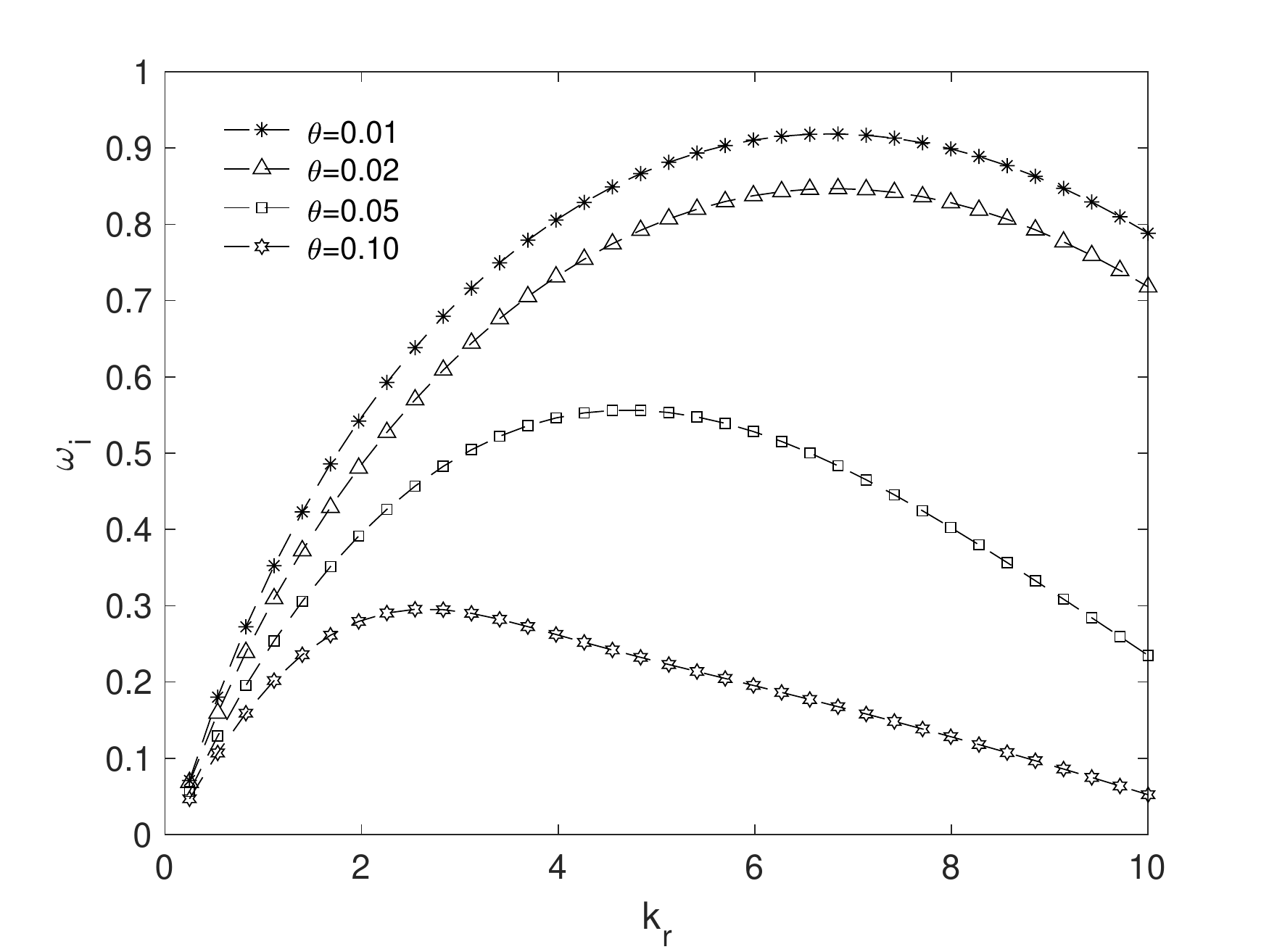}
\caption{}
\end{subfigure}
\caption{ Effects of the jet momentum thickness on the temporal stability for low viscosity jets ($M = 20$),  for a viscosity thickness of $\theta _\mu =0.01$. (a) The axisymmetric mode, $\beta =0$ and (b) the helical mode, $\beta =1$. Other relevant parameters are $Re=1000$, $Sch=100$.}
\label{fig:wi_k_theta_high_M}
\end{figure}

\begin{figure}
\centering
\begin{subfigure}{0.49\textwidth}
\includegraphics[width= \textwidth]{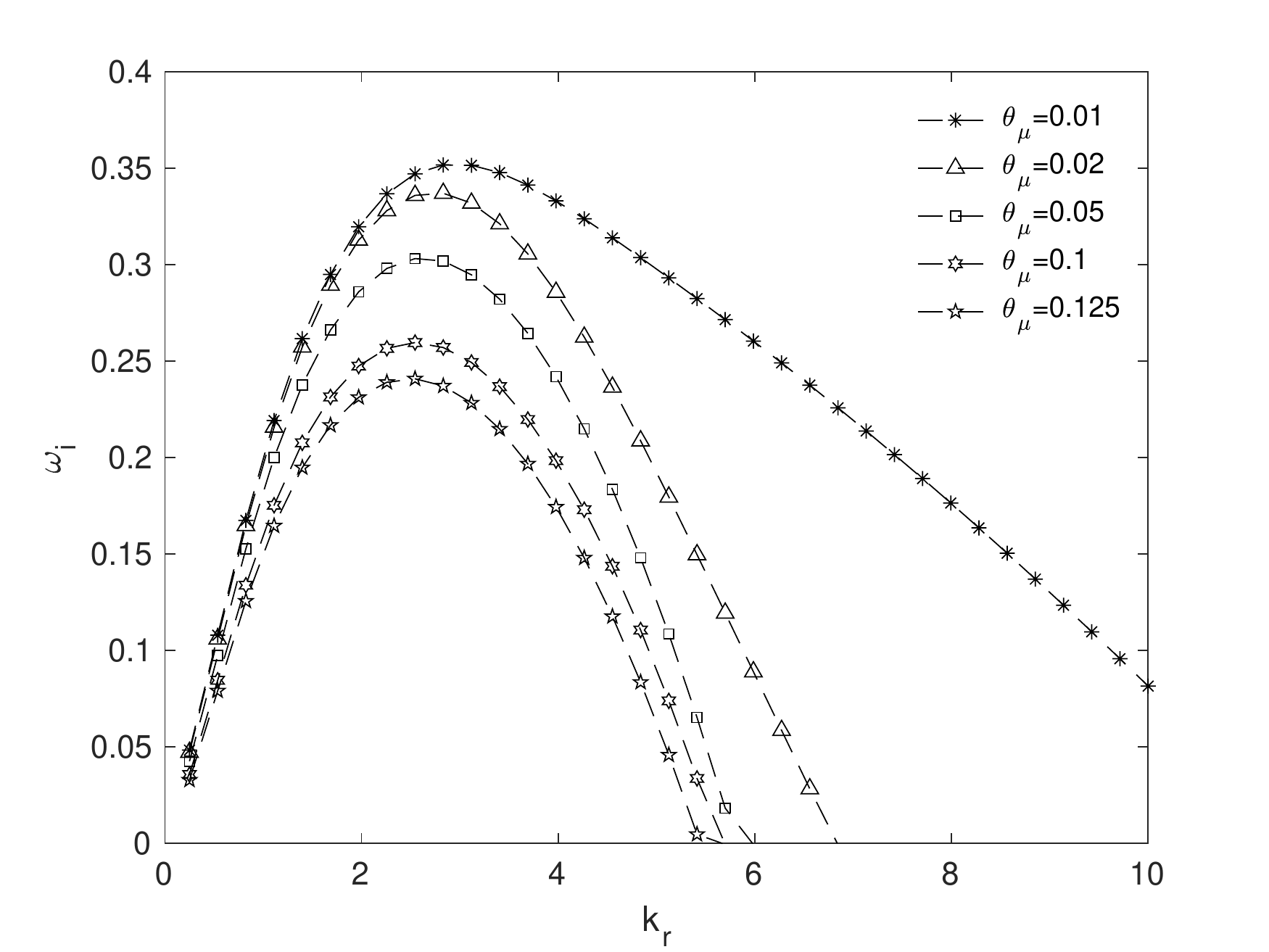}
\caption{}
\end{subfigure}
\begin{subfigure}{0.49\textwidth}
\includegraphics[width= \textwidth]{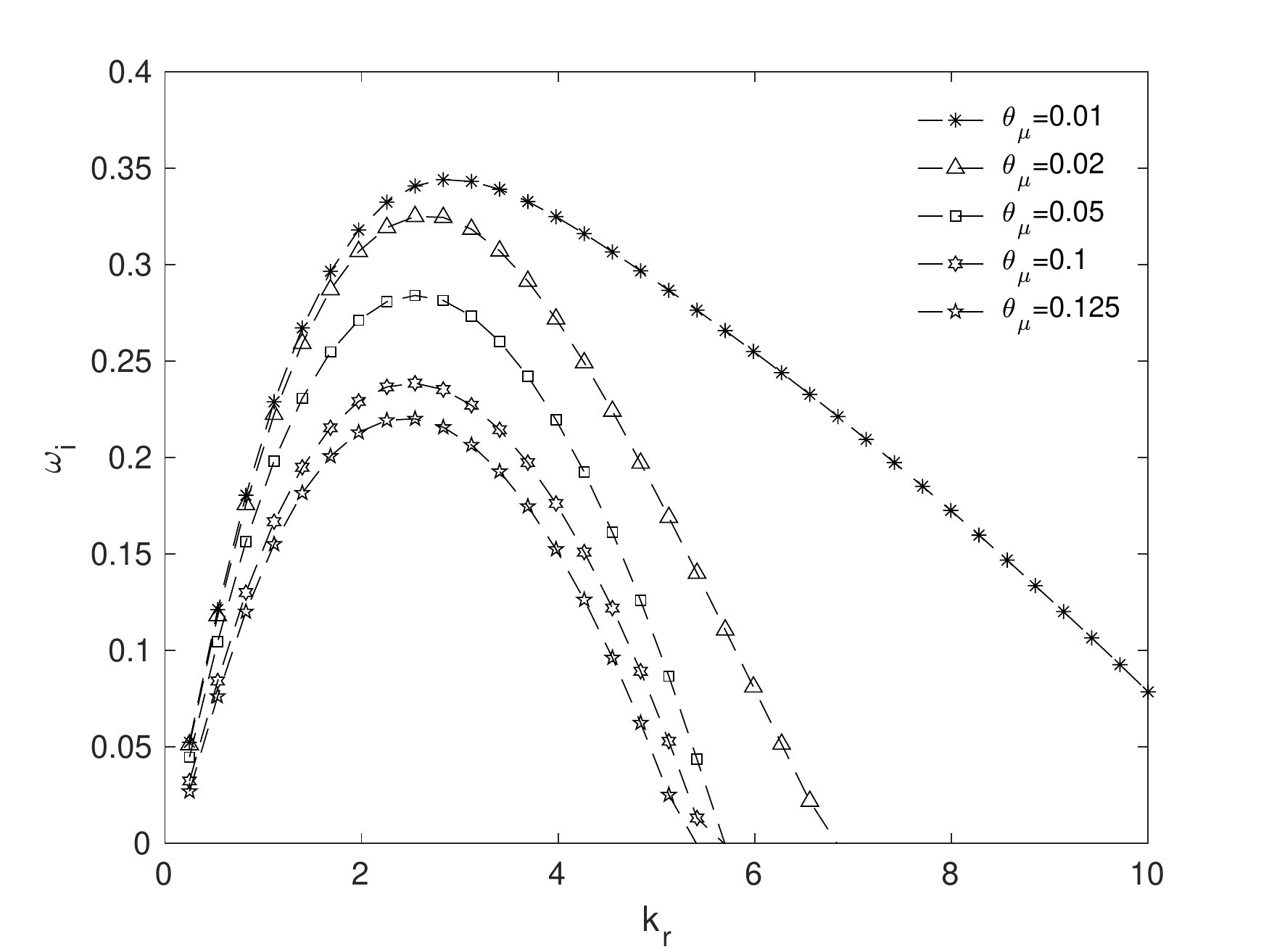}
\caption{}
\end{subfigure}
\caption{ Temporal Instability of low viscosity jets ($M = 20$, $\theta =0.1$) for different viscosity thickness. (a) The axisymmetric mode, $\beta =0$ and (b) the helical mode, $\beta =1$. Other relevant parameters are set to $Re=1000$, $Sch=100$.}
\label{fig:wi_k_thetamu_high_M}
\end{figure}

We also consider the influence of velocity and viscosity profile shapes, in terms of their respective regions of sharp variation ($\theta$ and $\theta _\mu$). One question that arises is regarding the relative influence of these gradient regions, and how they determine the dominant instability. Figure \ref{fig:wi_k_theta_high_M} plots growth rates as a function of the momentum thickness, for fixed viscosity ratio M=20, Reynolds number Re=1000 and Sc=100. The effects of decreasing momentum thickness is to make the jet more unstable, as expected while shifting the maximum growth rate to shorter wavelengths. Thus, the disturbance wavelength is seen to scale on the momentum thickness, at least for constant viscosity ratio. Similarly, when the momentum thickness is held constant and the viscosity gradient is increased at fixed M (Fig. \ref{fig:wi_k_thetamu_high_M}), the jet becomes more unstable, though the controlling wavelength does not vary. Thus, the mean velocity profile seems to be the controlling parameter.

\begin{figure}
\centering
\begin{subfigure}{0.49\textwidth}
\includegraphics[width= \textwidth]{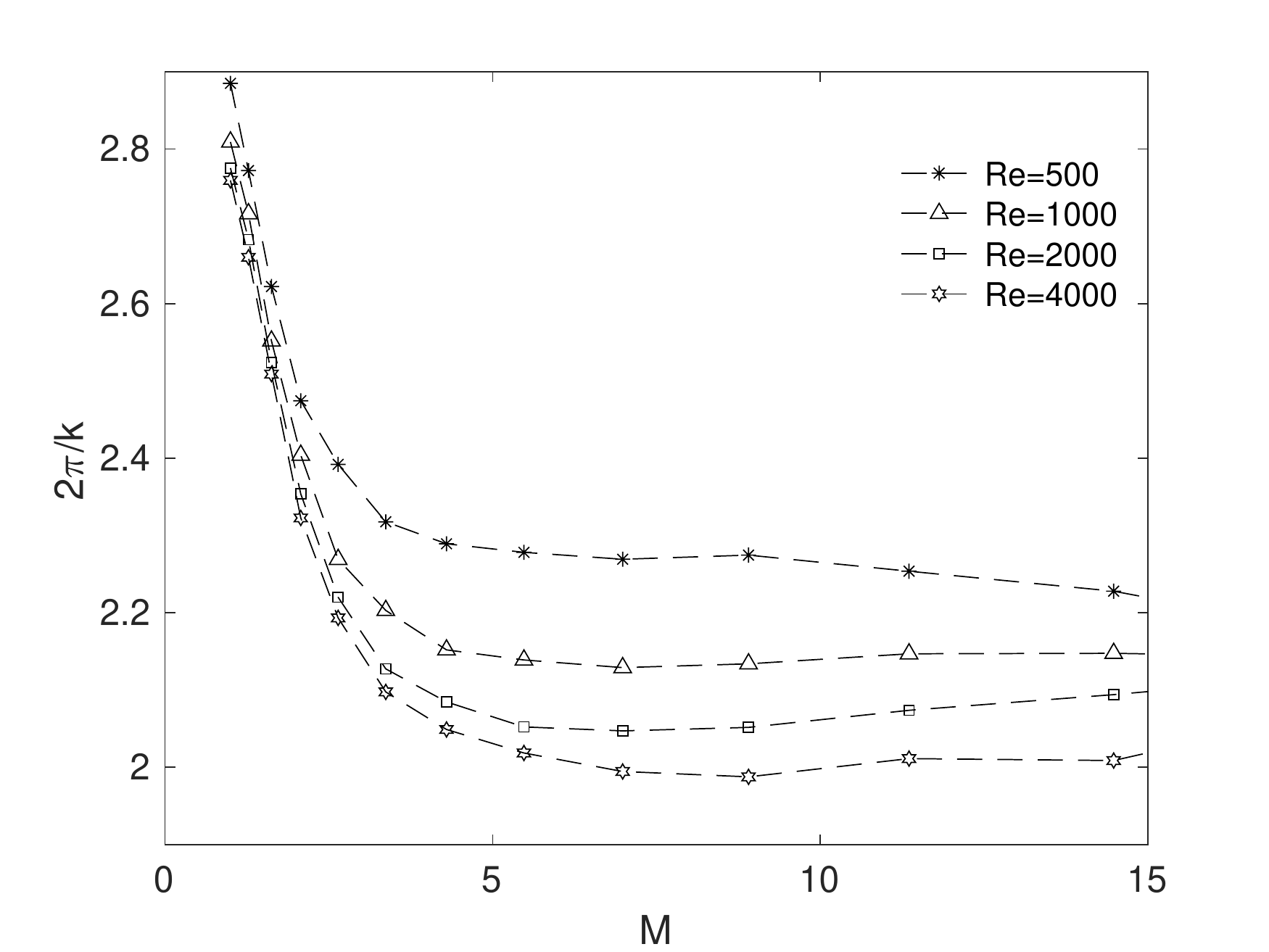}
\caption{}
\end{subfigure}
\begin{subfigure}{0.49\textwidth}
\includegraphics[width= \textwidth]{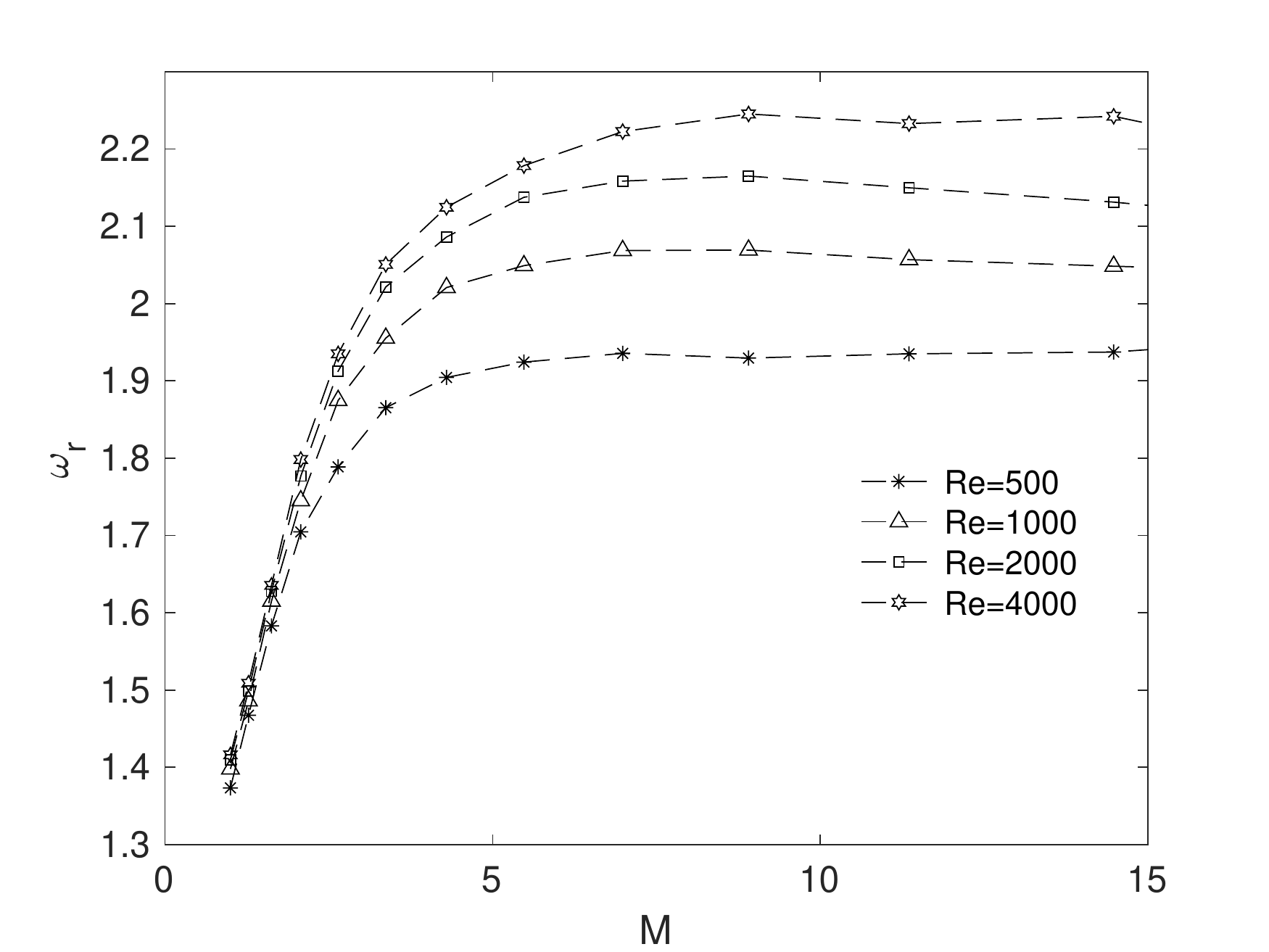}
\caption{}
\end{subfigure}
\caption{ Variation of wavelength $\frac{2\pi}{k}$ and frequency corresponding to the maximum temporal growth rate, as a function of viscosity ratio for the axisymmetric mode, $\beta=0$. (a) wavelength of the fastest growing mode (b) frequency of this mode. Other relevant parameters are: $\theta =0.1$, $\theta_\mu =0.01$, $Sc=100$, $Re=1000$.}
\label{fig:Re_high_M_0}
\end{figure}

\begin{figure}
\centering
\begin{subfigure}{0.49\textwidth}
\includegraphics[width= \textwidth]{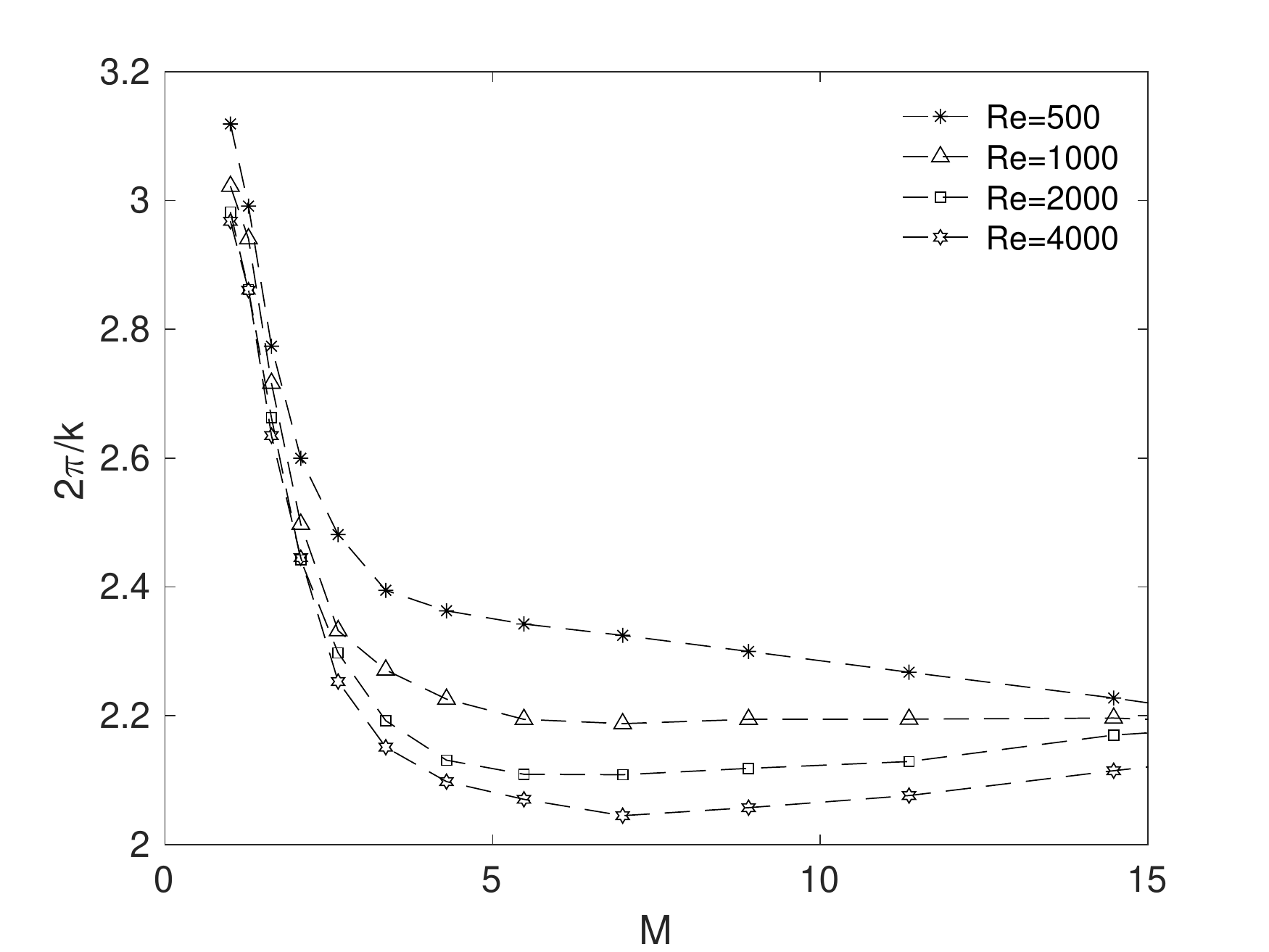}
\caption{}
\end{subfigure}
\begin{subfigure}{0.49\textwidth}
\includegraphics[width= \textwidth]{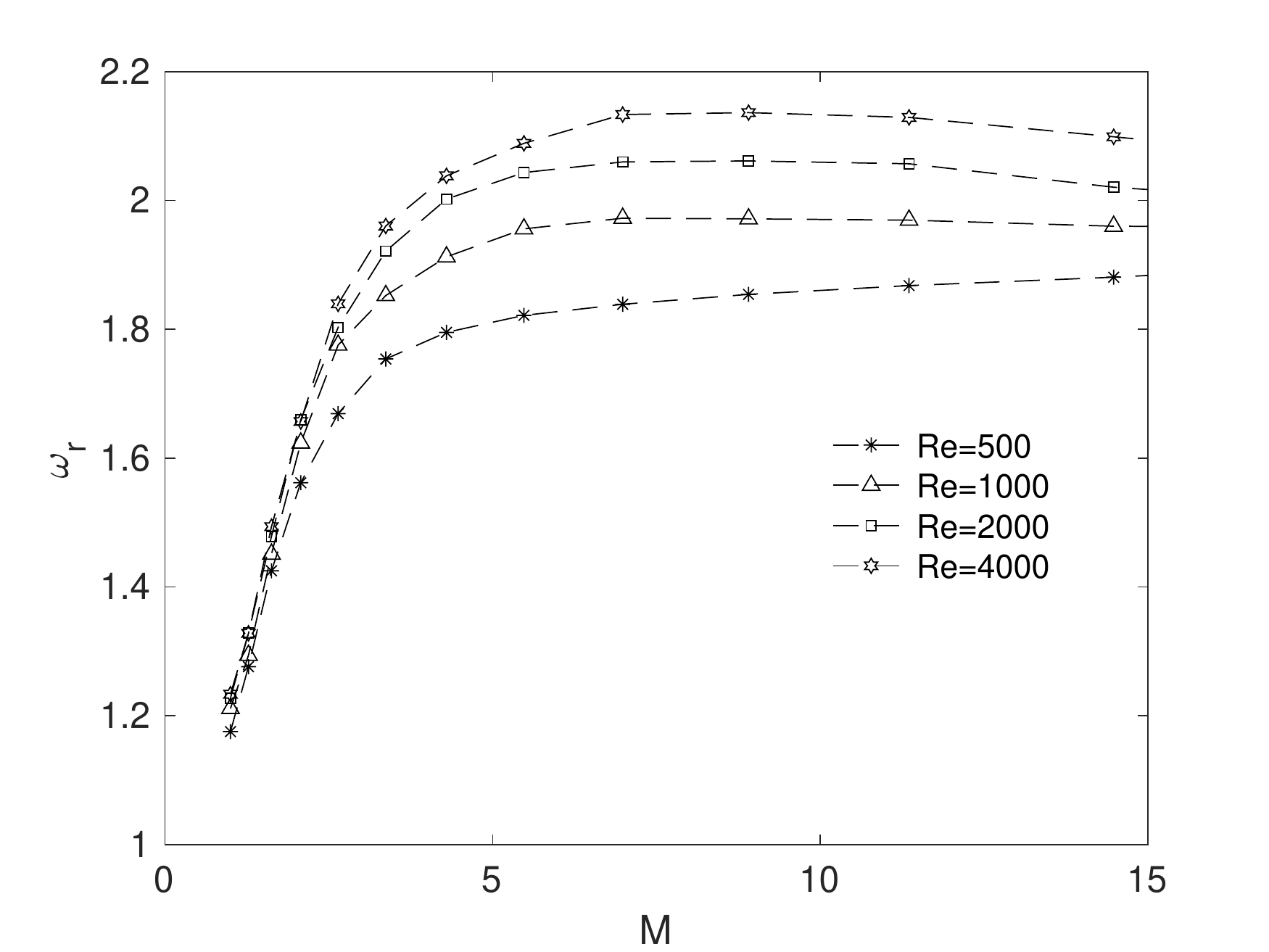}
\caption{}
\end{subfigure}
\caption{ Variation of wavelength $\frac{2\pi}{k}$ and frequency corresponding to the maximum temporal growth rate, as a function of viscosity ratio for the helical mode, $\beta=1$. (a) wavelength of the fastest growing mode (b) frequency of this mode. Other relevant parameters are: $\theta =0.1$, $\theta_\mu =0.01$, $Sch=100$, $Re=1000$..}
\label{fig:Re_high_M_1}
\end{figure}

\begin{figure}
\centering
\begin{subfigure}{0.49\textwidth}
\includegraphics[width= \textwidth]{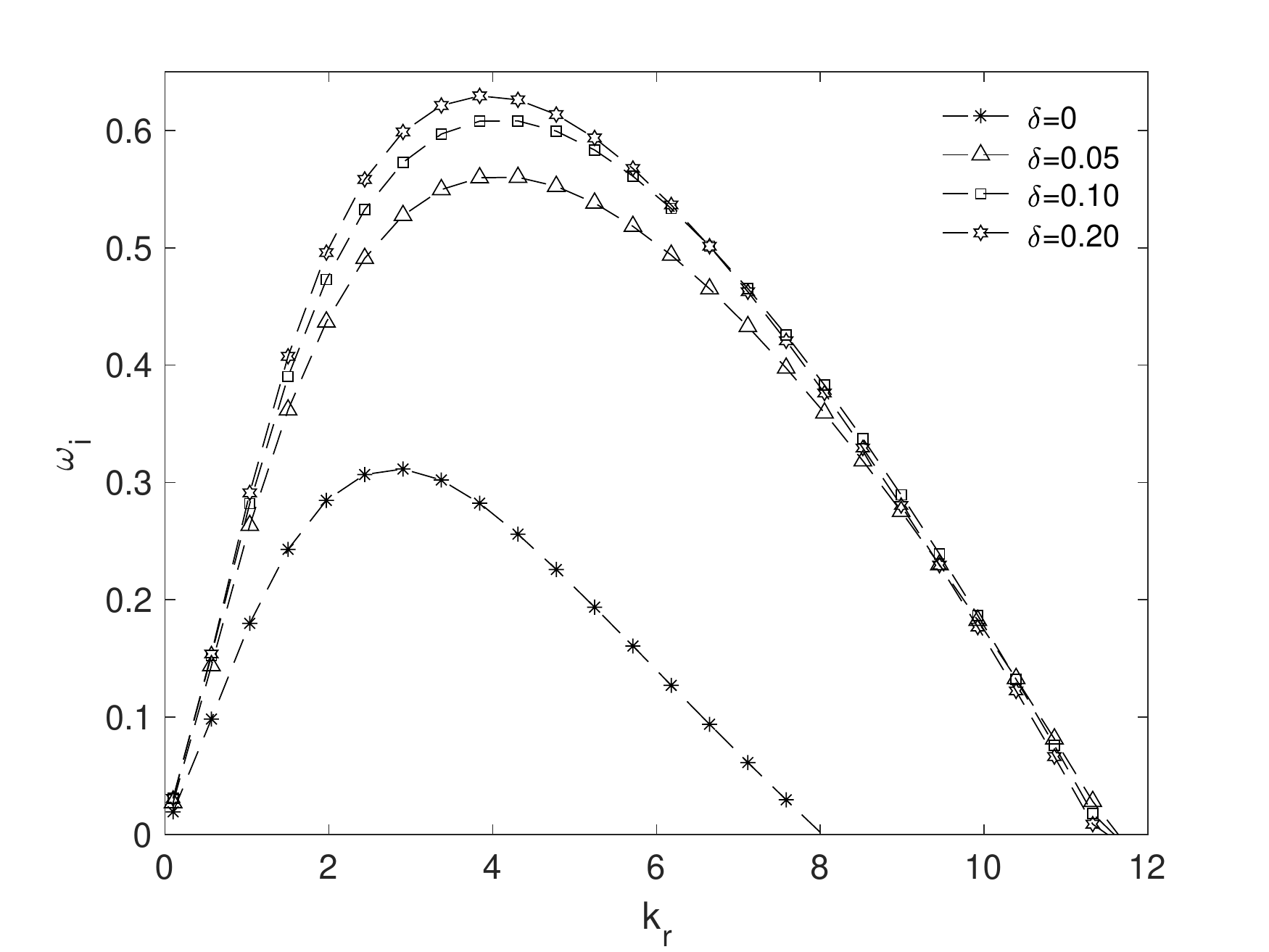}
\caption{}
\end{subfigure}
\begin{subfigure}{0.49\textwidth}
\includegraphics[width= \textwidth]{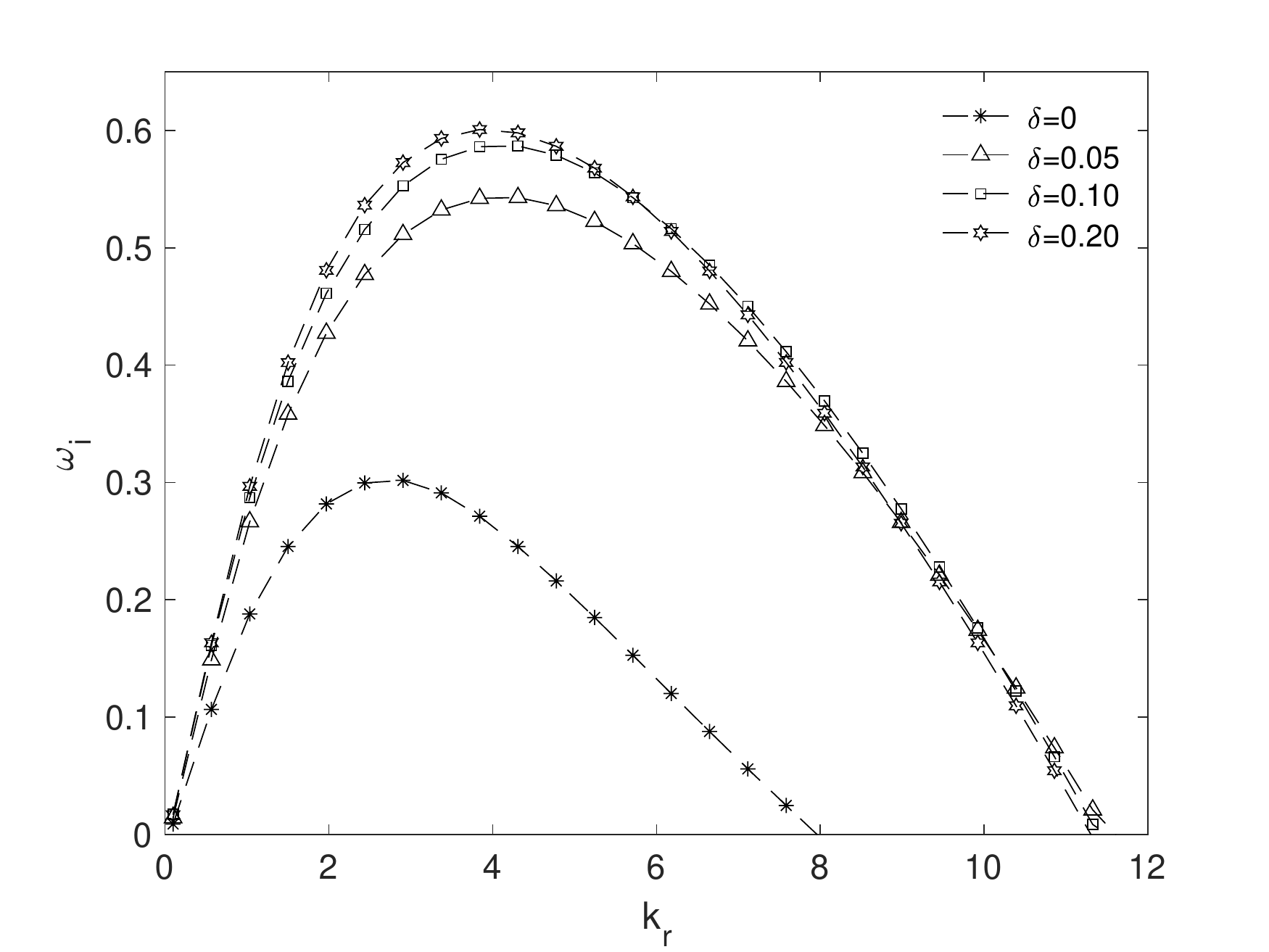}
\caption{}
\end{subfigure}
\caption{The effects of a shift in velocity profile relative to the region of viscosity gradient. (a)$\beta =0$, axisymmetric mode, (b) $\beta =1$, helical mode. The other parameters have values $\theta =0.1$, $\theta _\mu =0.01$, Sc=100, Re=1000.} 
\label{fig:delta_effects}
\end{figure}

 A compilation of the behavior of the wavelength and frequency of the fastest growing mode is shown in Fig. \ref{fig:Re_high_M_0} for axisymmetric and in Fig. \ref{fig:Re_high_M_1} for helical modes as a function of viscosity ratio. For both modes, the wavelength decreases sharply as M increases, before asymptoting to a constant Reynolds-number dependent value. The frequency increases until $M\sim 5$ before becoming insensitive to M. 

Finally, we consider the effect of a shift of velocity profiles by a distance $\delta$ in the radially inward direction, due to the retarding effects of a high viscosity ambient medium. From Fig. \ref{fig:delta_effects} it is apparent that increasing $\delta$ destabilizes the flow. A small shift of $\delta =0.05$, or half the momentum thickness, seems to cause a dramatic shift in growth rates, with little further increase for larger shifts. This is somewhat counter-intuitive since the viscosity gradient now occurs in a region of low values of velocity and velocity gradient. In other situations where there are two inflection points, such as low-density jets \citep{Srinivasan2010} or the planar miscible shear layers studied by \citet{Sahu2014}, the instability is strengthened when the regions strongly overlap. To understand this better, we first verified that such radially inward shifts of velocity profile stabilized the flow for M=1, which indicated that this was indeed a viscosity-linked effect. On closer examination, we find that the critical layer (the radial location where the wave speed matched the base velocity) shifts outward towards the viscosity gradient region, when a radial shift is introduced. Such destabilization when the critical layer overlaps the gradient region has been previously noted and explained by \citet{Ranganathan2001}. 


\subsection{Spatio-temporal Analysis}
The spatio-temporal analysis of the base profiles is carried out next, for a fixed set of conditions. The group velocity of disturbances is given by $d\omega /dk$; a saddle point ($d\omega /dk=0$) in the contours of the complex frequency $\omega$ in the complex wave-number plane is indicative of absolute instability, provided that the growth rate $\omega _{0i}>0$ at the saddle point location $\omega _0$, and the saddle point satisfies the `pinching' criterion \citep{Bers1983, Briggs1964}, which stipulates that the saddle point is formed by the merger of two waves, one traveling upstream and the other traveling downstream. The presence of absolute instability in local profiles has been strongly linked to experimental observations of global modes in the laboratory reference frame \citep{Huerre1990}. For this reason, it becomes interesting to explore whether certain variable viscosity jets might  support absolute instability. 

\begin{figure}
\centering
\begin{subfigure}{0.49\textwidth}
\includegraphics[width= \textwidth]{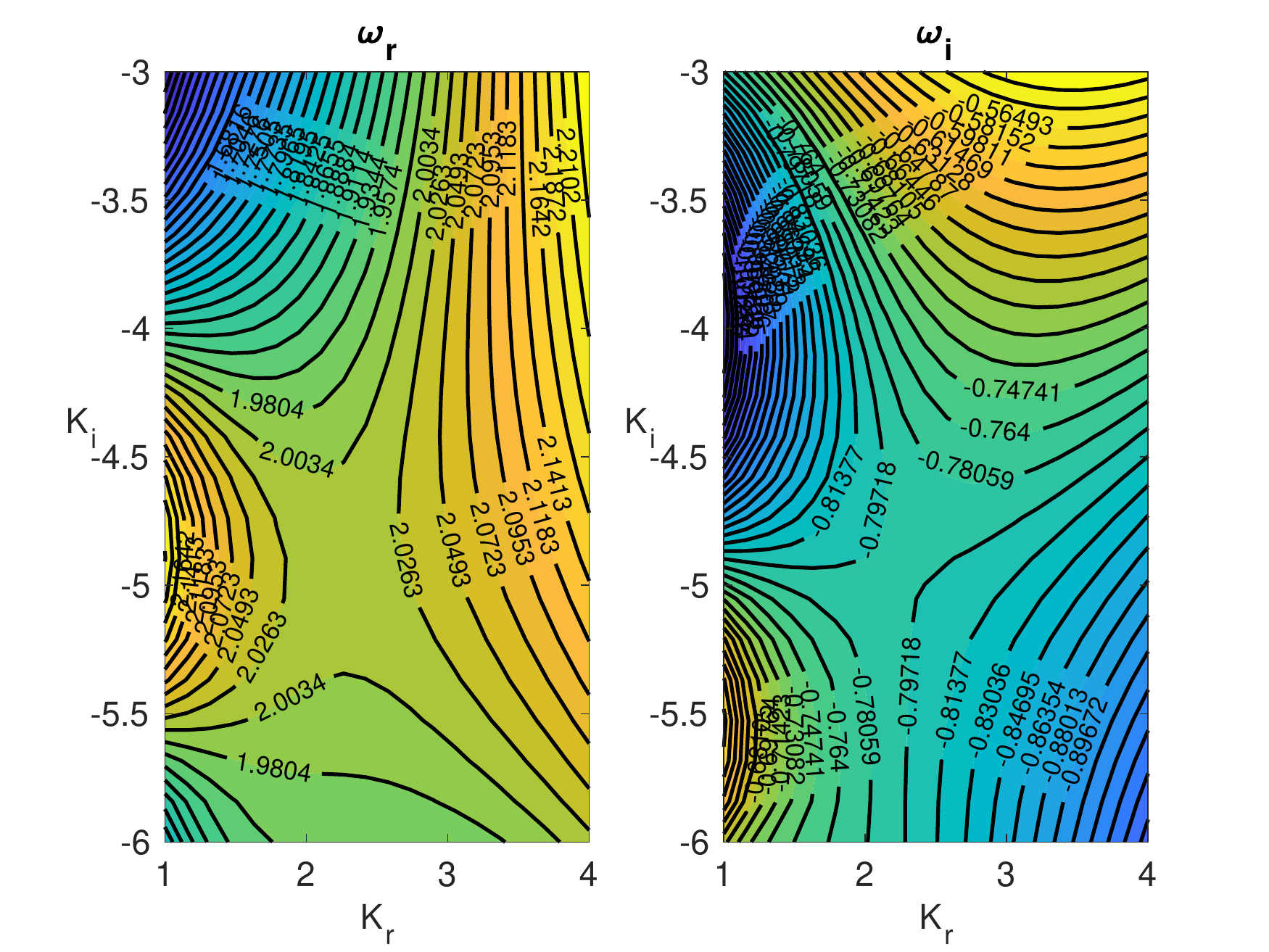}
\caption{}
\end{subfigure}
\begin{subfigure}{0.49\textwidth}
\includegraphics[width= \textwidth]{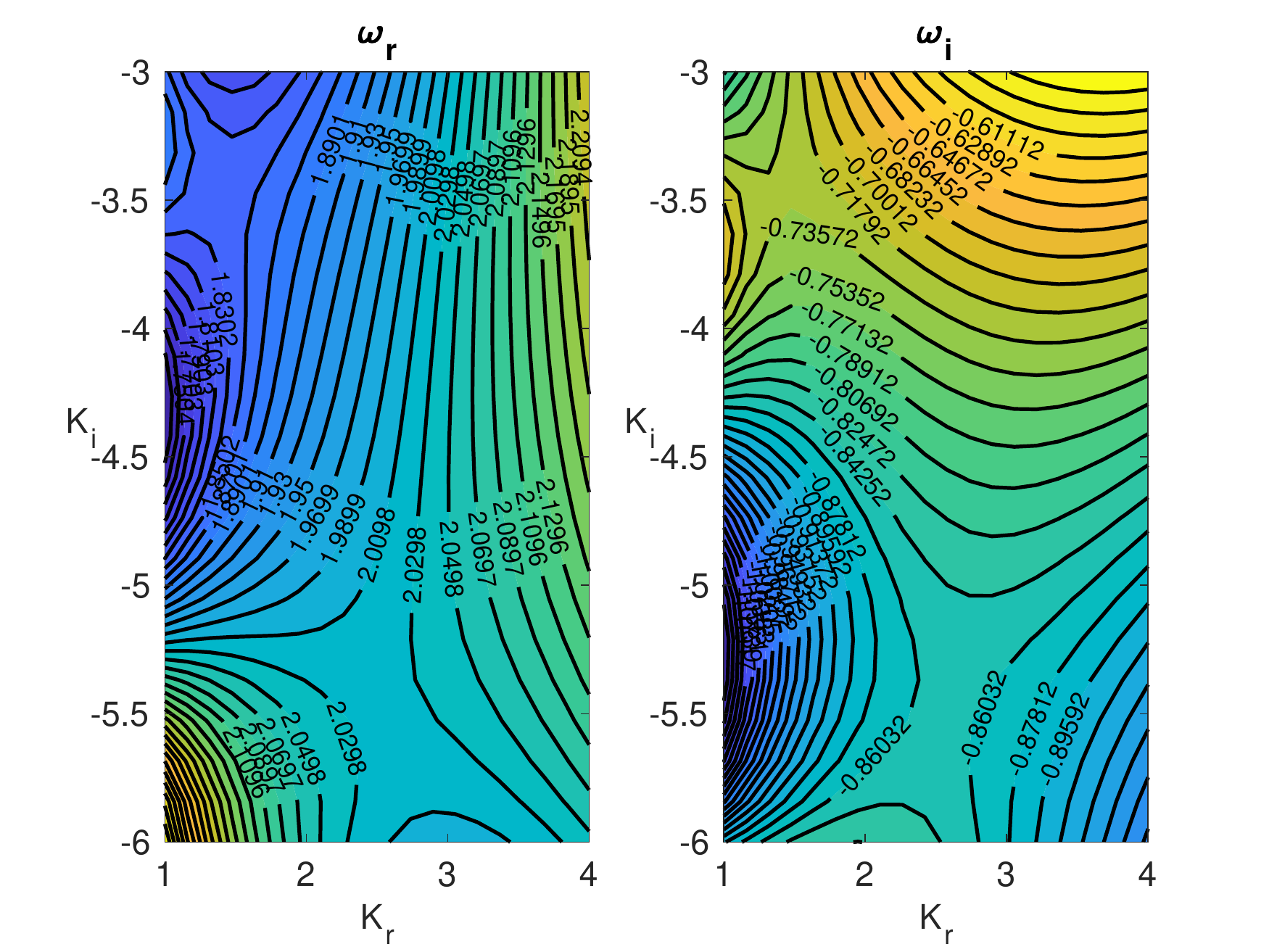}
\caption{}
\end{subfigure}
\caption{ Contours of frequency and growth rate on the complex wavenumber plane, showing convective instability of low viscosity jet with $\delta=0.05$ for profile I: (a) $\beta=0$ with the saddle point at $k_0=2.002-5.0461i$ and $\omega_0 = 2.0179-0.7750i$ (b) $\beta=1$ with the saddle point at $k_0=2.5488-5.4831i$ and $\omega_0=2.0255-0.8573i$ . Other relevant parameters are set to the values $(Re,M,Sc,\theta,\theta_\mu)=(1000,40,100,0.05,0.01)$.}\label{fig:AU_Profile_I}
\end{figure}

Figure \ref{fig:AU_Profile_I}(a) and (b) shows contours of $\omega$ for a low-viscosity jet (M=40, Re=1000) and a radially inward shifted velocity profile. The growth rate at the saddle point location is negative for both axisymmetric and helical modes. A similar set of parameters used in the experiments of \citet{Srinivasan2023} yielded self-sustained oscillations with helical modes, whose origin is tentatively assumed to lie in an initially linear perturbation. When attempting to reconcile these disparate findings, one notes that the stability analysis presented so far assumes an arbitrary value of $\delta$ that needs to be delineated more clearly, either through theory or experiment. 

In resolving this issue, two approaches can be followed: (i) one can recognize that in an experimental facility, the nozzle geometry controls the boundary layer at the nozzle exit plane for a given Re. Therefore, for a laminar exit profile, the momentum thickness $\theta$ has an inverse relationship with the square-root of the Reynolds number, which can be written as $D/\theta = a + b/\sqrt{Re}$, where $a$ and $b$ are experimentally derived constants, D is the nozzle diameter and $\theta$ is dimensional momentum boundary layer thickness; (ii) Alternatively, one can consider that for a given thickness over which the viscosity jump occurs, the boundary layer growth is controlled by the balance between advection and spatially varying momentum diffusion, so that the velocity profile can be deduced by solution of a Blasius-type equation with spatially varying viscosity. In this second case, the velocity profile is solved as a function of prescribed viscosity profile and Re, from which the momentum thickness is calculated. The second approach is followed henceforth, though results from the first approach are also presented in the following section. 

\begin{equation}
f(\eta)^{\prime \prime \prime}+\frac{1}{2 \mu_0} f(\eta) f(\eta)^{\prime \prime}+\frac{\mu_0^{\prime}}{\mu_0} f(\eta)^{\prime \prime}=0
\end{equation}

For prescribed values of the viscosity ratio, viscosity profile and Reynolds number, this nonlinear ordinary differential equation can be solved by imposing appropriate boundary conditions at infinity and ensuring continuity of velocity and shear stress across the diffusive interface (see Appendix for details). 

Figure~(\ref{fig:similarity}) shows the similarity solution for three different values of viscosity ratio. Also shown are tanh-profiles used in this study, whose parameter values ($\theta$, $\delta$) have been chosen to fit these similarity profiles. The solid line is the similarity solution and the discrete points indicate the $\tanh$ profiles with different shift distances of $\delta_U$ and momentum thickness of $\theta$. While there is a difference between the two profiles for large radial distances, these are unlikely to affect the stability characteristics, and we note that the shear layer region is captured accurately by a tanh-profile with the appropriate values of $\theta$ and $\delta$ (see Table \ref{tab:similaritytable}). Of course, the similarity solution is required in order to arrive at these values of $\theta$ and $\delta$,  however this does appear to justify the use of tanh-profiles for the preceding results of temporal and spatio-temporal analysis. 

\begin{figure}
\centerline{
\includegraphics[width= 0.5\textwidth]{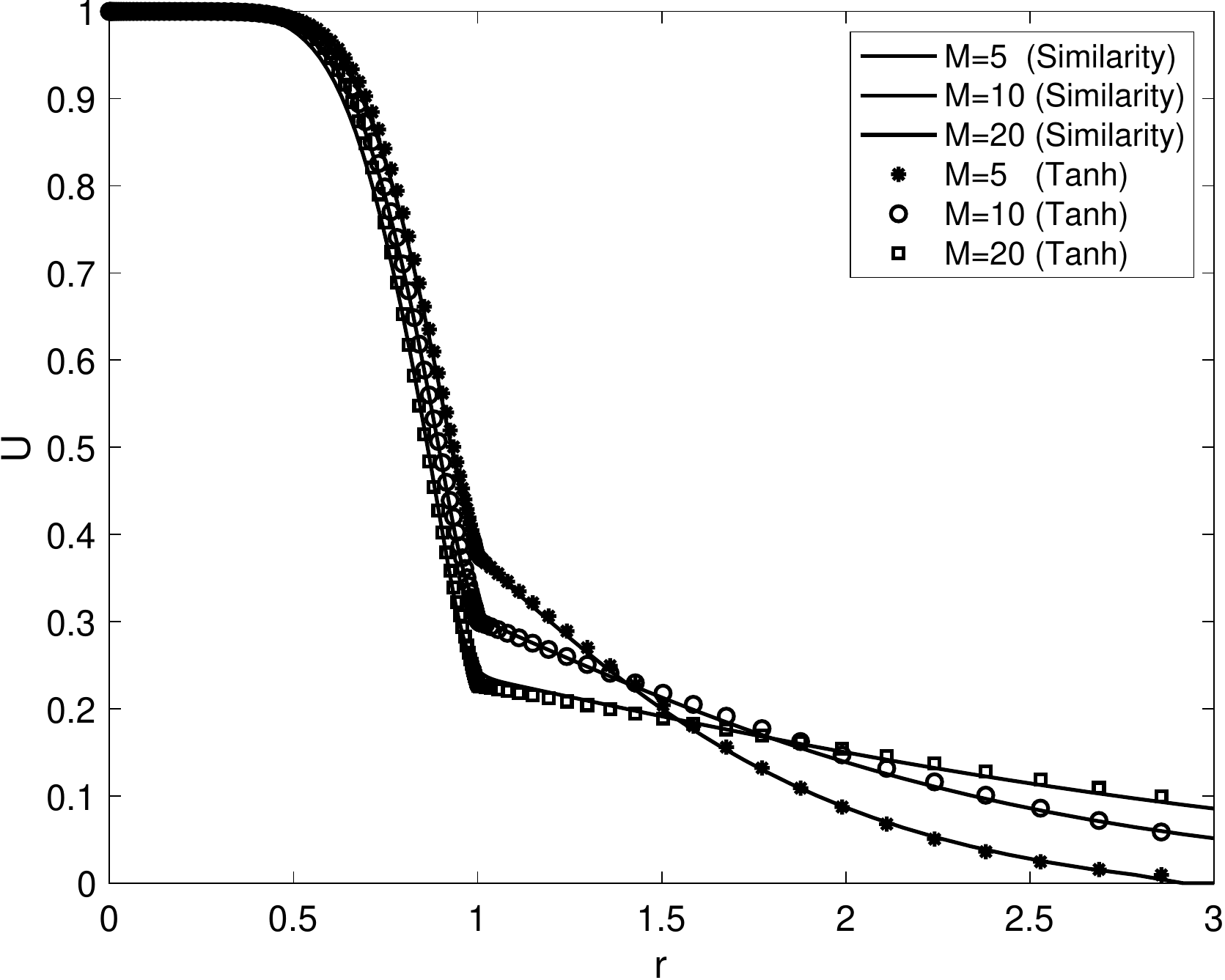}}
\caption{Similarity solution for a jet with variable viscosity.}
\label{fig:similarity}
\end{figure}

\begin{table}
\centering
\begin{tabular}{ccccccccc}
\hline$M$ & 60 & 56 & 48 & 40 & 32 & 25 & 17 & 9 \\
\hline $\delta$ & $0.185$ & $0.181$ & $0.177$ & $0.169$ & $0.158$  & $0.146$ & $0.131$ & $0.108$ \\
\hline $\theta$ & 0.096 & $0.096$  & $0.100$ & $0.104$ & $0.104$ & $0.108$ & $0.115$ & $0.115$ \\
\hline
\end{tabular}
\caption{Values of $\theta$ and $\delta$ obtained for various values of M from the boundary-layer equation when $\theta_mu = 0.01$.}
\label{tab:similaritytable}
\end{table}


\begin{equation}
U = \frac{1+\text{Tanh}\left.[\frac{1}{4\theta}(\frac{1}{r+\delta}-r-\delta)\right.]}{2}
\end{equation}

\begin{figure}
\centering
\begin{subfigure}{0.49\textwidth}
\includegraphics[width= \textwidth]{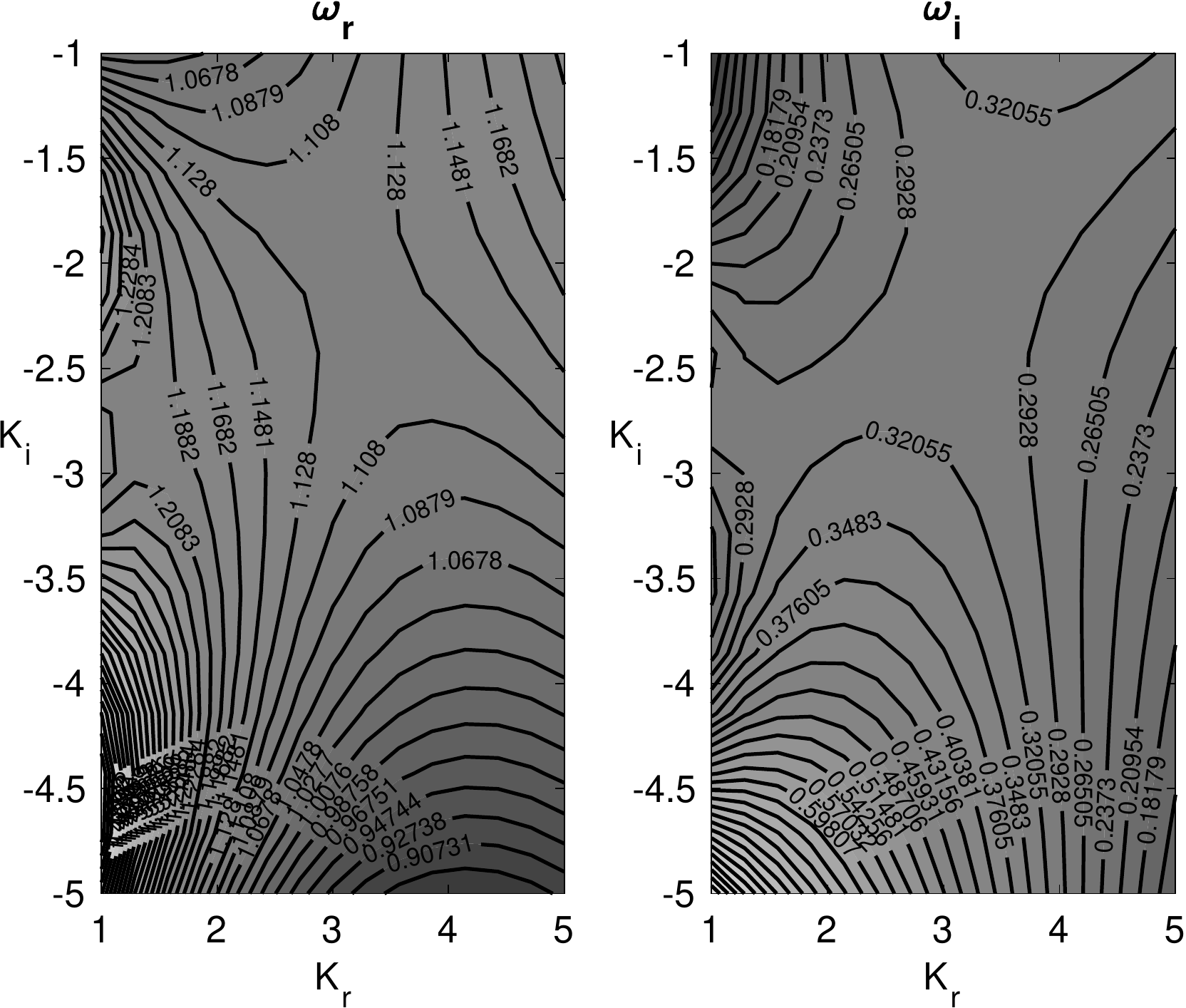}
\caption{}
\end{subfigure}
\begin{subfigure}{0.49\textwidth}
\includegraphics[width= \textwidth]{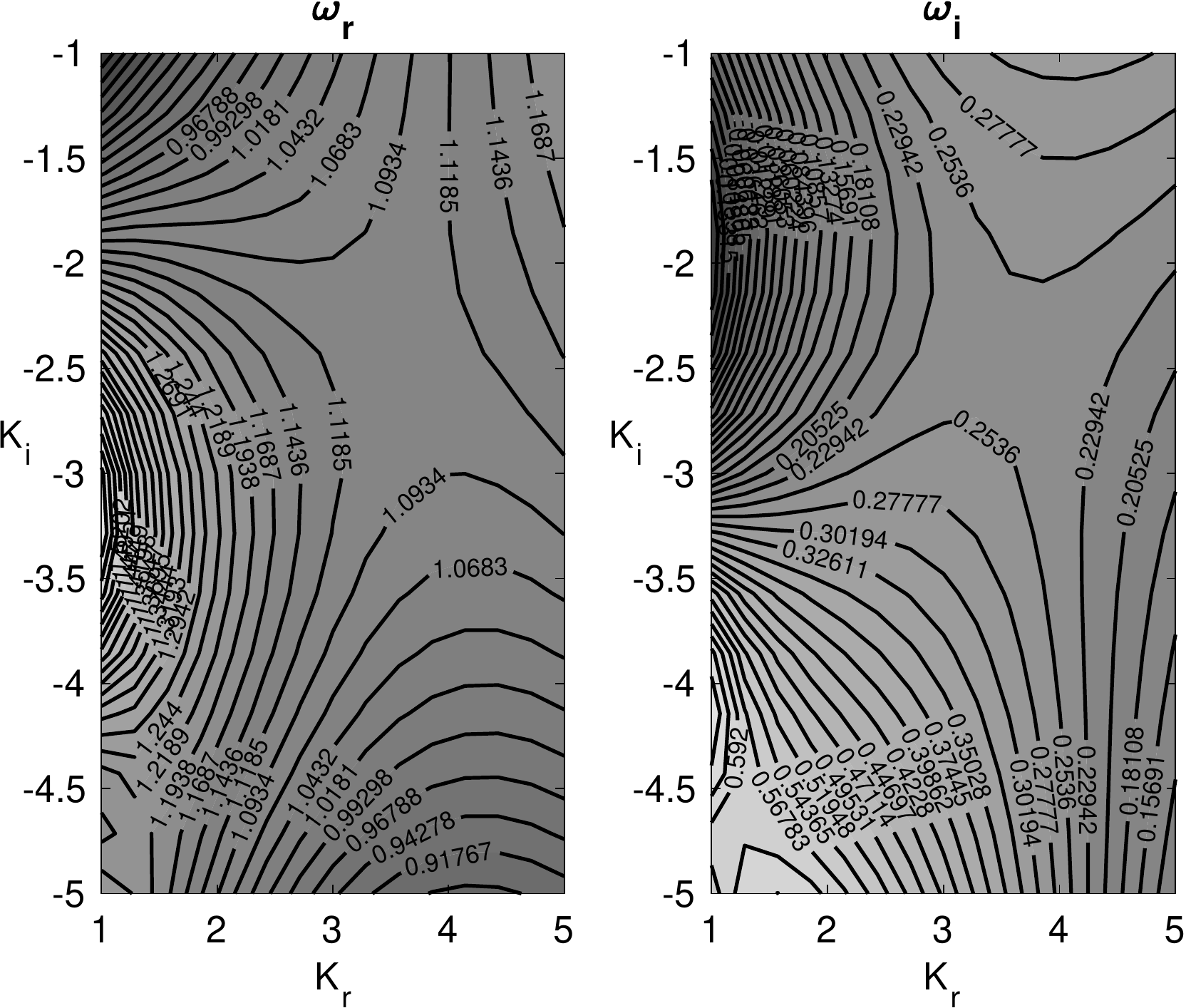}
\caption{}
\end{subfigure}
\caption{Contours of frequency and growth rate on the complex wavenumber plane, showing absolute instability in $\omega$ plane for both (a) $\beta=0$ and (b) $\beta=1$ when $(Re, M, Sc,\theta, \theta_\mu, \delta)=(750,40,100,0.104,0.01,0.169)$. The saddle points are $k_0=3.203-2.096i$, $\omega_0= 1.123+0.302i$ and $k_0=3.547-2.426i$, $\omega_0 = 1.107+0.249i$.}
\label{fig:AU_M60}
\end{figure}

Figure \ref{fig:AU_M60} shows the complex wavenumber plane for velocity profiles obtained from the similarity solution, for a Reynolds number of 750, M=40 and a prescribed value of viscosity thickness of 0.01, representative of the near-field of the jet.  A total of four saddle points with positive growth rates are now found for the axisymmetric mode as well as the helical mode. Upon further examination, it was found that the modes near the upper half of the complex plane were always more dominant than the two interior of the negative half-plane. These more unstable modes (one axisymmetric and one helical) are the focus of further examination. The presence of these saddle points in the low-viscosity jet configuration appears to confirm that it is possible to trigger absolute instability through viscosity contrast alone in a free shear layer. For the case of a planar shear layer in the vicinity of a species concentration gradient,  \citet{Sahu2014} did perform a spatio-temporal analysis, but their study presents contours of complex frequency  for the case of countercurrent shear layers with significant counterflow; which is a well-documented phenomenon for planar constant property mixing layers \citep{Huerre1985, Forliti2005}. It is possible that the low viscosity ratio (2.718 and 0.368) may also have been a factor in not observing absolute instability without counterflow. In the rest of this article, we focus on characterizing these two absolutely unstable modes and the parameter space delineating their transition from convective to absolute instability, with particular emphasis on replicating the helical modes observed by \citet{Srinivasan2023} in their low-viscosity jets with a single nozzle geometry.  

\begin{figure}
\centering
\begin{subfigure}{0.4\textwidth}
\includegraphics[width= \textwidth]{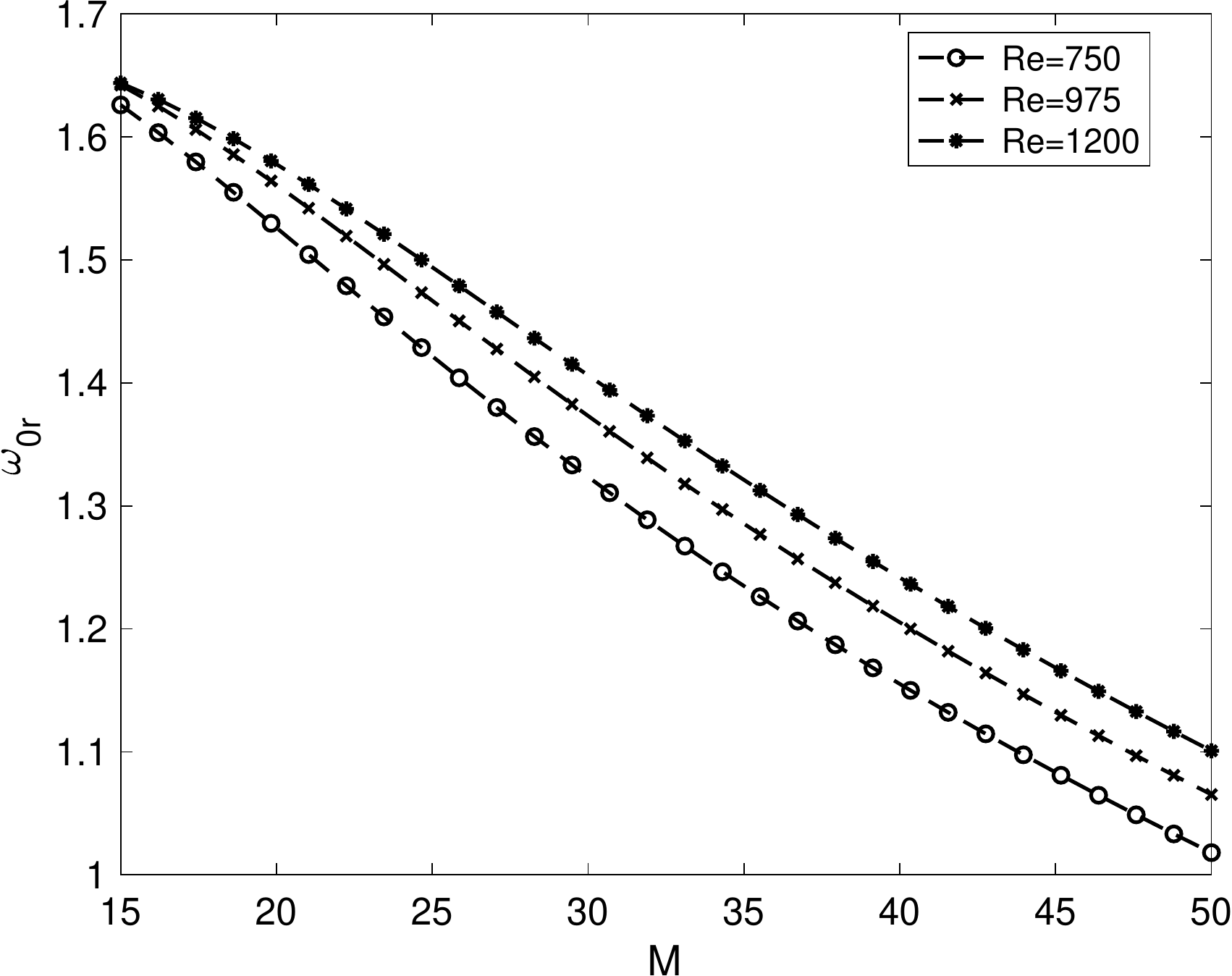}
\caption{}
\end{subfigure}
\begin{subfigure}{0.4\textwidth}
\includegraphics[width= \textwidth]{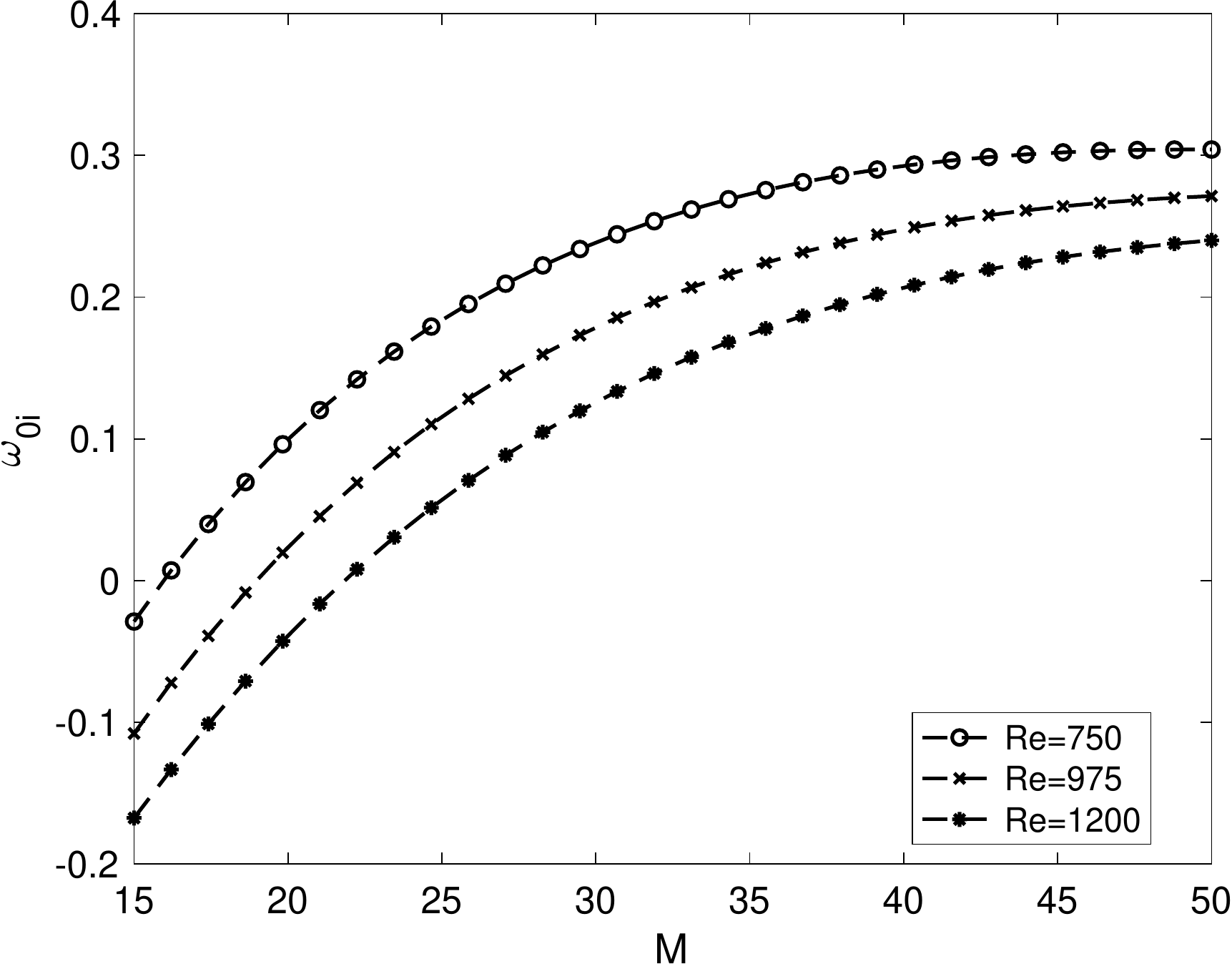}
\caption{}
\end{subfigure}
\begin{subfigure}{0.4\textwidth}
\includegraphics[width= \textwidth]{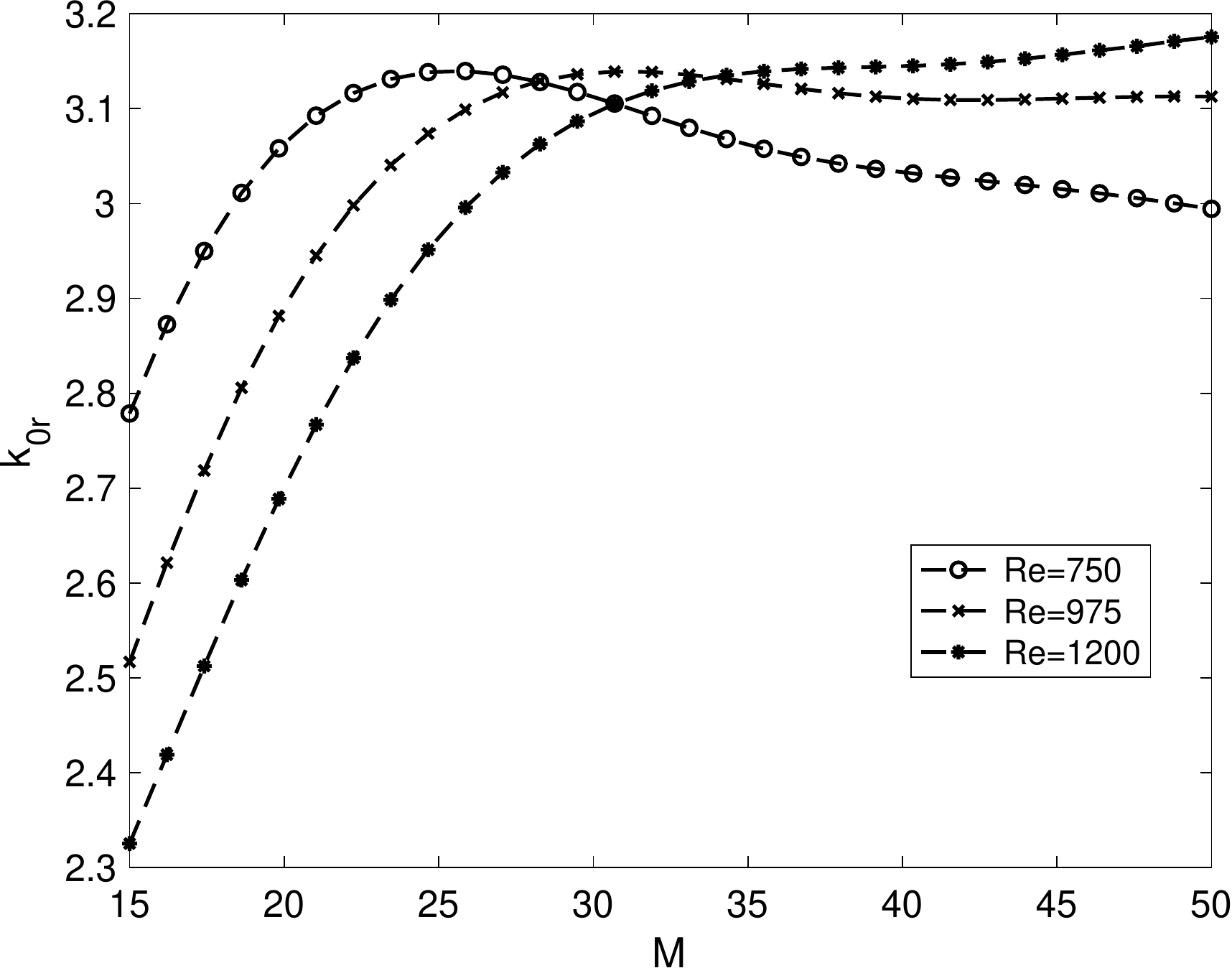}
\caption{}
\end{subfigure}
\begin{subfigure}{0.4\textwidth}
\includegraphics[width= \textwidth]{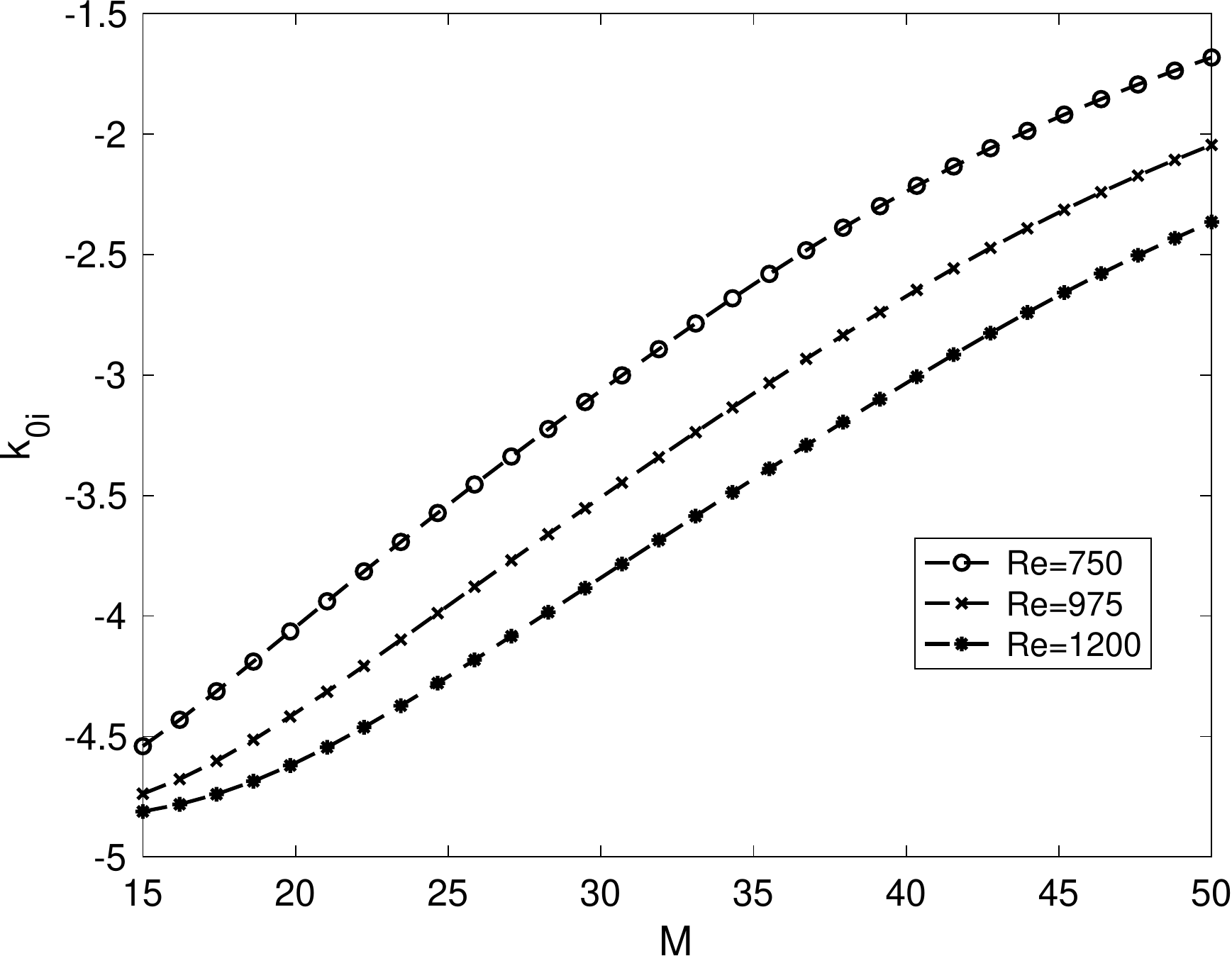}
\caption{}
\end{subfigure}
\caption{ Variation of $\omega$ and $k$ with viscosity for different $Re$ by tracking the saddle points for top left axisymmetric mode at $(Sc,\theta_\mu,\delta)=(100,0.01,0.06)$.}
\label{up_axi_mode}
\end{figure}

\begin{figure}
\centering
\begin{subfigure}{0.4\textwidth}
\includegraphics[width= \textwidth]{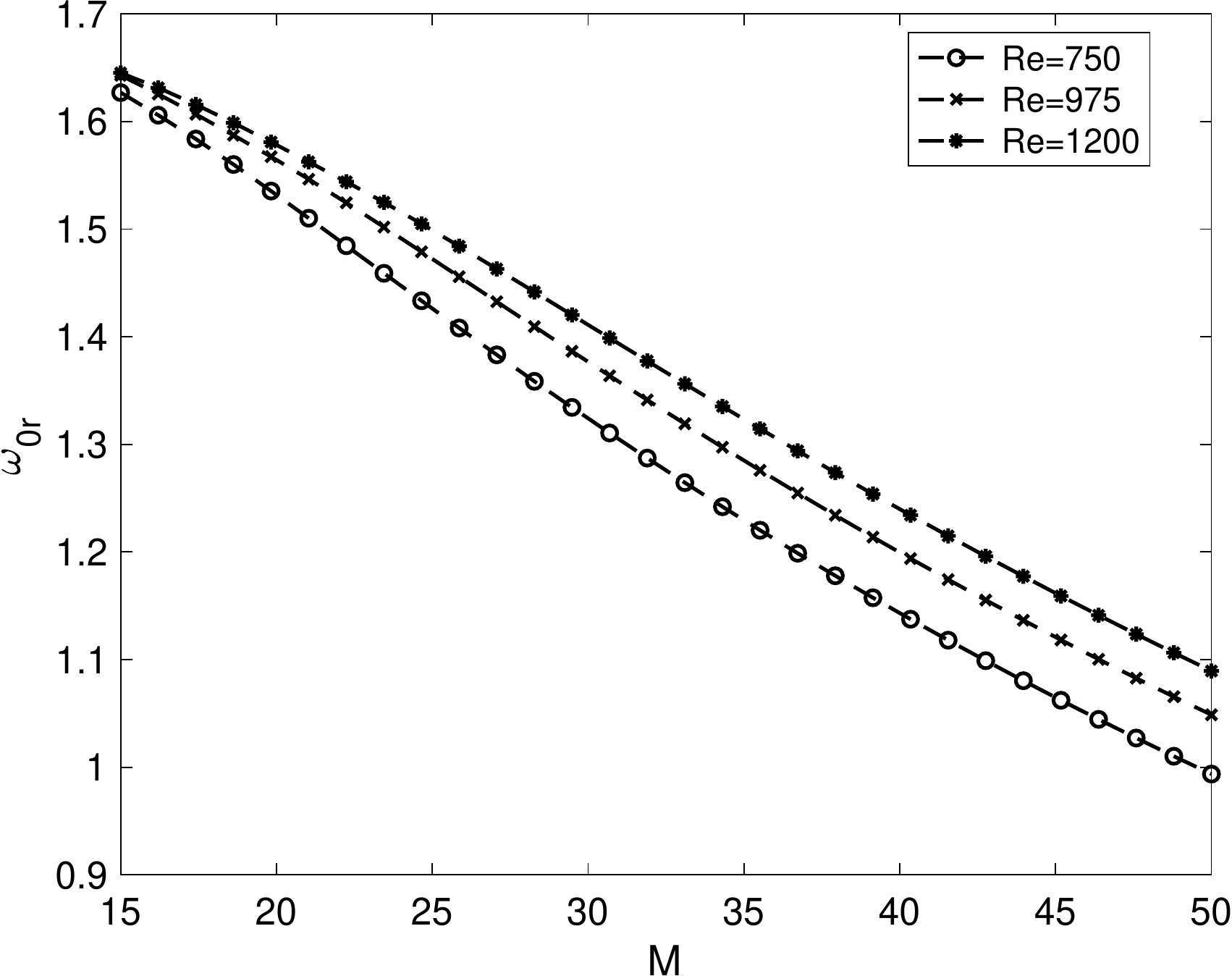}
\caption{}
\end{subfigure}
\begin{subfigure}{0.4\textwidth}
\includegraphics[width= \textwidth]{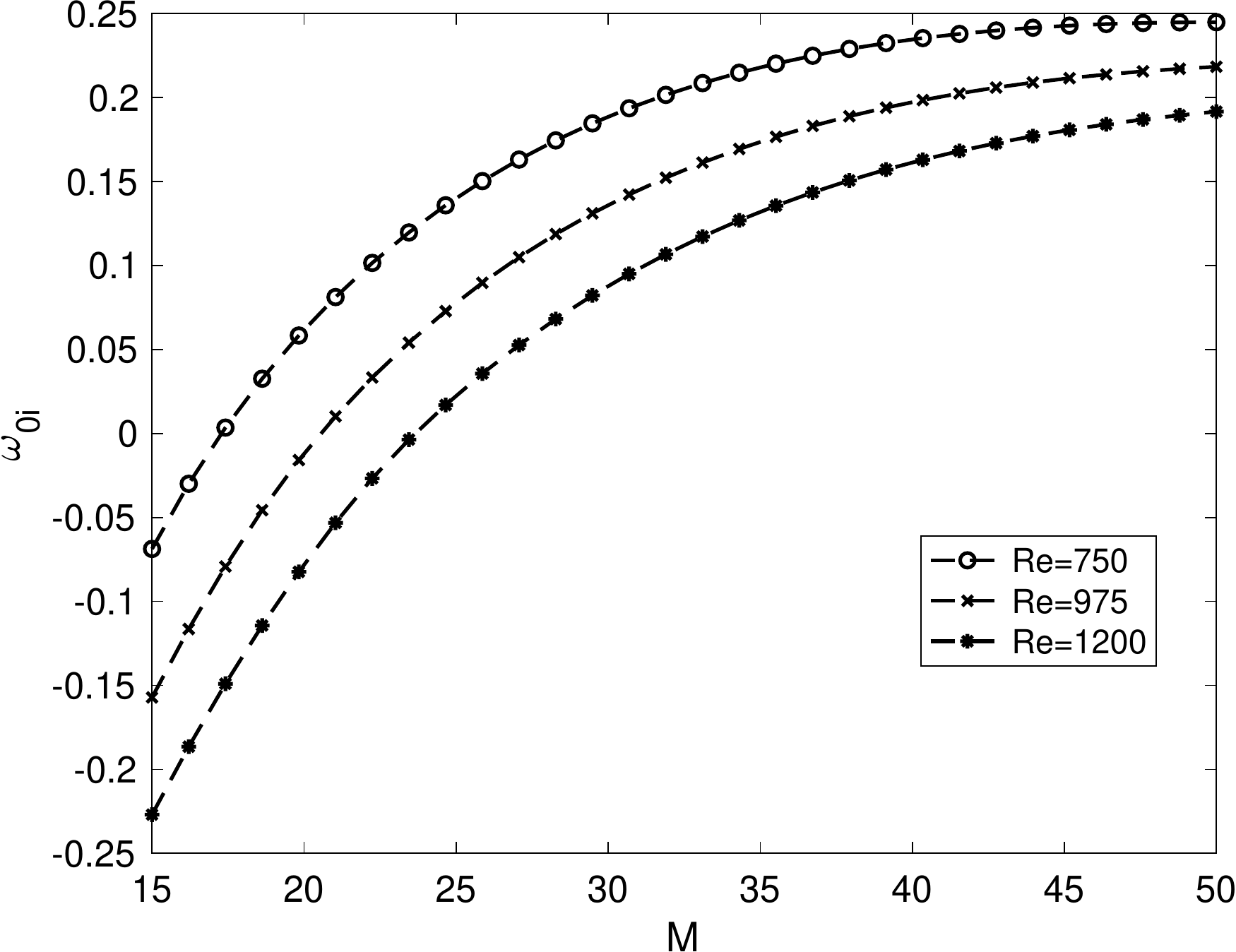}
\caption{}
\end{subfigure}
\begin{subfigure}{0.4\textwidth}
\includegraphics[width= \textwidth]{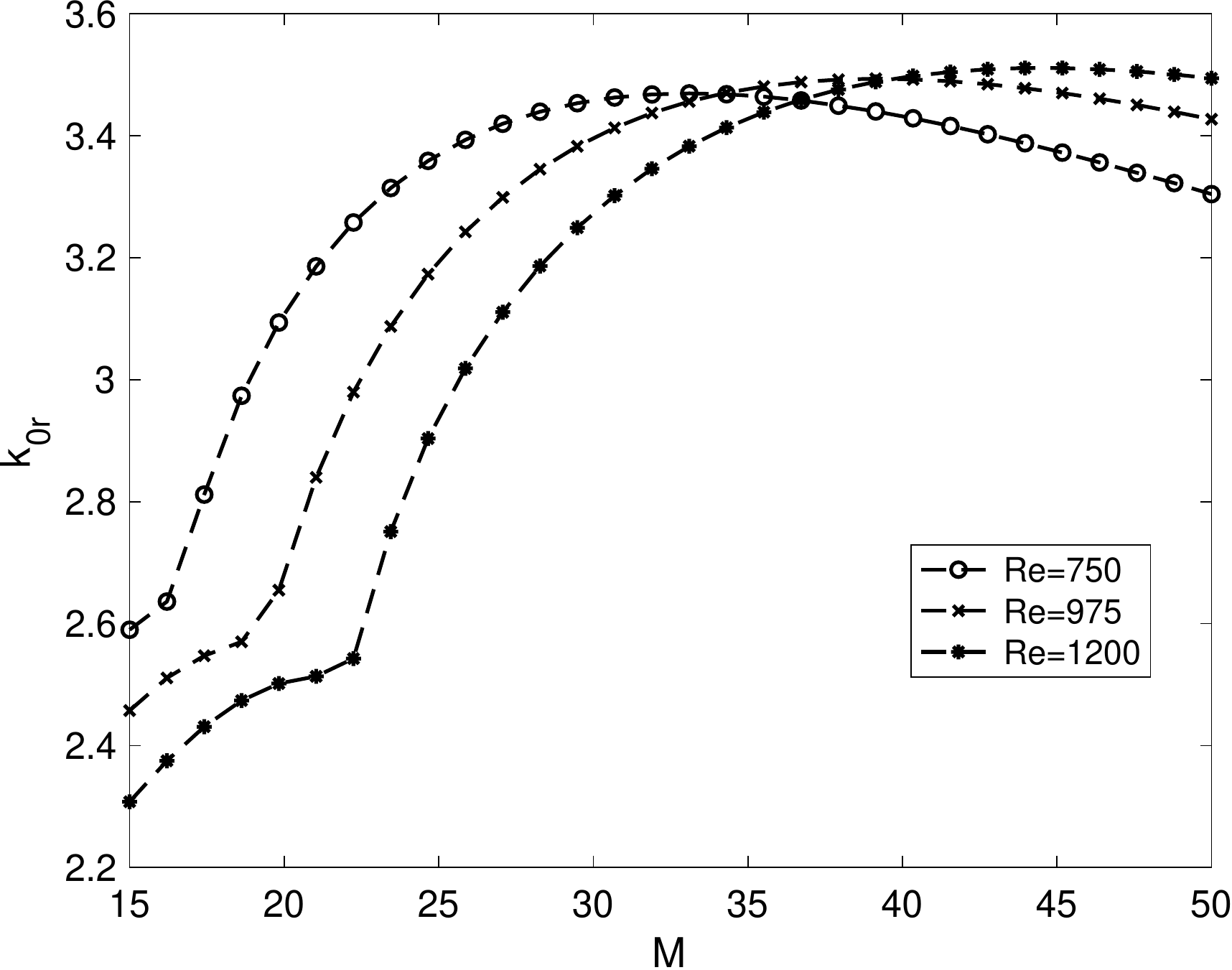}
\caption{}
\end{subfigure}
\begin{subfigure}{0.4\textwidth}
\includegraphics[width= \textwidth]{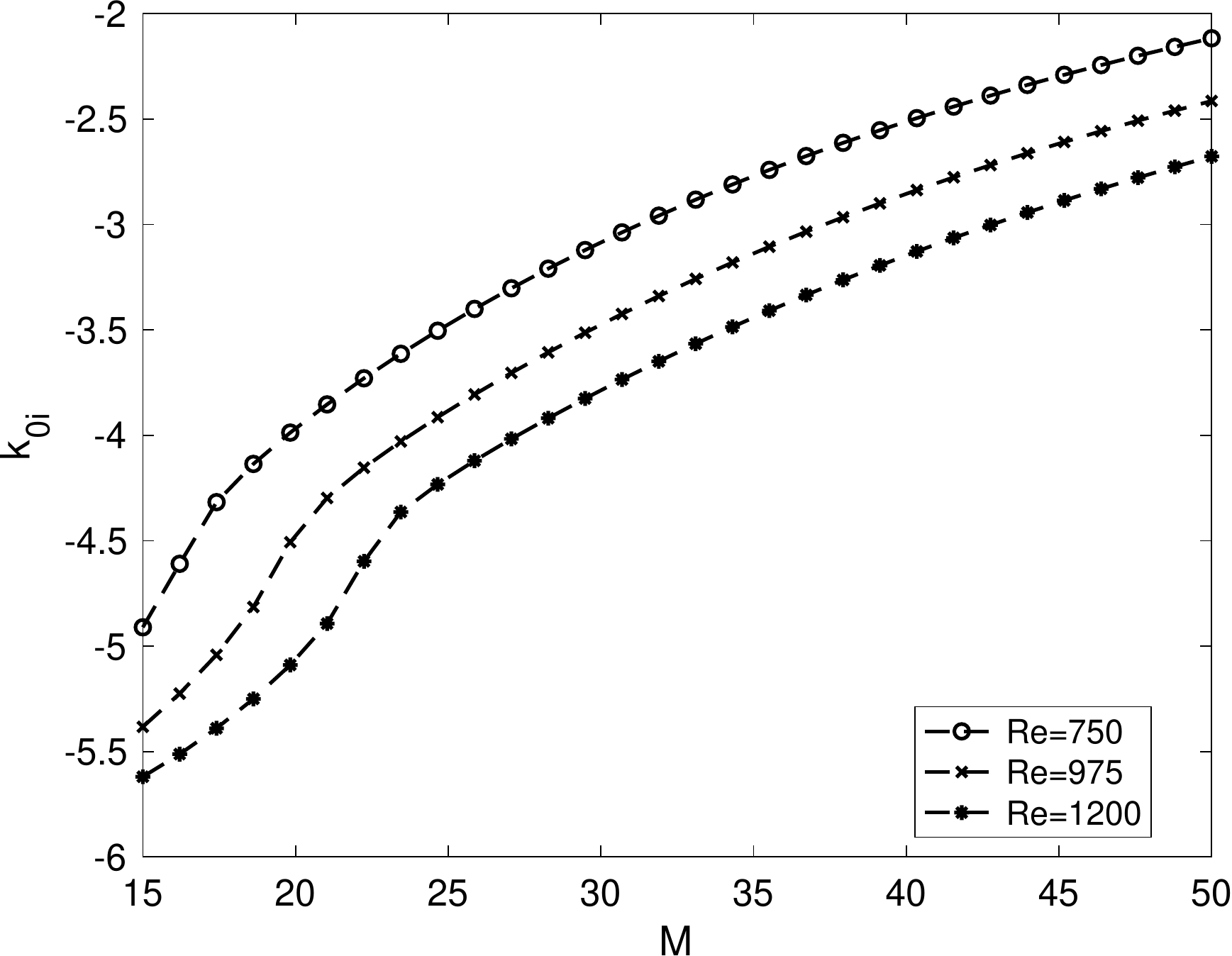}
\caption{}
\end{subfigure}
\caption{ Variation of $\omega$ and $k$ with viscosity for different $Re$ by tracking the saddle points for top right helical mode at $(Sc,\theta,\theta_\mu,\delta)=(100,0.08,0.005,0.06)$.}
\label{up_helical_mode}
\end{figure}

Figures \ref{up_axi_mode} and \ref{up_helical_mode} track the locus of saddle points for both axisymmetric and helical modes as a function of M for different Reynolds numbers and a fixed value of viscosity layer thickness, and velocity profiles based on the corresponding similarity solution.  The behavior of these parameters (wavenumber, spatial and temporal growth rates, and wavenumber) are largely the same for both modes, though differences exist in when the two modes become absolutely unstable. The growth rate is an increasing function of M, saturating at large N, with the transition to absolute instability occurring at lower M ($\sim $ 15-22) for the axisymmetric mode relative to the helical mode ($\sim $ 25-30). The real frequency  is a decreasing function of M for both  modes for $M >10$. From Fig. \ref{up_axi_mode}(b) and \ref{up_helical_mode}(b), one observes that the larger the Re, the smaller the $\omega_i$, implying that at large Re, the dominance of convective effects over diffusive transport may lead to weakening and potential suppression of this instability. The wavenumber is an increasing function of M at low M and eventually saturates, while the spatial growth rate appears to continuously increase with M.

\begin{figure}
\centering
\begin{subfigure}{0.4\textwidth}
\includegraphics[width= \textwidth]{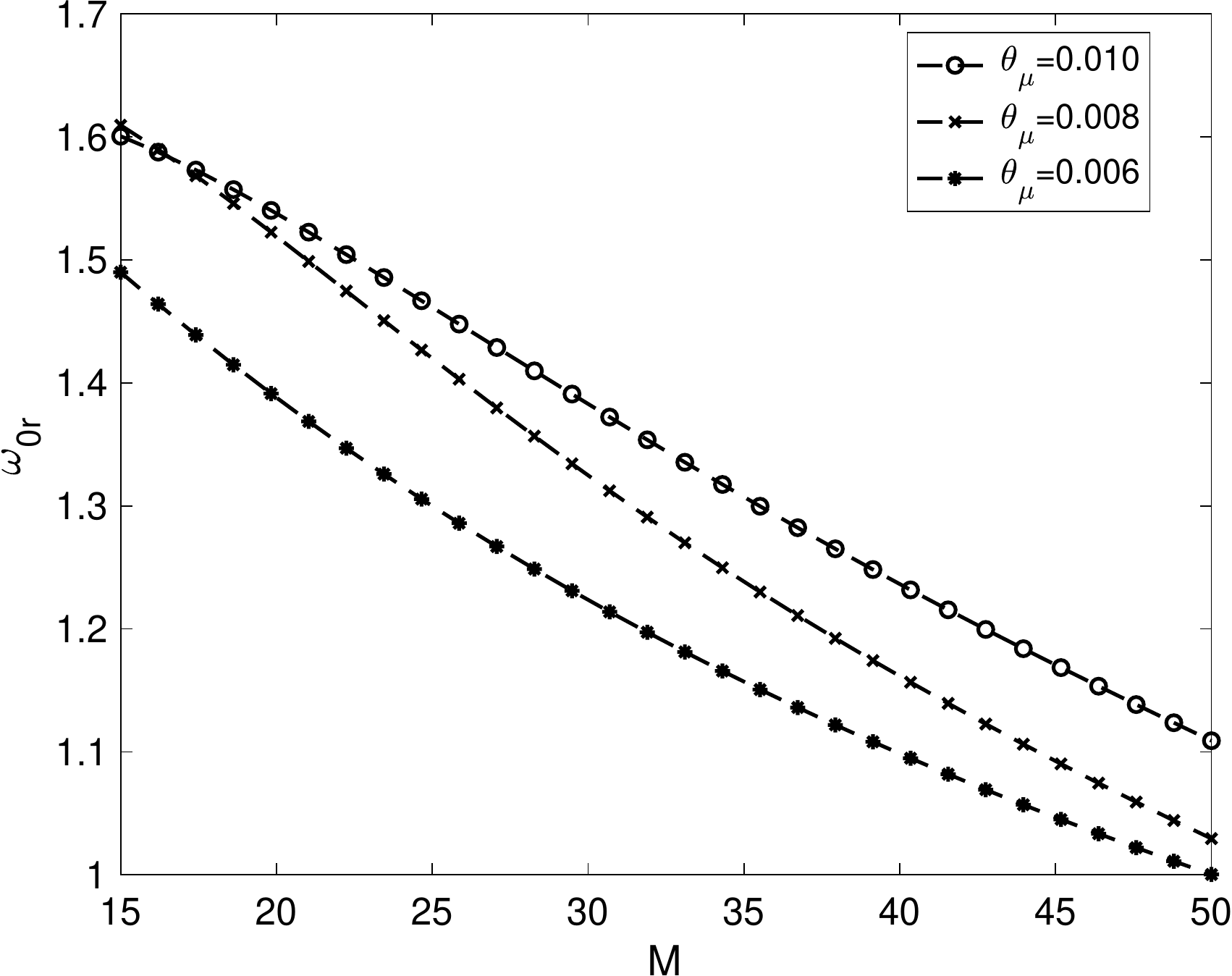}
\caption{}
\end{subfigure}
\begin{subfigure}{0.4\textwidth}
\includegraphics[width= \textwidth]{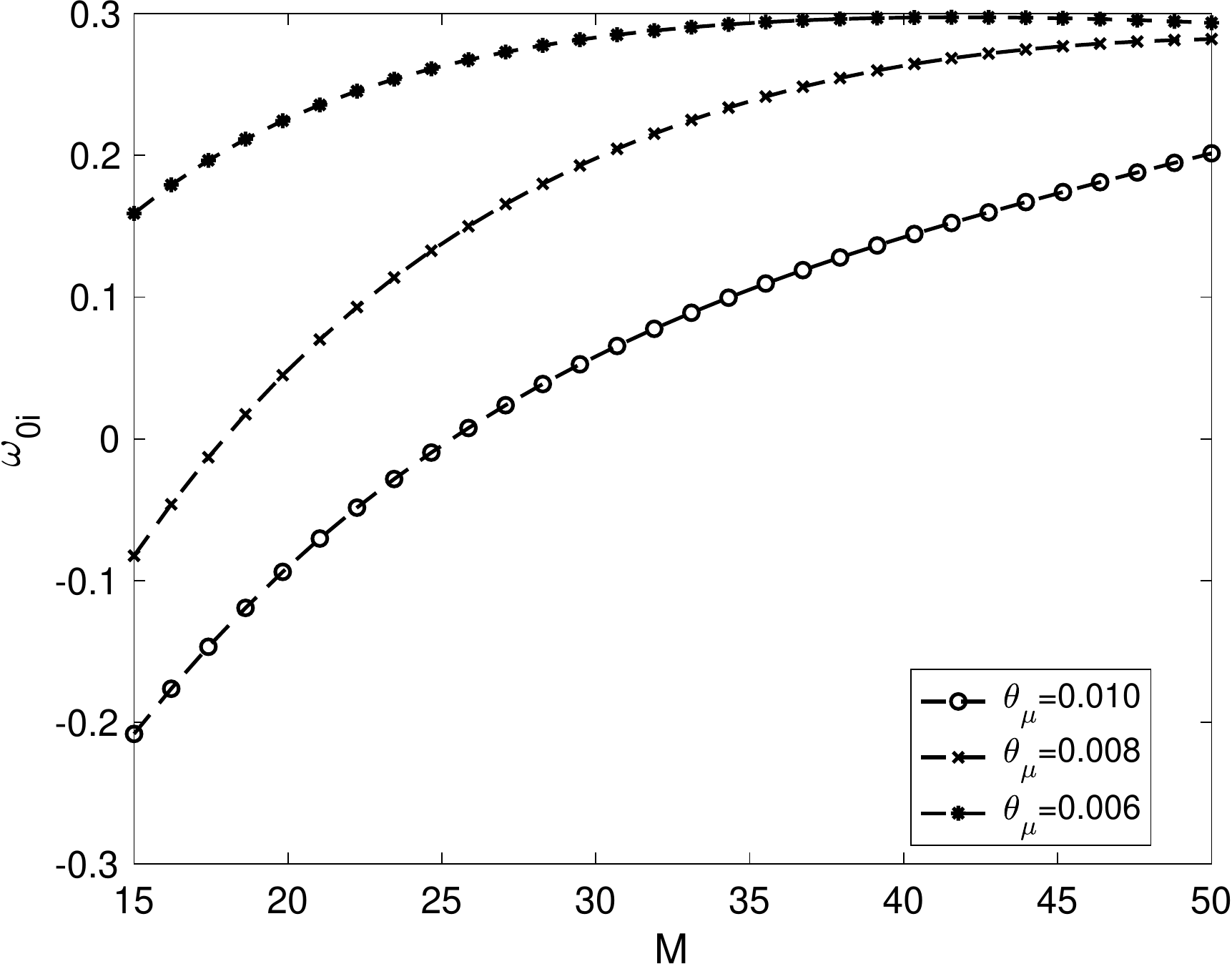}
\caption{}
\end{subfigure}
\begin{subfigure}{0.4\textwidth}
\includegraphics[width= \textwidth]{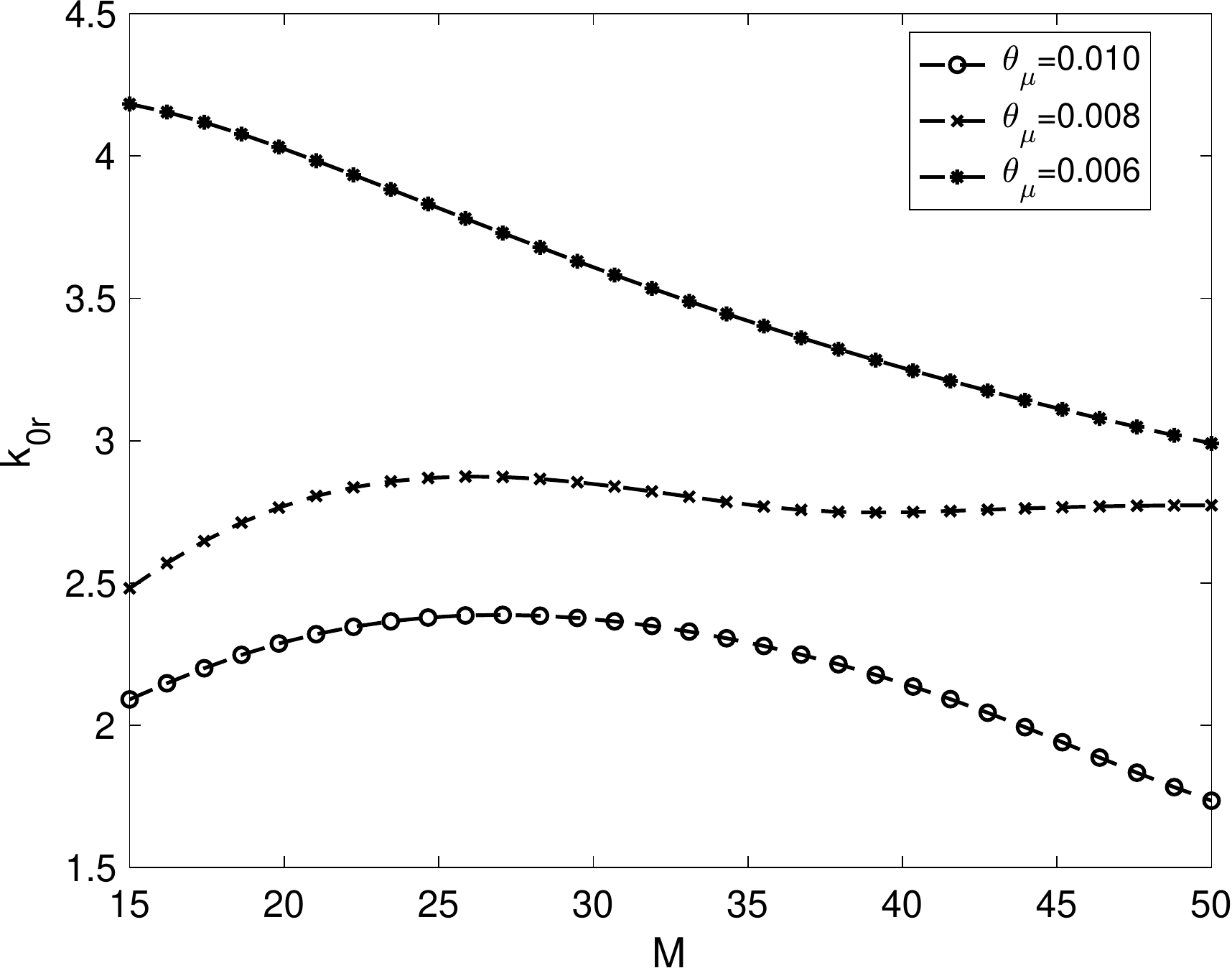}
\caption{}
\end{subfigure}
\begin{subfigure}{0.4\textwidth}
\includegraphics[width= \textwidth]{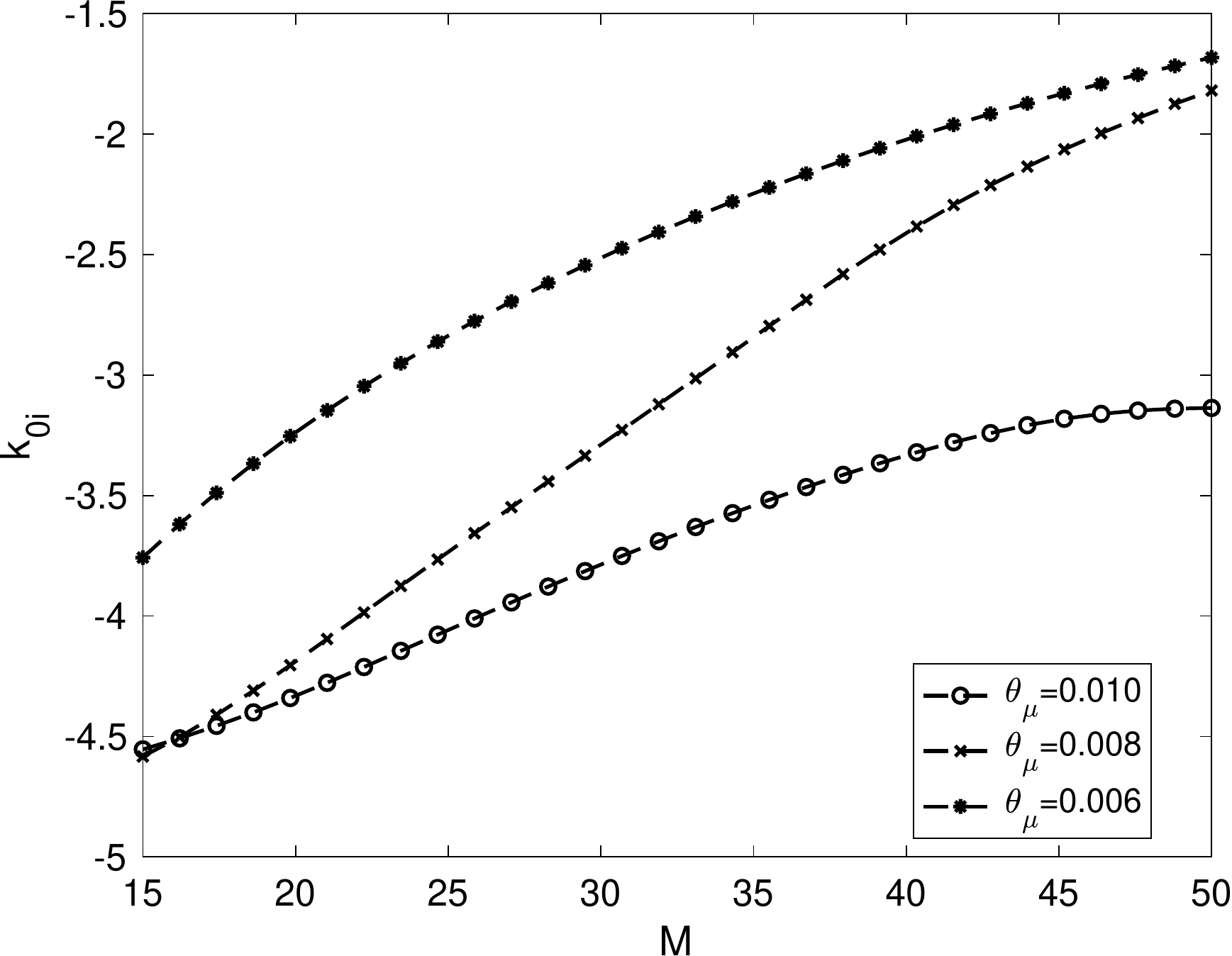}
\caption{}
\end{subfigure}
\caption{ Variation of saddle point $\omega$ and $k$ of the dominant axisymmetric mode  with viscosity ratio M and three values of viscosity thickness. Velocity profiles are based on the similarity solution at each condition. The fixed parameters are  $(Re, Sc,\theta_\mu)=(750, 100, 0.005)$.} 
\label{asy_up_mode_thetaM}
\end{figure}

\begin{figure}
\centering
\begin{subfigure}{0.4\textwidth}
\includegraphics[width= \textwidth]{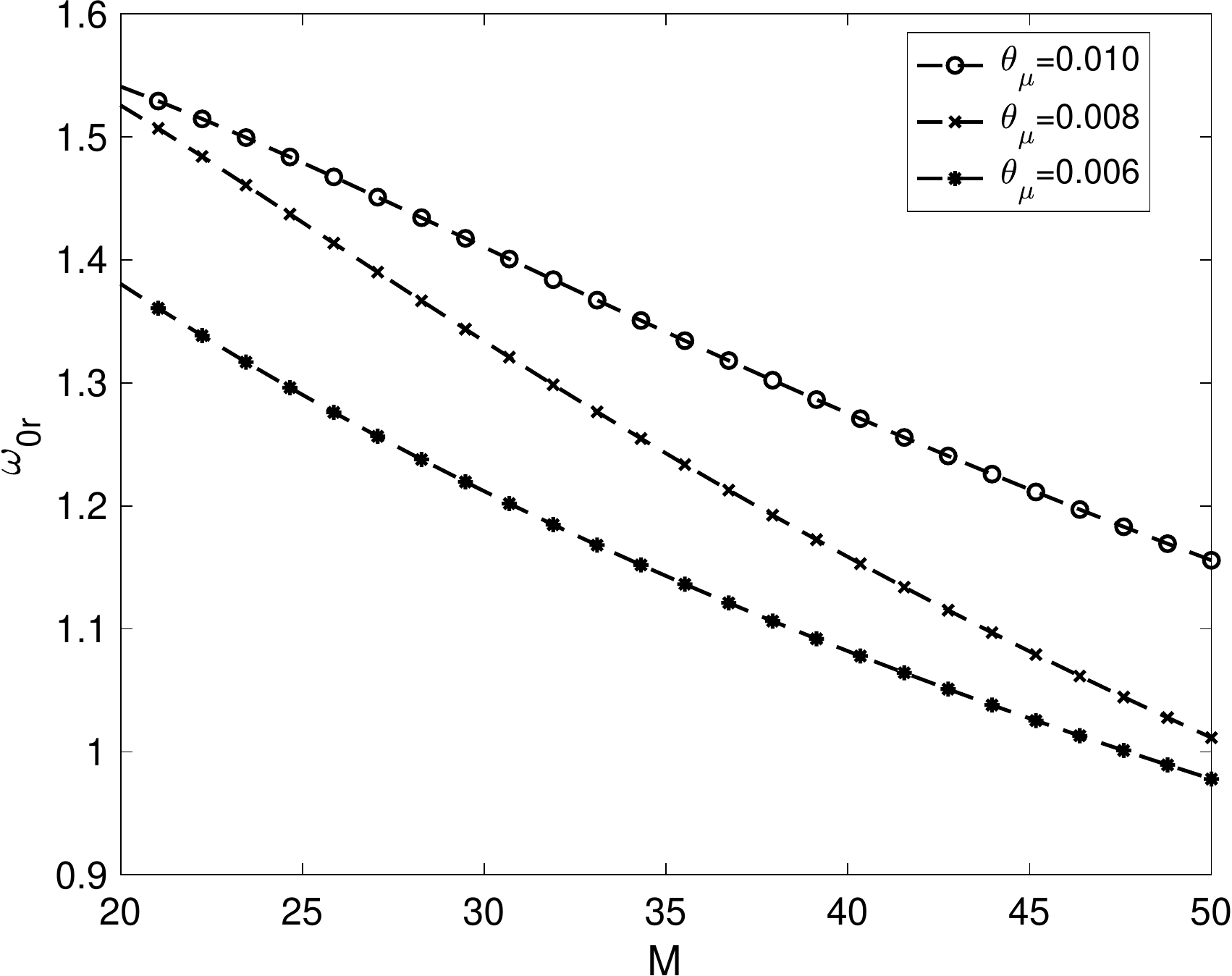}
\caption{}
\end{subfigure}
\begin{subfigure}{0.4\textwidth}
\includegraphics[width= \textwidth]{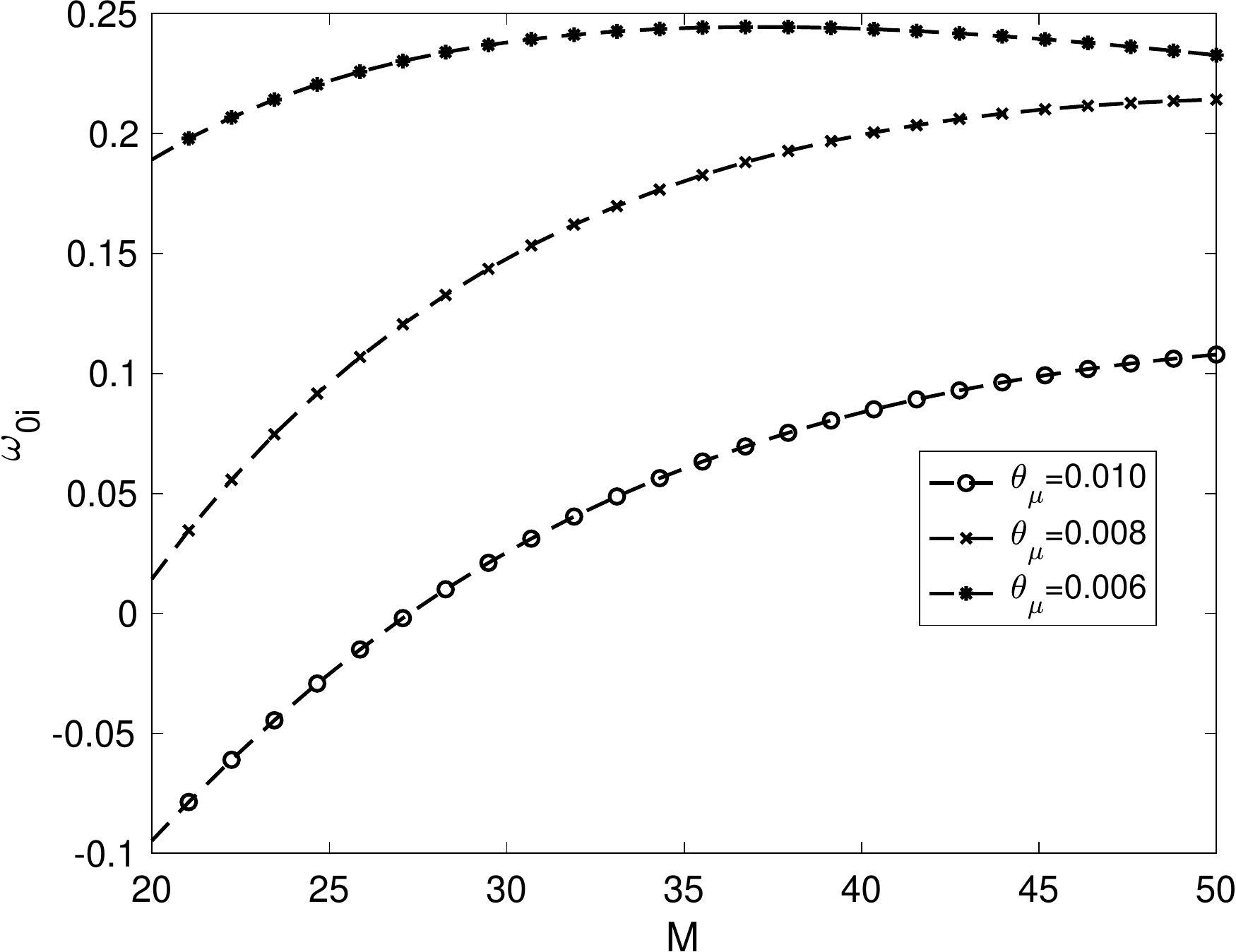}
\caption{}
\end{subfigure}
\begin{subfigure}{0.4\textwidth}
\includegraphics[width= \textwidth]{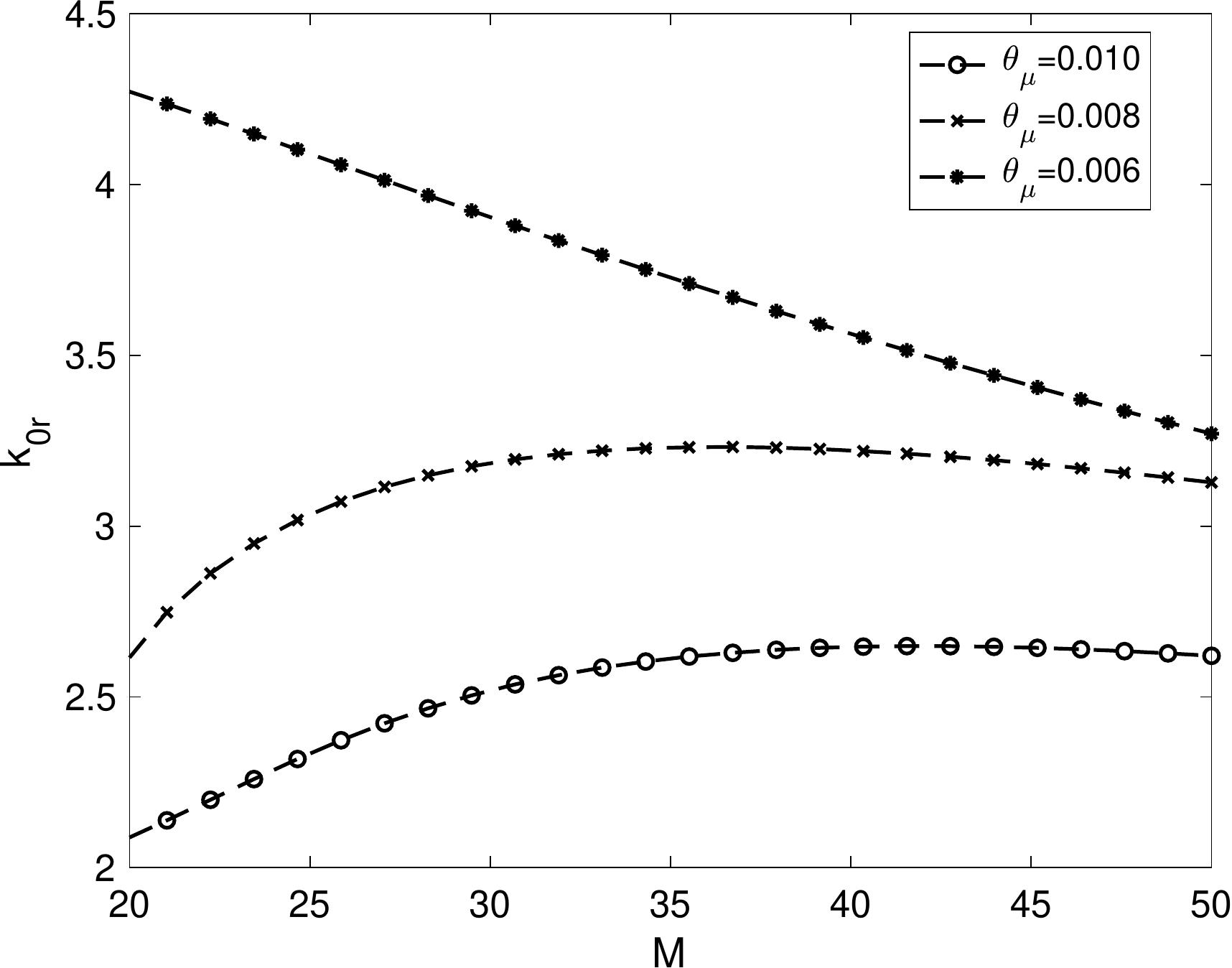}
\caption{}
\end{subfigure}
\begin{subfigure}{0.4\textwidth}
\includegraphics[width= \textwidth]{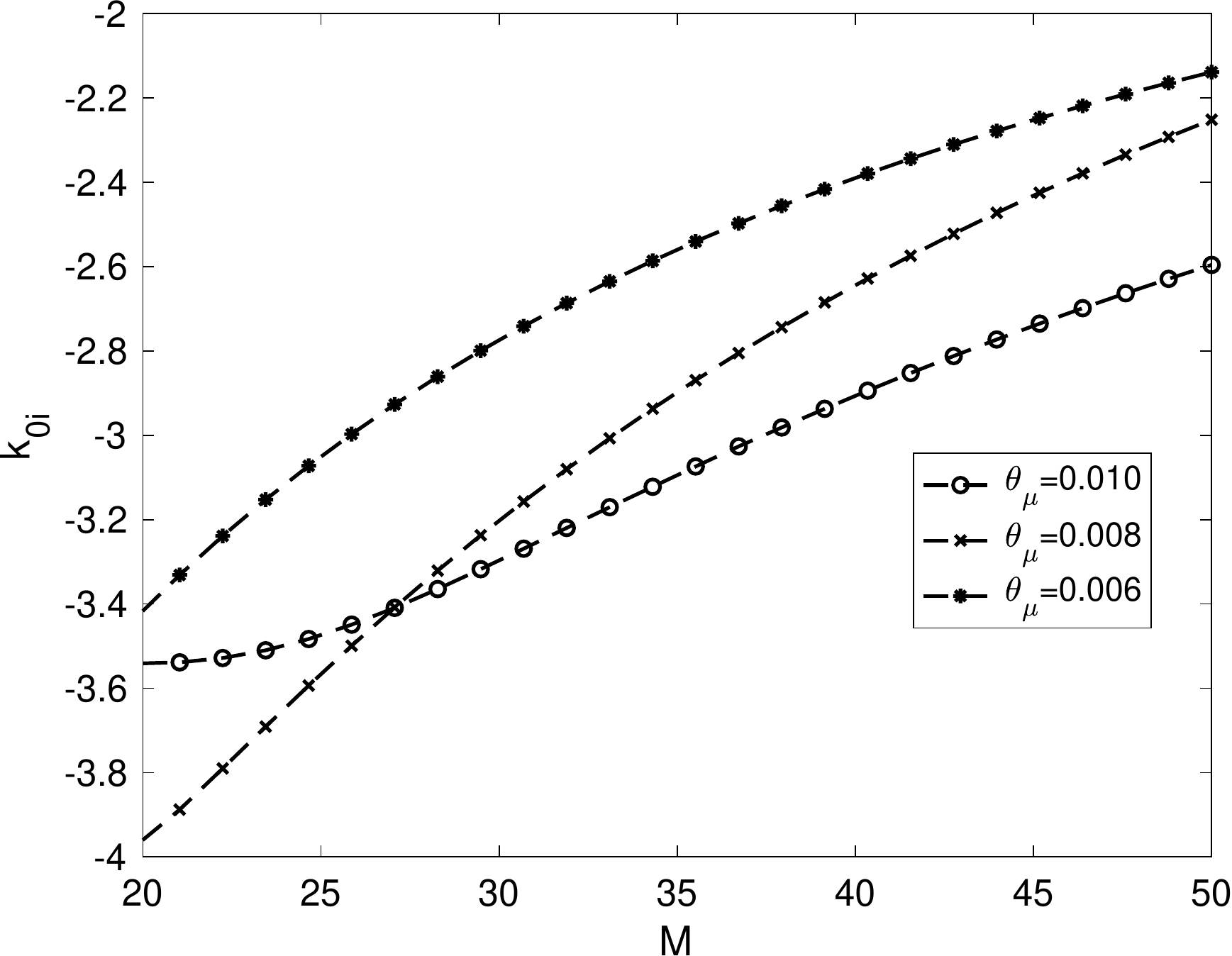}
\caption{}
\end{subfigure}
\caption{ Variation of saddle point $\omega$ and $k$ of the dominant helical mode  with viscosity ratio M and three values of viscosity thickness. Velocity profiles are based on the similarity solution at each condition. The fixed parameters are  $(Re,Sc,\theta_\mu)=(750, 100, 0.005)$.} 
\label{Helical_up_mode_thetaM}
\end{figure}

Figures \ref{asy_up_mode_thetaM}  and \ref{Helical_up_mode_thetaM} explore the behavior of these modes with respect to $\theta +\mu$ and M for fixed Re. As expected, sharper gradients in viscosity (smaller $\theta_mu$) lead to increased temporal growth rates for both axisymmetric and helical modes (Figs. \ref{asy_up_mode_thetaM}b  and \ref{Helical_up_mode_thetaM}b). Further, the wavenumber of the unstable mode is strongly influenced by $\theta_mu$, with higher wavenumbers at smaller $\theta_mu$ (Figs. \ref{asy_up_mode_thetaM}c  and \ref{Helical_up_mode_thetaM}c). Reduction in $\theta_mu$ also increases the spatial growth rate at fixed M (Figs. \ref{asy_up_mode_thetaM}d  and \ref{Helical_up_mode_thetaM}d).

\subsection{Comparison with experiments}

The above results strongly suggest that over a wide range of Re and  viscosity ratio M, the axisymmetric mode is more unstable than the helical mode. However, in the experiments of \citet{Srinivasan2023}, injection of salt water into propylene glycol invariably produced self-sustained helical modes oscillating at a discrete frequency, as confirmed by Fourier analysis of jet images and hot film anemometry. Disturbance frequencies of the helical mode qualitatively followed the trends in Fig. \ref{up_helical_mode}, with frequency increasing as the tank of glycol was diluted (M decreased). Once the viscosity ratio dropped below some critical (Re-dependent) value, axisymmetric modes were observed, though the nature of this mode (self-sustained or convective) was not examined. This raises the question of whether the velocity profiles being considered in the present study are the most appropriate for comparison with experiments. Despite the similarity solution yielding velocity profiles that are dependent on M, without any \textit{a priori} assumption of tanh-type behavior, it still differs from experiment, as discussed previously. Notably, the momentum thickness at the start of the species diffusion process is not zero and is determined by the jet Reynolds number. 

The transition from helical to axisymmetric modes observed by \citet{Srinivasan2023} is first attempted to be predicted by assuming that the momentum thickness $\theta$ and Re are linked, and M and $\theta_\mu$ can be prescribed as independent parameters, while $\delta$ is taken from Table \ref{tab:similaritytable}. With this, saddle points are sought in the complex plane with $\omega _{0i}=0$ and this absolute/convective transition boundary is mapped for helical and axisymmetric modes in the M-Re plane, corresponding to the two control parameters in the experiment. 

\begin{figure}
\centering
\includegraphics[width= 0.6\textwidth]{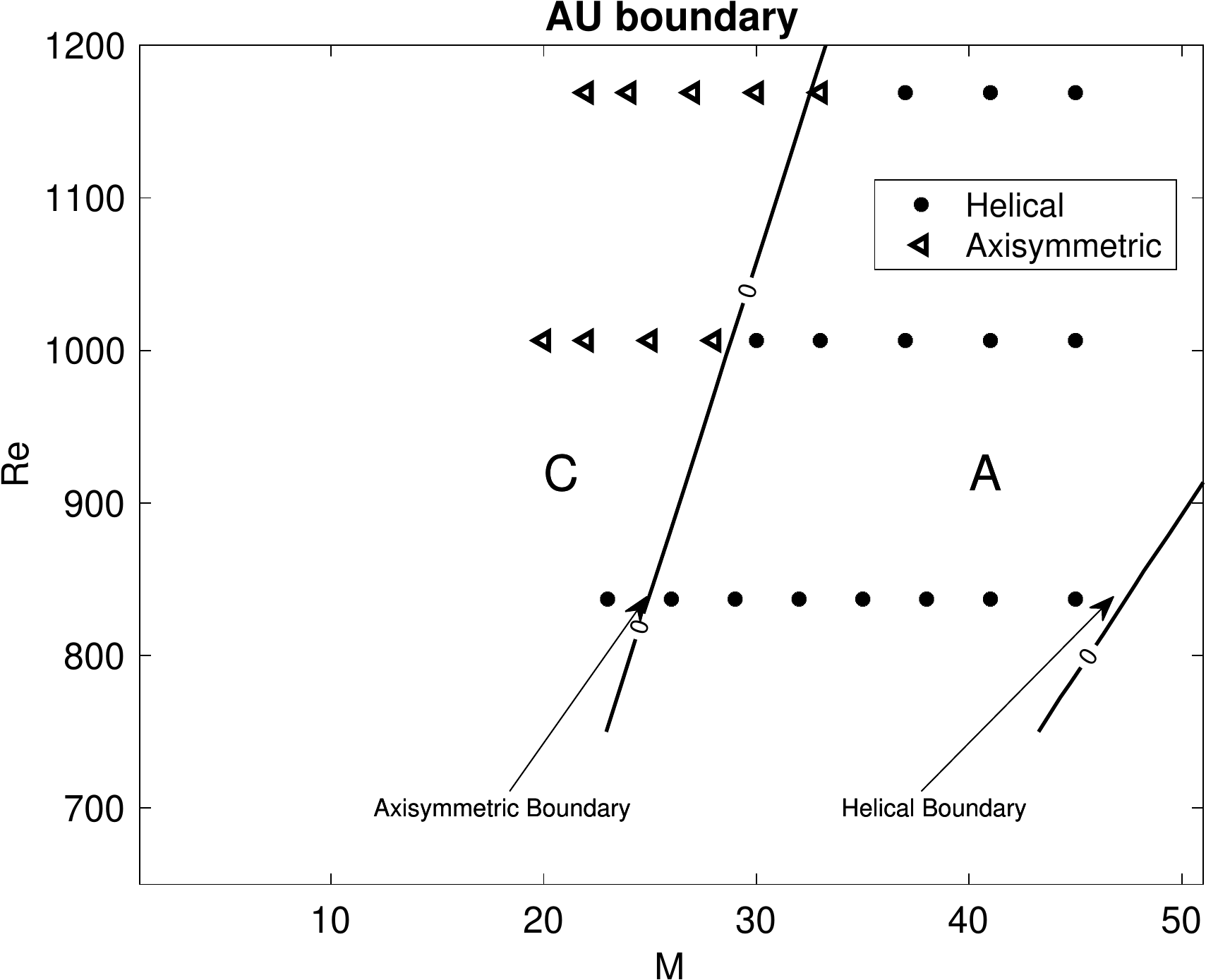}
\caption{ Absolute and convective instability transition boundaries in the M-Re plane when $(Sc,\theta_\mu)=(100,0.01)$. Velocity profiles are tanh- profiles with $\delta$ based on the corresponding similarity solution, and $\theta$ taken from experiments.}
\label{AC_boundary}
\end{figure}

The results are shown in Fig. \ref{AC_boundary}. Discrete symbols mark the values of Re and M where images were acquired in the experiment. The convective/absolute stability boundary still suggests that at high values of M, the axisymmetric mode is expected to be more unstable. Aside from this major drawback, it is also clear that the behavior of the transition boundary does not quite match the experiments.

\begin{figure}
\centering
\includegraphics[width= 0.6\textwidth]{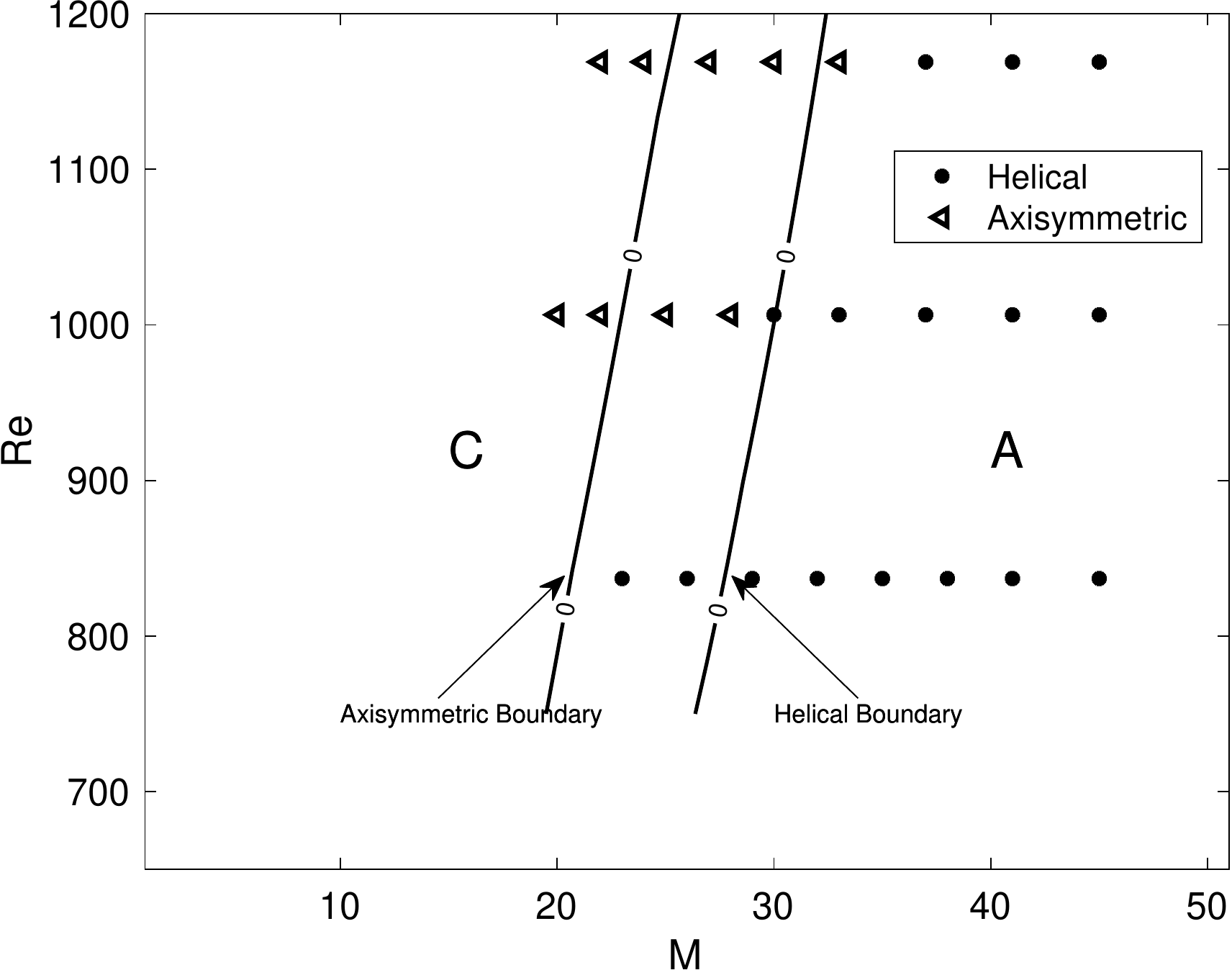}
\caption{ Absolute and convective instability transition boundaries compared to the experimentally observed transitions. Calculations are for velocity profiles taken from the similarity solution and $(Sc,\theta_\mu)=(100,0.01)$.}
\label{AC_boundary_shift}
\end{figure}

Since the use of experimentally measured values of $\theta$ fails to improve agreement with theory, we return to the similarity solution. When profiles entirely based on the similarity solution are used in order to calculate the absolute/convective transition boundary, the results shown in Fig. \ref{AC_boundary_shift} are obtained. As before, a major issue that remains is the prediction of a more unstable axisymmetric mode at large M and constant Re, that is not supported by the experiments. However, the transition boundary of the helical mode appears to match very well with that encountered in the experiments. This leads to the possibility that the experimental facility, through some unknown combination of boundary and/or inlet conditions, or for other unclear reasons, somehow favors the selection of the helical mode at the expense of the axisymmetric mode at high M. It should be noted that M was decreased below the transition value of M for the helical mode, axisymmetric modes were observed; but were not investigated in sufficient detail to understand their nature. 

\section{Summary and Discussion} 

A temporal stability analysis has been carried out for a round jet emerging into a medium of higher viscosity. The temporal stability analysis suggests that the near-critical behavior of the axisymmetric and helical modes are substantially different. However, at higher Re, for a broad range of conditions, the two modes are nearly equally unstable, and both are more unstable than the constant-viscosity jet. The additional destabilization is attributed to the presence of extra terms in the kinetic energy equation, which represents the interaction of the mean velocity gradient with the viscosity field; other source terms include the coupling of the velocity fluctuations to the mean viscosity gradient. The base profiles used in this study reflect an assumption of retardation of mean velocity by a more viscous ambient, leading to reduced gradients in the species diffusion layer, and making the apparent effect of M fairly weak when considering temporal growth rates. The axial disturbance velocity does not decay to zero at the centerline and therefore communicates across the jet diameter. The instability wavelength scales on the momentum thickness for low-viscosity jets. 

When accompanied by a radial inward shift, and sufficiently high values of M, velocity profiles representing the extreme near-field of the jet support absolute instability. The validity of the tanh-type velocity profiles used in the temporal analysis is checked by obtaining a similarity solution from the boundary layer equations for variable viscosity, and verifying that the resulting solution admit tanh- profiles with appropriate parameter values. Over a parameter space defined by (M,Re, $\theta_mu$), two absolutely unstable modes are predicted by the present analysis, with an axisymmetric mode being triggered at lower M and becoming progressively more unstable as M is increased, with the helical mode establishing itself at higher M. Both modes become more unstable as the viscosity gradients become sharper, with spatial and temporal growth rates increasing as M is increased and/or $\theta_mu$ is decreased. The wavelength of the unstable modes also follows the behavior of $\theta_mu$. 

As far as the authors are aware only \citet{Srinivasan2023} in the same group has experimentally investigated a similar configuration for viscosity ratios as high as 50, and at Reynolds numbers above 1000.  In that study, the experiments were carried out by progressively diluting a large tank of high-viscosity fluid, with initial measurements showing helical modes which later turned to axisymmetric modes. Discrete frequencies were observed in the helical mode spectra, but in some cases, they were also observed in the axisymmetric data, which were attributed to some unknown source of noise. 

Attempts to resolve this discrepancy by solving a Blasius-type equation for the velocity profile are successful in capturing the transition boundary for the helical mode. However, the use of such profiles does not explain why axisymmetric modes are not observed at large M.  Without an experimental measurement of the velocity profiles in the real flow, this question may be hard to answer. One possibility is that in the jet near-field, finite-thickness effects of the nozzle lip alter the velocity profile, adding a velocity defect. The boundary layer analysis presented here also does not consider the initial momentum thickness already established inside the nozzle, and therefore may not be representative of the real flow. These issues require significant consideration, with further measurements and/or high fidelity-computations, and are reserved for a future study.  \\

\paragraph{\textbf{Acknowledgements}} We gratefully acknowledge support for this work from the National Science Foundation (Grant CBET/2023932). We are also grateful for useful suggestions and feedback from D. Forliti during the preparation of this manuscript.  

\paragraph{\textbf{Declaration of Interests}} The authors report no conflicts of interest. 
\clearpage
\bibliographystyle{jfm}
\bibliography{refs}

\begin{thebibliography}{56}
\expandafter\ifx\csname natexlab\endcsname\relax\def\natexlab#1{#1}\fi
\def\au#1{#1} \def\ed#1{#1} \def\yr#1{#1}\def\at#1{#1}\def\jt#1{\textit{#1}}
  \def\bt#1{#1}\def\bvol#1{\textbf{#1}} \def\vol#1{#1} \def\pg#1{#1}
  \def\publ#1{#1}\def\arxiv#1{#1}\def\org#1{#1}\def\st#1{\textit{#1}}

\bibitem[Aiton \& Driscoll(2018)]{Aiton2018}
{\sc \au{Aiton, Kevin~W} \& \au{Driscoll, Tobin~A}} \yr{2018}  \at{An adaptive
  partition of unity method for chebyshev polynomial interpolation}.  \jt{SIAM
  Journal on Scientific Computing}  \bvol{40}~(1),  \pg{A251--A265}.

\bibitem[Bers(1983)]{Bers1983}
{\sc \au{Bers, A}} \yr{1983} {Space-time evolution of plasma
  instabilities-absolute and convective}.  \bt{In {\em Basic Plasma Physics:
  Selected Chapters, Handbook of Plasma Physics, Volume 1\/}},  \pg{p. 451}.

\bibitem[Boeck \& Zaleski(2005)]{Boeck2005}
{\sc \au{Boeck, Thomas} \& \au{Zaleski, St{\'e}phane}} \yr{2005}  \at{Viscous
  versus inviscid instability of two-phase mixing layers with continuous
  velocity profile}.  \jt{Physics of fluids}  \bvol{17}~(3),  \pg{032106}.

\bibitem[Boomkamp {\em et~al.\/}(1997)Boomkamp, Boersma, Miesen \&
  Beijnon]{Boomkamp1997a}
{\sc \au{Boomkamp, P.a.M.}, \au{Boersma, B~J}, \au{Miesen, R H~M} \&
  \au{Beijnon, G~V}} \yr{1997}  \at{{A Chebyshev Collocation Method for Solving
  Two-Phase Flow Stability Problems}}.  \jt{Journal of Computational Physics}
  \bvol{132}~(2),  \pg{191--200}.

\bibitem[Briggs(1964)]{Briggs1964}
{\sc \au{Briggs, R~J}} \yr{1964} {\em {Electron-stream interaction with
  plasmas}\/}.  \publ{Cambridge, MA: MIT Press}.

\bibitem[Cao {\em et~al.\/}(2003)Cao, Ventresca, Sreenivas \& Prasad]{Cao2003}
{\sc \au{Cao, Qing}, \au{Ventresca, Amy~L}, \au{Sreenivas, K~R} \& \au{Prasad,
  Ajay~K}} \yr{2003}  \at{{Instability due to Viscosity Stratification
  Downstream of a Centerline Injector}}.  \jt{The Canadian Journal of Chemical
  Engineering}  \bvol{81}~(October),  \pg{913--922}.

\bibitem[Chakravarthy {\em et~al.\/}(2015)Chakravarthy, Lesshafft \&
  Huerre]{Chakravarthy2015}
{\sc \au{Chakravarthy, RVK}, \au{Lesshafft, L} \& \au{Huerre, P}} \yr{2015}
  \at{Local linear stability of laminar axisymmetric plumes}.  \jt{Journal of
  Fluid Mechanics}  \bvol{780},  \pg{344--369}.

\bibitem[Chakravarthy {\em et~al.\/}(2018)Chakravarthy, Lesshafft \&
  Huerre]{Chakravarthy2018}
{\sc \au{Chakravarthy, RVK}, \au{Lesshafft, Lutz} \& \au{Huerre, P}} \yr{2018}
  \at{Global stability of buoyant jets and plumes}.  \jt{Journal of Fluid
  Mechanics}  \bvol{835},  \pg{654--673}.

\bibitem[Chomaz {\em et~al.\/}(1991)Chomaz, Huerre \& Redekopp]{Chomaz1991}
{\sc \au{Chomaz, Jean-Marc}, \au{Huerre, Patrick} \& \au{Redekopp, Larry~G}}
  \yr{1991}  \at{{A frequency selection criterion in spatially developing
  flows}}.  \jt{Studies in applied mathematics}  \bvol{84}~(2),  \pg{119--144}.

\bibitem[D'Olce {\em et~al.\/}(2008)D'Olce, Martin, Rakotomalala, Salin, Talon,
  D'Olce, Martin, Rakotomalala, Salin \& Talon]{DOlce2008}
{\sc \au{D'Olce, M}, \au{Martin, J}, \au{Rakotomalala, N}, \au{Salin, D},
  \au{Talon, L}, \au{D'Olce, M}, \au{Martin, J}, \au{Rakotomalala, N},
  \au{Salin, D} \& \au{Talon, L}} \yr{2008}  \at{{Pearl and mushroom
  instability patterns in two miscible fluids' core annular flows}}.
  \jt{Physics of Fluids}  \bvol{20}~(2),  \pg{24104}.

\bibitem[Driscoll \& Weideman(2014)]{Driscoll2014}
{\sc \au{Driscoll, Tobin~A} \& \au{Weideman, JAC}} \yr{2014}  \at{Optimal
  domain splitting for interpolation by chebyshev polynomials}.  \jt{SIAM
  Journal on Numerical Analysis}  \bvol{52}~(4),  \pg{1913--1927}.

\bibitem[Forliti {\em et~al.\/}(2005)Forliti, Tang \& Strykowski]{Forliti2005}
{\sc \au{Forliti, David~J}, \au{Tang, Brian~A} \& \au{Strykowski, Paul~J}}
  \yr{2005}  \at{An experimental investigation of planar countercurrent
  turbulent shear layers}.  \jt{Journal of Fluid Mechanics}  \bvol{530},
  \pg{241--264}.

\bibitem[Govindarajan(2004)]{Govindarajan2004}
{\sc \au{Govindarajan, Rama}} \yr{2004}  \at{{Effect of miscibility on the
  linear instability of two-fluid channel flow}}.  \jt{International Journal of
  Multiphase Flow}  \bvol{30}~(10),  \pg{1177--1192}.

\bibitem[Govindarajan \& Sahu(2014)]{Govindarajan2014b}
{\sc \au{Govindarajan, Rama} \& \au{Sahu, Kirti~Chandra}} \yr{2014}
  \at{{Instabilities in viscosity-stratified flow}}.  \jt{Annual Review of
  Fluid Mechanics}  \bvol{46}~(1),  \pg{331--353}.

\bibitem[Goyal \& Meiburg(2006)]{goyal2006miscible}
{\sc \au{Goyal, N} \& \au{Meiburg, E}} \yr{2006}  \at{Miscible displacements in
  hele-shaw cells: two-dimensional base states and their linear stability}.
  \jt{Journal of Fluid Mechanics}  \bvol{558},  \pg{329--355}.

\bibitem[Hallberg \& Strykowski(2006)]{Hallberg2006}
{\sc \au{Hallberg, MP} \& \au{Strykowski, PJ}} \yr{2006}  \at{On the
  universality of global modes in low-density axisymmetric jets}.  \jt{Journal
  of Fluid Mechanics}  \bvol{569},  \pg{493--507}.

\bibitem[Hickox(1971)]{Hickox1971}
{\sc \au{Hickox, Charles~E}} \yr{1971}  \at{Instability due to viscosity and
  density stratification in axisymmetric pipe flow}.  \jt{The physics of
  Fluids}  \bvol{14}~(2),  \pg{251--262}.

\bibitem[Hinch(1984)]{Hinch1984}
{\sc \au{Hinch, E~John}} \yr{1984}  \at{A note on the mechanism of the
  instability at the interface between two shearing fluids}.  \jt{Journal of
  Fluid Mechanics}  \bvol{144},  \pg{463--465}.

\bibitem[Ho \& Huerre(1984)]{Ho1984a}
{\sc \au{Ho, C-M} \& \au{Huerre, P}} \yr{1984}  \at{{Perturbed Free Shear
  Layers}}.  \jt{Annual Review of Fluid Mechanics}  \bvol{16}~(1),
  \pg{365--422}.

\bibitem[Hooper(1987)]{Hooper1985}
{\sc \au{Hooper, A~P}} \yr{1987}  \at{{Shear-flow instability due to a wall and
  a viscosity discontinuity at the interface}}.  \jt{Journal of Fluid
  Mechanics}  \bvol{179},  \pg{201}.

\bibitem[Hooper \& Boyd(1983)]{Hooper1983a}
{\sc \au{Hooper, A~P} \& \au{Boyd, W G~C}} \yr{1983}  \at{{Shear flow
  instability at the interface between two viscous fluids}}.  \jt{J. Fluid
  Mech.}  \bvol{128},  \pg{507--528}.

\bibitem[Hu \& Joseph(1989)]{Hu1989}
{\sc \au{Hu, Howard~H} \& \au{Joseph, Daniel~D}} \yr{1989}  \at{Lubricated
  pipelining: stability of core-annular flow. part 2}.  \jt{Journal of fluid
  mechanics}  \bvol{205},  \pg{359--396}.

\bibitem[Huerre \& Monkewitz(1985)]{Huerre1985}
{\sc \au{Huerre, P} \& \au{Monkewitz, P~A}} \yr{1985}  \at{{Absolute and
  convective instabilities in free shear layers}}.  \jt{Journal of Fluid
  Mechanics}  \bvol{159},  \pg{151}.

\bibitem[Huerre \& Monkewitz(1990)]{Huerre1990}
{\sc \au{Huerre, Patrick} \& \au{Monkewitz, Peter~A}} \yr{1990}  \at{Local and
  global instabilities in spatially developing flows}.  \jt{Annual review of
  fluid mechanics}  \bvol{22}~(1),  \pg{473--537}.

\bibitem[Joseph {\em et~al.\/}(1984)Joseph, Renardy \& Renardy]{Joseph1984}
{\sc \au{Joseph, Daniel~D}, \au{Renardy, Michael} \& \au{Renardy, Yuriko}}
  \yr{1984}  \at{Instability of the flow of two immiscible liquids with
  different viscosities in a pipe}.  \jt{Journal of Fluid Mechanics}
  \bvol{141},  \pg{309--317}.

\bibitem[Juniper(2006)]{Juniper2006a}
{\sc \au{Juniper, Matthew~P}} \yr{2006}  \at{{The effect of confinement on the
  stability of two-dimensional shear flows}}.  \jt{Journal of Fluid Mechanics}
  \bvol{565},  \pg{171}.

\bibitem[Khorrami {\em et~al.\/}(1989)Khorrami, Malik \&
  Ash]{khorrami1989application}
{\sc \au{Khorrami, Mehdi~R}, \au{Malik, Mujeeb~R} \& \au{Ash, Robert~L}}
  \yr{1989}  \at{Application of spectral collocation techniques to the
  stability of swirling flows}.  \jt{Journal of Computational Physics}
  \bvol{81}~(1),  \pg{206--229}.

\bibitem[Lesshafft \& Huerre(2007)]{Lesshafft2007}
{\sc \au{Lesshafft, Lutz} \& \au{Huerre, Patrick}} \yr{2007}  \at{Linear
  impulse response in hot round jets}.  \jt{Physics of Fluids}  \bvol{19}~(2),
  \pg{024102}.

\bibitem[Matas {\em et~al.\/}(2011)Matas, Marty \& Cartellier]{Matas2011}
{\sc \au{Matas, Jean-Philippe}, \au{Marty, Sylvain} \& \au{Cartellier, Alain}}
  \yr{2011}  \at{Experimental and analytical study of the shear instability of
  a gas-liquid mixing layer}.  \jt{Physics of fluids}  \bvol{23}~(9),
  \pg{094112}.

\bibitem[Mattingly \& Chang(1974)]{Mattingly1974}
{\sc \au{Mattingly, GE} \& \au{Chang, CC}} \yr{1974}  \at{Unstable waves on an
  axisymmetric jet column}.  \jt{Journal of Fluid Mechanics}  \bvol{65}~(3),
  \pg{541--560}.

\bibitem[Michalke(1984)]{Michalke1984}
{\sc \au{Michalke, Alfons}} \yr{1984}  \at{{Survey on jet instability theory}}.
   \jt{Progress in Aerospace Sciences}  \bvol{21},  \pg{159--199}.

\bibitem[Mohammadi \& Smits(2017)]{Mohammadi2017}
{\sc \au{Mohammadi, Alireza} \& \au{Smits, Alexander~J}} \yr{2017}  \at{Linear
  stability of two-layer couette flows}.  \jt{Journal of Fluid Mechanics}
  \bvol{826},  \pg{128--157}.

\bibitem[Monkewitz(1988)]{Monkewitz1988}
{\sc \au{Monkewitz, Peter~A}} \yr{1988}  \at{{The absolute and convective
  nature of instability in two-dimensional wakes at low Reynolds numbers}}.
  \jt{Physics of Fluids}  \bvol{31}~(5),  \pg{999}.

\bibitem[Monkewitz {\em et~al.\/}(1990)Monkewitz, Bechert, Barsikow \&
  Lehmann]{Monkewitz1990a}
{\sc \au{Monkewitz, Peter~A}, \au{Bechert, Dietrich~W}, \au{Barsikow, Bernd} \&
  \au{Lehmann, Bernhard}} \yr{1990}  \at{{Self-excited oscillations and mixing
  in a heated round jet}}.  \jt{Journal of Fluid Mechanics}  \bvol{213},
  \pg{611}.

\bibitem[Morris(1976)]{Morris1976}
{\sc \au{Morris, Philip J P J~J}} \yr{1976}  \at{{The spatial viscous
  instability of axisymmetric jets}}.  \jt{Journal of Fluid Mechanics}
  \bvol{7}~(3 , Oct. 8, 1976),  \pg{511--529}.

\bibitem[Pathikonda {\em et~al.\/}(2021)Pathikonda, Usta, Ahmad, Khan, Gillis,
  Dhodapkar, Jain, Ranjan \& Aidun]{Pathikonda2021}
{\sc \au{Pathikonda, Gokul}, \au{Usta, Mustafa}, \au{Ahmad, Michael~C},
  \au{Khan, Irfan}, \au{Gillis, Paul}, \au{Dhodapkar, Shrikant}, \au{Jain,
  Pradeep}, \au{Ranjan, Devesh} \& \au{Aidun, Cyrus~K}} \yr{2021}  \at{Mixing
  behavior in a confined jet with disparate viscosity and implications for
  complex reactions}.  \jt{Chemical Engineering Journal}  \bvol{403},
  \pg{126300}.

\bibitem[Pier \& Huerre(2001)]{Pier2001}
{\sc \au{Pier, B} \& \au{Huerre, P}} \yr{2001}  \at{{Nonlinear self-sustained
  structures and fronts in spatially developing wake flows}}.  \jt{Journal of
  Fluid Mechanics}  \bvol{435},  \pg{145--174}.

\bibitem[Ranganathan \& Govindarajan(2001)]{Ranganathan2001}
{\sc \au{Ranganathan, Balaji~T} \& \au{Govindarajan, Rama}} \yr{2001}
  \at{{Stabilization and destabilization of channel flow by location of
  viscosity-stratified fluid layer}}.  \jt{Physics of Fluids}  \bvol{13}~(1),
  \pg{1--3}.

\bibitem[Raynal {\em et~al.\/}(1996)Raynal, Harion, Favre-Marinet, Binder,
  J-L., Favre-Marinet \& Binder]{Raynal1996}
{\sc \au{Raynal, L}, \au{Harion, J-L.~L}, \au{Favre-Marinet, M}, \au{Binder,
  G}, \au{J-L., Harion}, \au{Favre-Marinet, M} \& \au{Binder, G}} \yr{1996}
  \at{{The oscillatory instability of plane variable-density jets}}.
  \jt{Physics of Fluids}  \bvol{8}~(4),  \pg{993--1006}.

\bibitem[Sahu \& Govindarajan(2014)]{Sahu2014}
{\sc \au{Sahu, Kirti~Chandra} \& \au{Govindarajan, Rama}} \yr{2014}
  \at{{Instability of a free-shear layer in the vicinity of a
  viscosity-stratified layer}}.  \jt{Journal of Fluid Mechanics}  \bvol{752},
  \pg{626--648}.

\bibitem[Salin \& Talon(2019)]{Salin2019}
{\sc \au{Salin, D} \& \au{Talon, L}} \yr{2019}  \at{Revisiting the linear
  stability analysis and absolute--convective transition of two fluid core
  annular flow}.  \jt{Journal of Fluid Mechanics}  \bvol{865},  \pg{743--761}.

\bibitem[Selvam {\em et~al.\/}(2007{\natexlab{{\em a\/}}})Selvam, Merk,
  Govindarajan \& Meiburg]{selvam2007stability}
{\sc \au{Selvam, Balakrishnan}, \au{Merk, S}, \au{Govindarajan, Rama} \&
  \au{Meiburg, E}} \yr{2007{\natexlab{{\em a\/}}}}  \at{Stability of miscible
  core--annular flows with viscosity stratification}.  \jt{Journal of Fluid
  Mechanics}  \bvol{592},  \pg{23--49}.

\bibitem[Selvam {\em et~al.\/}(2007{\natexlab{{\em b\/}}})Selvam, Merk,
  Govindarajan \& Meiburg]{Selvam2007}
{\sc \au{Selvam, B}, \au{Merk, S}, \au{Govindarajan, Rama} \& \au{Meiburg, E}}
  \yr{2007{\natexlab{{\em b\/}}}}  \at{{Stability of miscible core–annular
  flows with viscosity stratification}}.  \jt{Journal of Fluid Mechanics}
  \bvol{592},  \pg{23--49}.

\bibitem[Selvam {\em et~al.\/}(2009)Selvam, Talon, Lesshafft \&
  Meiburg]{Selvam2009}
{\sc \au{Selvam, B}, \au{Talon, Laurent}, \au{Lesshafft, L} \& \au{Meiburg, E}}
  \yr{2009}  \at{Convective/absolute instability in miscible core-annular flow.
  part 2. numerical simulations and nonlinear global modes}.  \jt{Journal of
  Fluid Mechanics}  \bvol{618},  \pg{323--348}.

\bibitem[Sreenivasan {\em et~al.\/}(1989)Sreenivasan, Raghu \&
  Kyle]{Sreenivasan1989}
{\sc \au{Sreenivasan, K~R}, \au{Raghu, S} \& \au{Kyle, D}} \yr{1989}
  \at{{Absolute instability in variable density round jets}}.  \jt{Experiments
  in Fluids}  \bvol{7}~(5),  \pg{309--317}.

\bibitem[Srinivasan {\em et~al.\/}(2010)Srinivasan, Hallberg \&
  Strykowski]{Srinivasan2010}
{\sc \au{Srinivasan, V}, \au{Hallberg, M~P} \& \au{Strykowski, P~J}} \yr{2010}
  \at{{Viscous linear stability of axisymmetric low-density jets: Parameters
  influencing absolute instability}}.  \jt{Physics of Fluids}  \bvol{22}~(2),
  \pg{24103}.

\bibitem[Srinivasan {\em et~al.\/}(2023)Srinivasan, Tan, Wright \&
  Dhotre]{Srinivasan2023}
{\sc \au{Srinivasan, Vinod}, \au{Tan, Xijun}, \au{Wright, Ian} \& \au{Dhotre,
  Akash}} \yr{2023} Global instabilities and mode transitions in a low
  viscosity jet emerging into a high viscosity medium.

\bibitem[Strykowski \& Niccum(1991)]{Strykowski1991a}
{\sc \au{Strykowski, P~J} \& \au{Niccum, D~L}} \yr{1991}  \at{{The stability of
  countercurrent mixing layers in circular jets}}.  \jt{Journal of Fluid
  Mechanics}  \bvol{227},  \pg{309--343}.

\bibitem[Tan \& Homsy(1986)]{tan1986stability}
{\sc \au{Tan, CT} \& \au{Homsy, GM0608}} \yr{1986}  \at{Stability of miscible
  displacements in porous media: Rectilinear flow}.  \jt{The Physics of fluids}
   \bvol{29}~(11),  \pg{3549--3556}.

\bibitem[Tomotika(1935)]{Tomotika1935}
{\sc \au{Tomotika, S}} \yr{1935}  \at{On the instability of a cylindrical
  thread of a viscous liquid surrounded by another viscous fluid}.
  \jt{Proceedings of the Royal Society of London. Series A-Mathematical and
  Physical Sciences}  \bvol{150}~(870),  \pg{322--337}.

\bibitem[Trefethen(2000)]{trefethen2000spectral}
{\sc \au{Trefethen, Lloyd~N}} \yr{2000} {\em Spectral methods in MATLAB\/}.
  \publ{SIAM}.

\bibitem[Valluri {\em et~al.\/}(2010)Valluri, N{\'{a}}raigh, Ding \&
  Spelt]{Valluri2010}
{\sc \au{Valluri, P}, \au{N{\'{a}}raigh, L O~{\'{O}}}, \au{Ding, H} \&
  \au{Spelt, P D M~M}} \yr{2010}  \at{{Linear and nonlinear spatio-temporal
  instability in laminar two-layer flows}}.  \jt{Journal of Fluid Mechanics}
  \bvol{656},  \pg{458--480}.

\bibitem[Yang {\em et~al.\/}(2021)Yang, Strykowski \& Srinivasan]{Yang2021}
{\sc \au{Yang, Jinwei}, \au{Strykowski, Paul~John} \& \au{Srinivasan, Vinod}}
  \yr{2021}  \at{{Effects of confinement on absolute and convective
  instabilities for momentum-driven counter-current shear layers}}.
  \jt{Physical Review Fluids}  \bvol{6}~(7),  \pg{073901}.

\bibitem[Yecko {\em et~al.\/}(2002)Yecko, Zaleski \& Fullana]{Yecko2002}
{\sc \au{Yecko, Philip}, \au{Zaleski, St{\'e}phane} \& \au{Fullana,
  Jose-Maria}} \yr{2002}  \at{Viscous modes in two-phase mixing layers}.
  \jt{Physics of Fluids}  \bvol{14}~(12),  \pg{4115--4122}.

\bibitem[Yiantsios \& Higgins(1988)]{Yiantsios1988}
{\sc \au{Yiantsios, Stergios~G} \& \au{Higgins, Brian~G}} \yr{1988}  \at{Linear
  stability of plane poiseuille flow of two superposed fluids}.  \jt{The
  Physics of fluids}  \bvol{31}~(11),  \pg{3225--3238}.

\bibitem[Yih(1967)]{Yih1967}
{\sc \au{Yih, C~S}} \yr{1967}  \at{{Instability due to viscosity
  stratification}}.  \jt{Journal of Fluid Mechanics}  \bvol{27},
  \pg{337--352}.

\end{thebibliography}


\section{Appendix}
\subsection{Appendix A: Semi-global Sharp Interface Capturing Technique}

In this section, we discuss some details of the numerical procedure used to solve the perturbation equations, starting with the need to go beyond the traditional collocation approach that employs a single mapping to transform the physical domain to [-1,1]. A mapping used in other low-density jet studies \citep{Lesshafft2007, Srinivasan2010} to transform from [0, $\infty$] to [-1,1] while clustering points near r=1, the location of the shear layer, yields only one points in the radial region [0.99 1.01] when N=250 polynomials are used. This is acceptable when dealing with relatively diffusive flows with low Sc $\sim 1$ are examined, as was the case in the above studies. High Sc flows produce interfaces that are not sharp but are quite thin relative to other scales, such as the jet diameter in the present study. In order to get at least 6 points (say) in this region, we need N= 8858 polynomials. 

An alternative approach is to split the domain into two subdomains with their adjacent boundary chosen near the point of (weak) discontinuity, and populate the subdomains with standard Chebyshev grids and implementation, which will automatically cluster points near the common boundary. This is a `semi-global' method in that function values at all points in the domain are not used to evaluate derivatives at any location; only values in the same subdomain are used. Nevertheless, this would be expected to preserve exponential convergence in each subdomain; however, the jumps in higher order derivatives at the interface effectively prevents  exponential convergence.

Recently, \citet{Aiton2018} proposed an alternative approach, which employs two overlapping subdomains, with the overlap region covering the region of sharp gradients. Since the overlapped region is near the boundary of both domains, high spatial resolution in point distribution can be achieved. Within each subdomain, a high-order representation of the function is possible.  \citet{Aiton2018} propose weighting the two different representations of the function in the overlapping region to come up with an interpolated value, which is accurate to high order. This is the scheme we follow, with equal weighting from both sides. 

The matrix formulation is outlined below. Suppose $D_1$ and $D_2$ denote the first and second-order derivative matrix operators for the standard Chebyshev method on domain $x=[-1,1]$, so that the velocity derivative is $\frac{dU}{dy}=DU$. Now suppose  the calculation domain is $[0,L]$, $L>1$. We split the domain $[0,L]$ into $[0,L_1]$, $[L_1,L_2]$ and $[L_2,L]$, respectively. For each domain, we can introduce the nonlinear mappings as follows.

On domain $r = [0,L_1]$, we use the following nonlinear mapping function 
\begin{equation}
r_1 = \frac{R_c L_1(1-x)}{2 R_c+L_1(1-x^2)},
\end{equation}
where $R_c$ is a free parameter which can be used to adjust the nodes distribution on this domain.

For domain $[L_1,L_2]$ and $[L_2,L]$, we use other two nonlinear mappings as follows,
\begin{equation}
r_2 = \frac{L_1-L_2}{2}x + \frac{L_1+L_2}{2},
\end{equation}
and 
\begin{equation}
r_3 = \frac{L(1-x)}{b+x} + L_2,
\end{equation}
where $b$ is also a free parameter using the adjust the nodes distribution.
The first order derivative matrix for the three domain can be derived via the chain rules as follows:
\begin{equation}
d_1 = -\frac{[2R_c+L_1(1-x^2)]^2}{R_c L_1 (2R_c+R_1(x-1)^2)}D,
\end{equation}

\begin{equation}
d_2 = \frac{2}{L_1+L_2}D,
\end{equation}

\begin{equation}
d3 = -\frac{(b+x)^2}{L(b+1)}D.
\end{equation}

From the matrices for each subdomain, we assemble the global matrix $\hat{D}$. If $u_1$, $u_2$ and $u_3$ are the velocity profiles for each domain, we can write:
\begin{equation}
\left[\begin{array}{l}
\frac{d u_{1}}{d r} \\
\frac{d u_{2}}{d r} \\
\frac{d u_{3}}{d r}
\end{array}\right]=\left[\begin{array}{lll}
d_{1} & 0 & 0 \\
0 & d_{2} & 0 \\
0 & 0 & d_{3}
\end{array}\right]\left[\begin{array}{l}
u_{1}(r) \\
u_{2}(r) \\
u_{3}(r)
\end{array}\right] 
\end{equation}
These lead to two points of overlap --- $r=L_1$ and $r=L_2$ which need to be reconciled in order to obtain a square global matrix. We  use the average values of last row of matrix $d_1$ and the first row of matrix $d_2$ to replace the row corresponding to $r=L_1$, the same as $r=L_2$. 

\begin{figure}
\centering
\begin{subfigure}{0.49\textwidth}
\includegraphics[width= \textwidth]{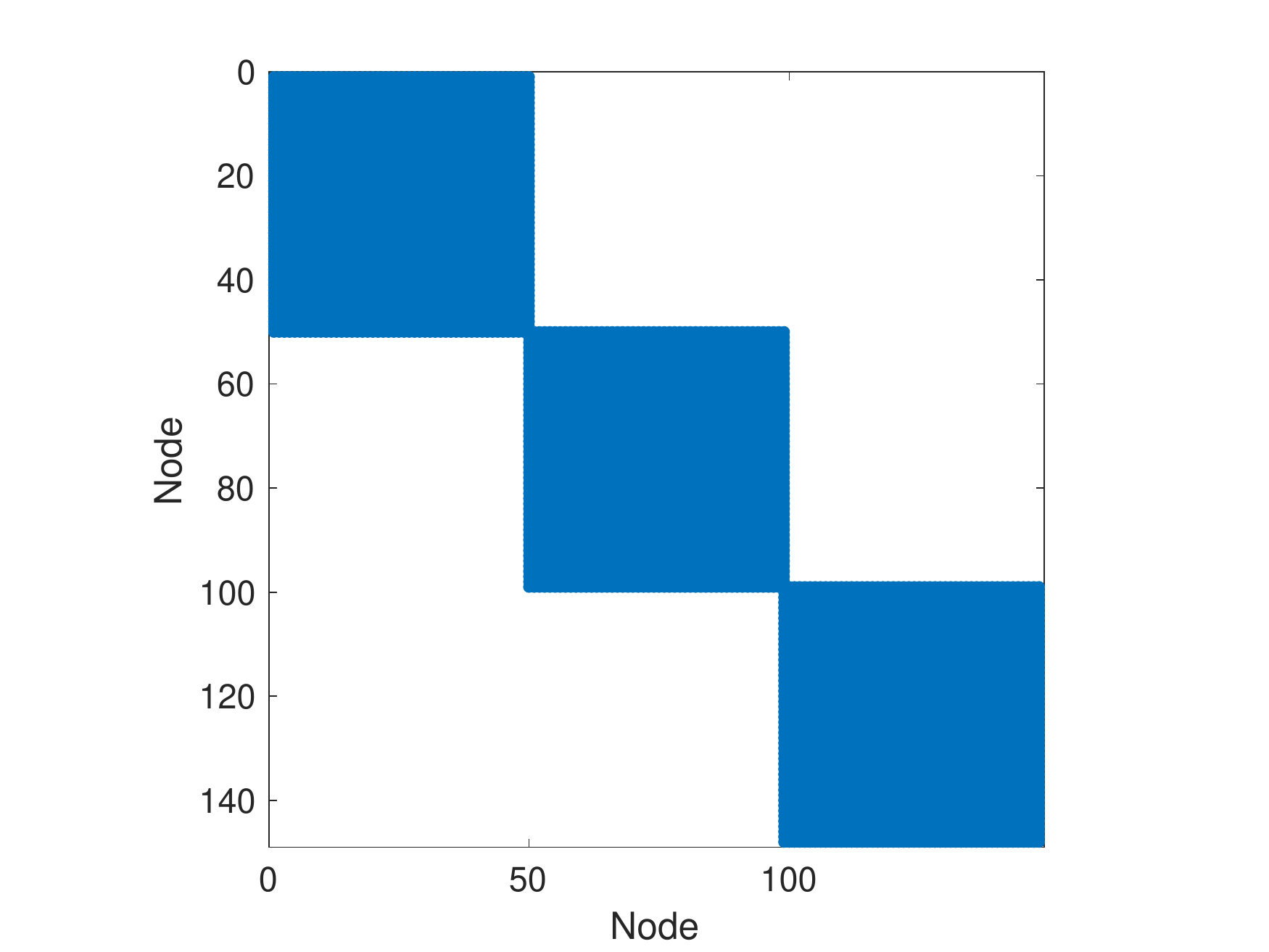}
\caption{}
\end{subfigure}
\begin{subfigure}{0.49\textwidth}
\includegraphics[width= \textwidth]{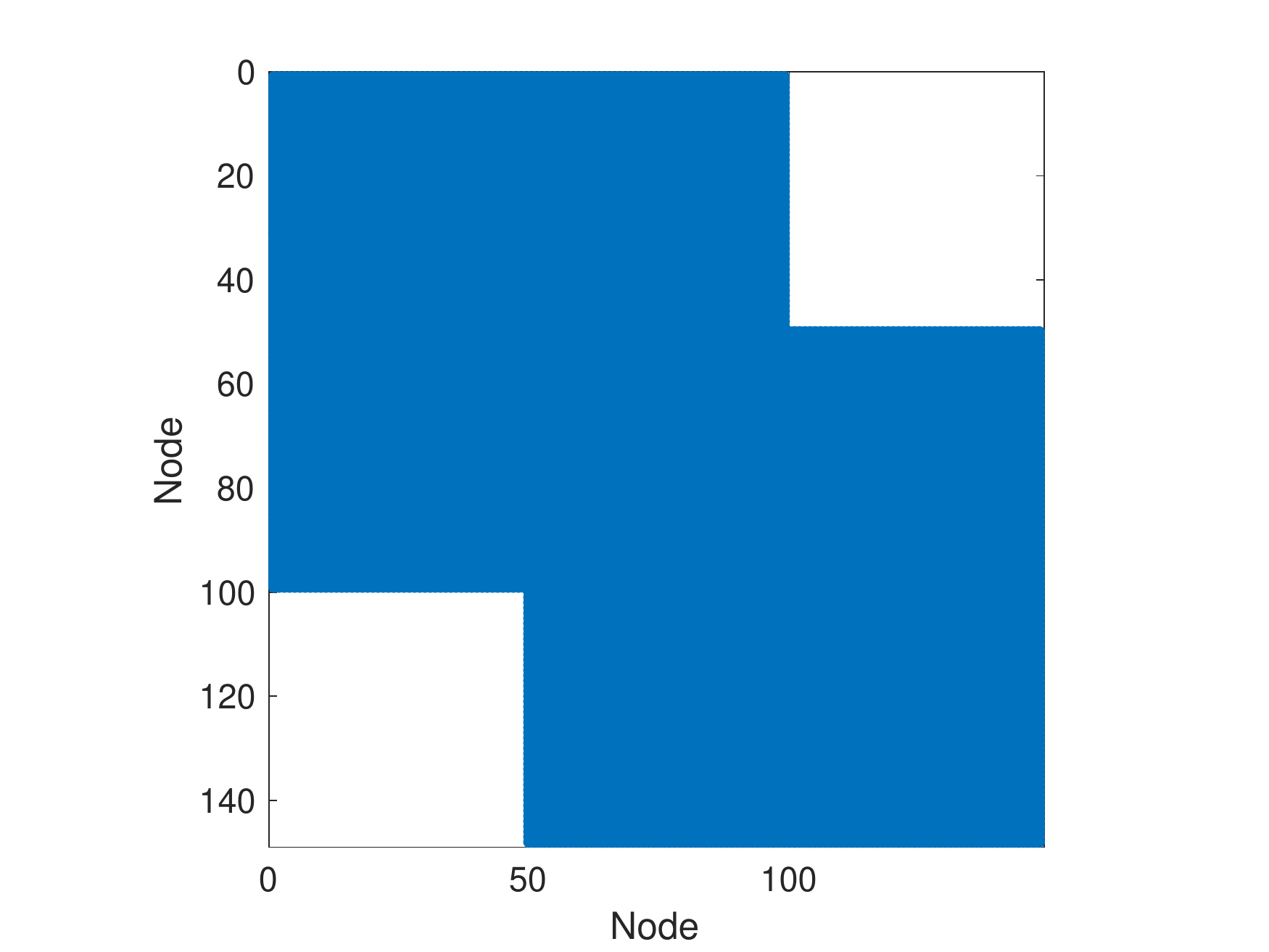}
\caption{}
\end{subfigure}
\caption{Non-zero nodes distribution of the semi-global method: (1) The first order derivative matrix and (b) the second order derivative matrix.}
\label{d1d2}
\end{figure}

\begin{figure}
\centerline{
\includegraphics[width= 0.5\textwidth]{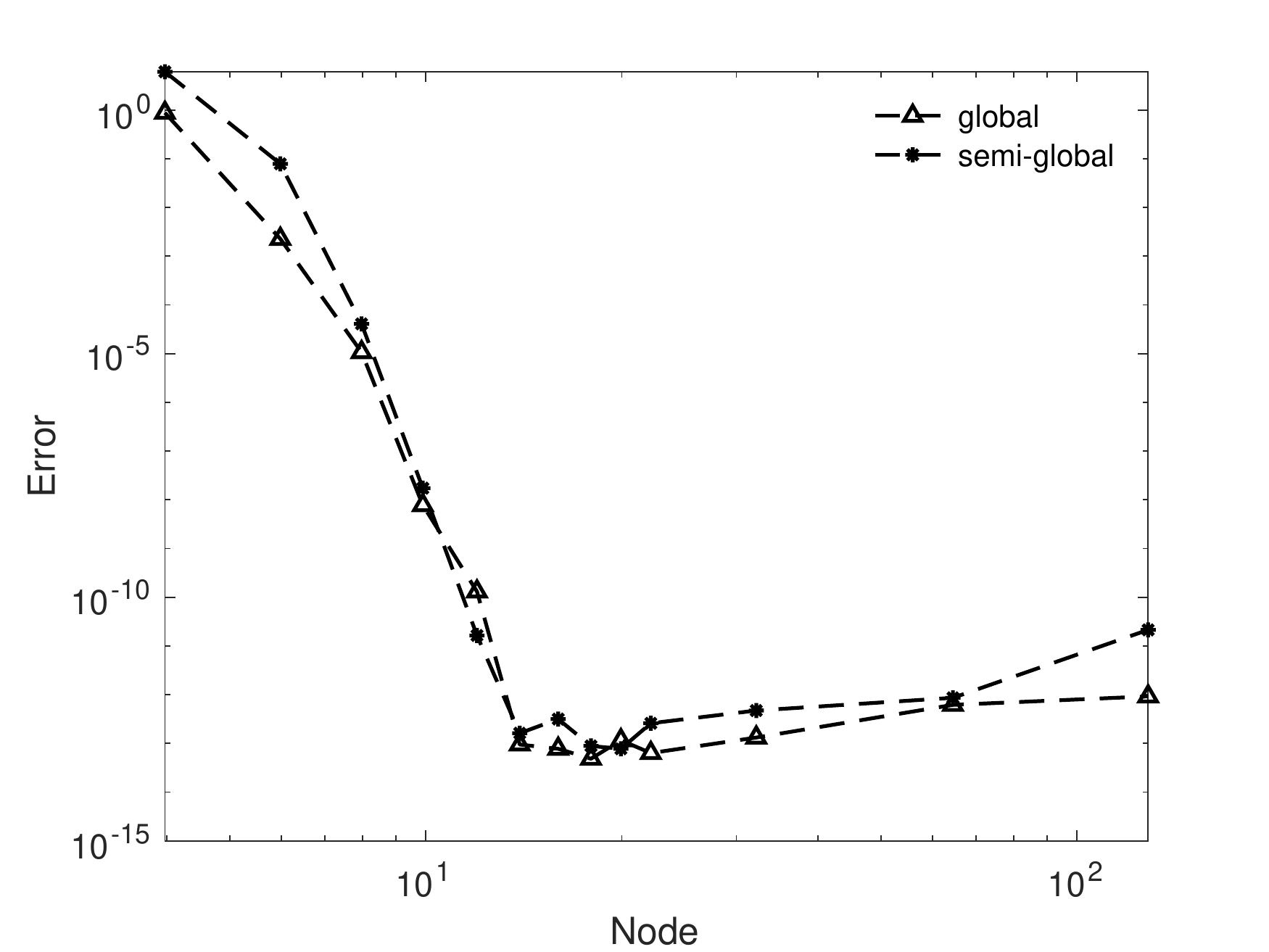}}
\caption{Convergence rate comparison of the global and semi-global Chebyshev spectral methods.}
\label{convergence}
\end{figure}

In Figure~(\ref{convergence}), we show the convergence comparison between a global Chebyshev spectral method and the semi-global Chebyshev spectral method for the boundary value problem tested by \citet{Aiton2018}.

\begin{equation}
y''+\lambda^2 y = 0
\end{equation}
where the eigenvalue $\lambda=n\pi$ and $n = 1,2,3 ...$. It indicates that they have nearly the same order of convergence rate. 

\subsection{Appendix B: Base state velocity profile and the numerical similarity solution for variable viscosity jet.}

Here we state the nonlinear ordinary differential equation from the steady-state, boundary layer equations given by
\begin{equation}
\begin{gathered}
u \frac{\partial u}{\partial x}+v \frac{\partial u}{\partial r}=\frac{\partial}{\partial r}\left(v \frac{\partial u}{\partial r}\right) \\
\frac{\partial u}{\partial x}+\frac{\partial v}{\partial r}=0
\end{gathered}
\end{equation}
where $x$ and $r$ are the original horizontal and transverse directions, respectively. Then one can transform the equations to a single-dimensionless variable, $\eta$, given by
\begin{equation}
\eta=\frac{y}{\sqrt{Re_x}}
\end{equation}
where $Re_x$ is the Reynolds number defined as $\frac{\mu_{0}x}{\rho_0 U_{0}}$. Here $\mu_0$ and $\rho_0$ are the viscosity and density at the center of the jet, respectively. Then once obtains
\begin{equation}
\label{ode_s}
f(\eta)^{\prime \prime \prime}+\frac{1}{2 \mu_0} f(\eta) f(\eta)^{\prime \prime}+\frac{\mu_0^{\prime}}{\mu_0} f(\eta)^{\prime \prime}=0
\end{equation}

where $f' = U(r)$. The boundary conditions used to solve equation~(\ref{ode_s}) are as follows if we shift down the coordinate system by $R_0$.
\begin{equation}
\begin{gathered}
f=0, \quad f^{\prime}= 1 \quad \text { at } y=-R_0, \\
f^{\prime}= 0 \quad \text { at } y= 10R_0,
\end{gathered}
\end{equation}
where we choose a large ratio of 10 to make sure the results are independent.
Then we use modified shooting methods to get the numerical solution with the above boundary condition. We first split the domain into two components and integrate them to get solutions for each part. Then, we balance the derivative of the velocity at the domain-splitting interface to get the numerical solutions. 
\end{document}